\documentclass{article}
\usepackage{here, subfigure, bm, url, amsmath, amssymb, authblk, rotating}

\def \figref  #1{Figure \ref{#1}}
\def \tabref  #1{Table \ref{#1}}
\def \eqref  #1{Equation (\ref{#1})}
\def \secref  #1{Section \ref{#1}}
\newcommand{\vect}[1]{\mbox{\boldmath $#1$}}
\newcommand{\mat}[1]{\mbox{{\boldmath $#1$}}}

\newcommand{\argmax}{\mathop{\rm argmax}\limits}

\begin{document}
\title{Causal Patterns:
Extraction of multiple causal relationships by
Mixture of Probabilistic Partial Canonical Correlation Analysis}

\author[1]{Hiroki Mori \thanks{mori@idr.ias.sci.waseda.ac.jp}}
\author[2]{Keisuke Kawano \thanks{kskkwn@gmail.com}}
\author[3]{Hiroki Yokoyama \thanks{hiroki.yokoyama@okadanet.org}}
\affil[1]{Future robotics organization,
    School of Fundamental Science and Engineering,
    Waseda University.}
\affil[2]{Department of Adaptive Machine Systems,
    Graduate School of Engineering,
    Osaka University.}
\affil[3]{Tamagawa University Brain Science Institute}
\date{}

\maketitle

\begin{abstract}
	In this paper, we propose a mixture of probabilistic partial canonical correlation analysis (MPPCCA)
	that extracts the \textit{Causal Patterns} from two multivariate time series.
	Causal patterns refer to the signal patterns within interactions of two elements
	having multiple types of mutually causal relationships, 
	rather than a mixture of simultaneous correlations or
	the absence of presence of a causal relationship between the elements.
	In multivariate statistics, partial canonical correlation analysis (PCCA) evaluates the correlation between two multivariates
	after subtracting the effect of the third multivariate.
	PCCA can calculate the Granger Causality Index (which tests whether a time-series can be predicted from another time-series), but is not applicable 
	to data containing multiple partial canonical correlations.
	After introducing the MPPCCA,
	we propose an expectation-maxmization (EM) algorithm that estimates the parameters and latent variables of the MPPCCA.
	The MPPCCA is expected to extract multiple partial canonical correlations from data series without any supervised signals
	to split the data as clusters.
	The method was then evaluated in synthetic data experiments.
	In the synthetic dataset, our method estimated the multiple partial canonical correlations more accurately than the existing method.
	To determine the types of patterns detectable by the method, experiments were also conducted on real datasets. The method estimated the communication patterns In motion-capture data.
	The MPPCCA is applicable to various type of signals such as brain signals,
	human communication and nonlinear complex multibody systems.
\end{abstract}

\maketitle

\vspace{0mm}
\section{Introduction}
\vspace{0mm}
Many everyday events are not causally related,
except in specific cases.
In human communication, certain patterns of body movements
such as speech (which combines various sound patterns by movements of a mouth and a throat),
and sign and body languages,
elicit respective responses from the opponent.
When one participant moves the right hand slowly forward,
the opponent copies the action to execute a handshake.
Alternatively, when one participant moves the right fist forward rapidly,
the opponent moves backward to avoid the blow.
Idle hands elicit no reaction from others.
Such changes of causality
depend on the interaction patterns within the time series.
In this paper, we refer to such interaction patterns as ``\textit{causal patterns }''.
In this context, a communication is a mosaic of multiple causal patterns.
However, can statistical methods extract causal patterns from data?

Among the statistical methods for finding causal relationships　within two time series,
there are Granger Causality (GC) and Transfer Entropy (TE), which are based on
prediction errors and Kullback-Leibler divergence of
conditional probabilities, respectively.
The two methods are equivalent when the data exhibit a Gaussian distribution
\cite{barnett2009granger}.
To analyze the mixed causal relationships in time series data,
we can split the data
into multiple categories or situations
based on the experimental or observational conditions, and apply a causality measure to the split data.
However, this type of analysis assumes
that the time series in each category has
a consistent stationary process.
Thus, when the time series inherently switches among different causal relationships at
different timings,
these methods cannot detect the relationships
because they assume only one type of dynamics between the target elements in the time-series. 

In brain science,
causality analyses determine the \textit{functional connectivity}
among different situations or tasks.
According to these analyses, the functional connectivities of individual brains depend on the conditions,
such as resting state, sleep stages
and cognitive tasks \cite{sato2010analyzing}.
If time-series data
could be separated based on their multiple causal relationships without any reference signals,
the causal patterns could be extracted
from the data streams used in the communication of the elements.
In human communications,
the causal patterns might constitute the
words in sign language (body movement patterns) and
spoken language (sound patterns).
Likewise, in a brain science context,
the causal patterns in the brain signals obtained by functional magnetic resonance imaging or electroencephalography might embody the
functional connectivities
among different information processes or
different states (such as sleep/wake states or cognitive tasks).

GC can be calculated by partial canonical correlation analysis (PCCA) with embedded vectors of time series \cite{mukuta2014probabilistic}.
If the GC detects multiple causal relationships between two time series,
the data can be represented by multiple PCCA models.
Here
we propose a mixture of probabilistic partial canonical correlation analyses (MPPCCA),
a GC-based approach that extracts the causal patterns from time-series data.
The method and its learning algorithm are examined
on synthetic data generated by a time-series model
and on real data of movements between two persons
measured by a motion capture system.

\vspace{0mm}
\section{Previous works}
\vspace{0mm}
Before describing our MPPCCA method,
we introduce four previous approaches; 
PCCA \cite{rao1969partial},
a probabilistic interpretation of PCCA (PPCCA) \cite{mukuta2014probabilistic},
the calculation of GC by PPCCA \cite{mukuta2014probabilistic},
and combined probabilistic models and EM algorithm \cite{bishop:2006:PRML}.

\vspace{0mm}
\subsection{Partial canonical correlation analysis}
\vspace{0mm}
PCCA \cite{rao1969partial} performs a canonical correlation analysis (CCA) between two multivariate variables
after eliminating the influence of the third multivariate variable.
CCA is widely used for calculating correlations between two multivariate variables.
The method seeks the linear transformations from two original variables
to the spaces exhibiting the highest correlation between the two variables.
In PCCA, 
the objective variables predicted by the third multivariate variables
are subtracted from the objective variables before computing the CCA.

Let us consider the partial canonical correlation of
 $\vect{y}^{(1)} = (y_1^{(1)},y_2^{(1)},...,y_{d_1}^{(1)})^T  \in \mathbb{R}^{d_1}$ and
 $\vect{y}^{(2)} = (y_1^{(2)},y_2^{(2)},...,y_{d_2}^{(2)})^T  \in \mathbb{R}^{d_2}$
after eliminating the influence from
$\vect{x} = (x_1,x_2,...,x_d)^T  \in \mathbb{R}^{d_x}$.
To determine the influence of $\vect{x}$,
PCCA calculates two linear regressions; one from
$\vect{x}$ to $\vect{y}^{(1)}$, the other from $\vect{x}$ to $\vect{y}^{(2)}$. The regression equations are given by Eqs. (1) and (2), respectively.
\begin{eqnarray}
	\vect{y}^{(1)} &=  \mat{A}^{(1)}\vect{x} + \vect{e}^{(1)}\\
	\vect{y}^{(2)} &=  \mat{A}^{(2)}\vect{x} + \vect{e}^{(2)}.
 \end{eqnarray}
To minimize the errors $\vect{e}^{(1)}$ and $\vect{e}^{(2)}$,
$\mat{A}^{(1)}$ and $\mat{A}^{(2)}$ are respectively solved as
\begin{eqnarray}
	\mat{A}^{(1)} &= \Sigma_{1x}\Sigma_{xx}^{-1}\\
	\mat{A}^{(2)} &= \Sigma_{2x}\Sigma_{xx}^{-1} ,
\end{eqnarray}
where $\Sigma_{1x}\in \mathbb{R}^{d_1 \times d_x}$ 
and $\Sigma_{2x}\in \mathbb{R}^{d_2 \times d_x}$ 
are the covariance matrices between
$\vect{y}^{(1)}$ and $\vect{x}$
and between $\vect{y}^{(2)}$  and $\vect{x}$, respectively.
$\Sigma_{xx}\in \mathbb{R}^{d_x \times d_x}$ is
the covariance matrix of $\vect{x}$.
After eliminating the influence of $\vect{x}$, the multivariate variables $\vect{y}^{(1)}$ and $\vect{y}^{(2)}$ transform to $\hat{\vect{y}^{(1)}}$ and $\hat{\vect{y}^{(2)}}$, respectively:
\begin{eqnarray}
	\vect{\hat{y}}^{(1)} &=  \vect{y}^{(1)} - \mat{\hat{A}}^{(1)}\vect{x}\\
	\vect{\hat{y}}^{(2)} &=  \vect{y}^{(2)} - \mat{\hat{A}}^{(2)}\vect{x}.
\end{eqnarray}
 The partial canonical correlation defines the canonical correlation between
 $\vect{\hat{y}}^{(1)}$ and $\vect{\hat{y}}^{(2)}$, and is solved by the generalized eigenvalue problem as follows
\cite{yamashita2011causal}.
\begin{eqnarray}
	\vect{0} &=& (\mat{\Sigma}_{12|x}^T \mat{\Sigma}_{11|x}^{-1} \mat{\Sigma}_{12|x} - \rho^2\mat{\Sigma}_{22|x}) \vect{u}^{(2)}  \\
	\vect{0} &=& (\mat{\Sigma}_{21|x}^T \mat{\Sigma}_{22|x}^{-1} \mat{\Sigma}_{21|x} - \rho^2\mat{\Sigma}_{11|x}) \vect{u}^{(1)} \\
	\mat{\Sigma}_{12|x} &=& \mat{\Sigma}_{1x}-\mat{\Sigma}_{1x}\mat{\Sigma}_{xx}^{-1}\mat{\Sigma}_{x2} \\
	\mat{\Sigma}_{21|x} &=& \mat{\Sigma}_{2x}-\mat{\Sigma}_{2x}\mat{\Sigma}_{xx}^{-1}\mat{\Sigma}_{x1}.
	\label{eq:pcca}
\end{eqnarray}
$\rho$ is the partial correlation coefficient, which represents
the strength of the correlation between $\vect{\hat{y}}^{(1)}$ and $\vect{\hat{y}}^{(2)}$.

Mukuta and Harada \cite{mukuta2014probabilistic} proposed
PPCCA as a generative model of causal relationships.
we introduce PPCCA as a part of the formulation of the MPPCCA
\secref{sec:generative_model}.

\vspace{0mm}
\subsection{Granger causality calculated by PCCA}
\vspace{0mm}
The GC index can be calculated by PCCA \cite{fujita2010identification,sato2010analyzing}.
The CG, which represents the causal relationship between two time series,
is commonly applied in economics \cite{stock1999business}
and neuroscience
\cite{roebroeck2005mapping}
analyses.

Given two time series $x$ and $y$,
the GC from $y$ to $x$
is defined as the ratio of two prediction errors:
(1) The prediction of the current $\vect{Y}$ from the past information of $\vect{Y}$, and
(2) The prediction of the current $\vect{Y}$ from the past information of both $\vect{Y}$ and $\vect{X}$.
This method predicts the current state from past information by linear regression, as formulated below.
\begin{eqnarray}
	\vect{x}_t &=& \mat{A}^T \vect{X}_{t-1}^{(m)} + \epsilon_{x_t|x_{t-1}^{(m)},t}
	\label{eq:granger_causality1} \\
	\vect{x}_t &=& \mat{B}^T \vect{X}_{t-1}^{(m)} + \mat{C}^T \vect{Y}_{t-1}^{(m)}
		+ \epsilon_{x_t|x_{t-1}^{(m)}y_{t-1}^{(m)},t}
	\label{eq:granger_causality2}
\end{eqnarray}
where $\mat{X}=(\vect{x}_1, \vect{x}_2,...,\vect{x}_t,...,\vect{x}_T)^T$, with
$\vect{x}_t=(x_1,x_2,...,x_{d_x})_t^T \in R^{d_x}$.
Similarly, $\mat{Y}=(\vect{x}_1, \vect{x}_2,...,\vect{x}_t,...,\vect{Y}_T)^T$, with
$\vect{y}_t=(y_1,y_2,...,y_{d_y})_t^T\in R^{d_y}$.
The embedding vectors $\vect{X}_{t-1}^{(m)}$ and $\vect{Y}_{t-1}^{(m)}$ are defined as
\begin{eqnarray}
	\vect{X}_{t-1}^{(m)} &= (\vect{X}_{t-1}^T \vect{X}_{t-2}^T ... \vect{X}_{t-m}^T)^T \in R^{md_x}\\
	\vect{Y}_{t-1}^{(m)} &= (\vect{Y}_{t-1}^T \vect{Y}_{t-2}^T ... \vect{Y}_{t-m}^T)^T \in R^{md_y}.
\end{eqnarray}
The embedding vector $\vect{X}_{t-1}^{(m)}$ is the time series
of $\vect{x}$ from $t-m$ to $t-1$. The prediction coefficients are given by
$A\in \mathbb{R}^{md_x\times d_x}$, 
$B\in \mathbb{R}^{md_x\times d_x}$ and
$C\in \mathbb{R}^{md_y\times d_x}$.
In \eqref{eq:granger_causality1}, the current state is predicted from the self-dynamics of $\vect{x}$ alone; in
\eqref{eq:granger_causality2}, it is predicted from the self-dynamics of $\vect{x}$
and the external input $\vect{y}$.
The GC is then defined by 
\begin{eqnarray}
G_{y \rightarrow x} = \ln \frac{\mathrm{tr}(\Sigma_{x_tx_t|x_{t-k}^{(m)}})}{\mathrm{tr}(\Sigma_{x_tx_t|x_{t-k}^{(m)}y_{t-k}^{(m)}})} \nonumber
\end{eqnarray}
where $\mathrm{tr}(\cdot)$ is the trace of the matrix, and
$\Sigma_{x_tx_t|x_{t-k}^{(m)}}$ and $\Sigma_{x_tx_t|x_{t-k}^{(m)}y_{t-k}^{(m)}}$
are the covariance matrices of
$\epsilon_{x_t|x_{t-1}^{(m)},t}$ and
$\epsilon_{x_t|x_{t-1}^{(m)}y_{t-1}^{(m)}, t}$, respectively \cite{ladroue2009beyond}.
Whether or not it uses the past information of the causative side, GC improves the predictability of future effects.

A PCCA formulation of GC is given in \cite{yamashita2011causal}.
Denoting the two target multivariate variables of the PCCA are denoted by $\vect{X}_{t}$ and $\vect{Y}_{t-1}^{(m)}$,
and the third multivariate variable (whose influence is to be eliminated from the target variables)
by $\vect{X}_{t-1}^{(m)}$, the GC is solved by the following generalized eigenvalue problem based on PCCA.
\begin{eqnarray}
	\begin{aligned}
		&\vect{0} = \left(\mat{\Sigma}_{x_t y_{t-k}^{(m)}|x_{t-k}^{(m)}}^T
		\mat{\Sigma}_{x_t x_t|x_{t-k}^{(m)}}^{-1} \mat{\Sigma}_{x_ty_{t-k}^{(m)}|x_{t-k}^{(m)}} \right. \nonumber \\
			& \qquad \qquad \qquad \qquad\left. - \rho^2\mat{\Sigma}_{y_{t-k}^{(m)}y_{t-k}^{(m)}|x_{t-k}^{(m)}}\right) \vect{a}\\
		&\mat{\Sigma}_{ab|c} = \mat{\Sigma}_{ac}-\mat{\Sigma}_{ac}\mat{\Sigma}_{cc}^{-1}\mat{\Sigma}_{cb}.
	\end{aligned}
\end{eqnarray}
The GC index is then defined by
\begin{eqnarray}
	G_{y \rightarrow x} &=& \frac{1}{2} \log_2 \frac{1}{1-\rho_1^2}.
	\label{eq:PCCA_granger_causality}
\end{eqnarray}
The larger the eigenvalue, the larger the GC index.
$\rho_1$ represents the maximum value of eigenvalues.

\vspace{0mm}
\subsection{Mixture of probabilistic models}
\vspace{0mm}
A complex probabilistic model
can be constructed by combining multiple probabilistic models with latent variables.
The EM algorithm is a maximum likelihood method
that estimates the latent variables and the model parameters
from observed samples.
The expectation and maximization steps (E-step and
M-step, respectively) are executed sequentially.
The E-step estimates the latent variables using the current parameter guesses of each model.
The M-step then estimates the parameters of each model
by maximizing the likelihood using the current latent variables.
The two steps are iterated
until the estimation converges.

In the next section,
we combine PPCCA with the EM algorithm.

\vspace{0mm}
\section{Formulation of mixed probabilistic partial canonical correlation analysis}
\vspace{0mm}
In this section,
we propose a mixture of probabilistic partial canonical correlation analysis (MPPCCA)
and an estimation method for the parameters and latent variables. The former is a generative model and the latter is based on the EM algorithm.

\vspace{0mm}
\subsection{Generative model}
\label{sec:generative_model}
\begin{figure}[tbp]
	\begin{center}
		\includegraphics[width=0.5\columnwidth]{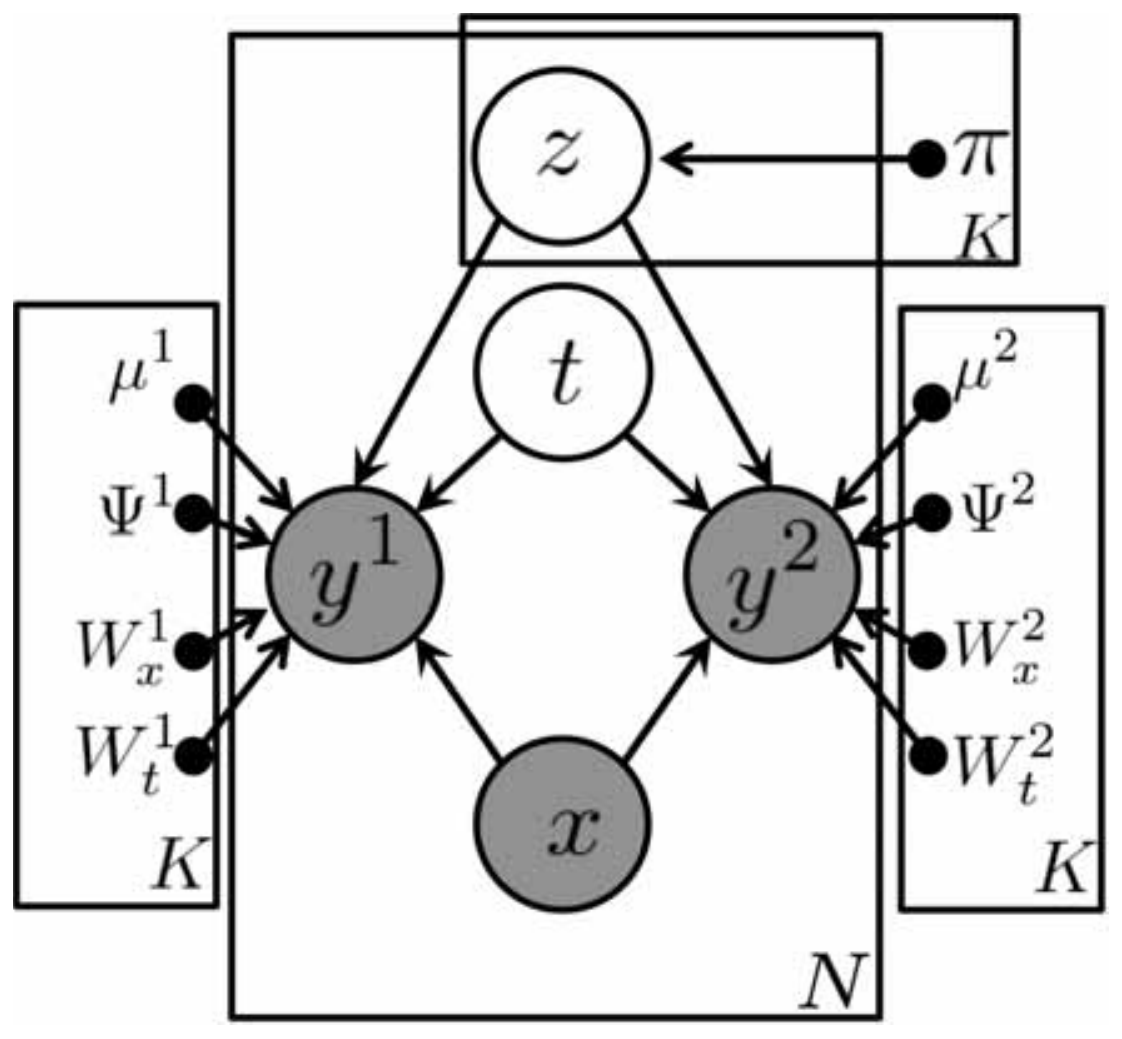}
		\caption{Graphical representation of the mixed probabilistic partial canonical correlation analysis.
			$N$ and $K$ denote the sample and mixture sizes, respectively.}
		\label{fig:The graphical model of MPPCCA}
	\end{center}
\end{figure}

\vspace{0mm}
The MPPCCA is graphically conceptualized in
\figref{fig:The graphical model of MPPCCA}.
The two target multivariate variables in the partial canonical correlation are defined as
$\vect{y}^{(1)} \in \mathbb{R}^{m_1}$ and $\vect{y}^{(2)} \in \mathbb{R}^{m_2}$.
The third variable, 
whose effect should be eliminated from $\vect{y}^{(1)}$ and $\vect{y}^{(2)}$, is denoted by $x \in \mathbb{R}^{d_x}$,.
The latent variables are $\vect{ t}_n \in \mathbb{R}^{d_t}$ and $z_{nk}\in \{0,1\}$ ($\min(d_1,d_2)\geq d_t$).
The variable $\vect{t}_n$ represents the common factor between $\vect{y}^{(1)}$ and $\vect{y}^{(2)}$, and $z_{nk}$ is a 1-out-of-K representative of the sample $n$. This
means that
element $z_k=1$ and all other elements are 0.
The element $z_{nk}$ indicates which PPCCA model generates the sample $n$.

\begin{eqnarray}
	\begin{aligned}
		&p(\vect{t}_n ) = \mathcal{N} (\vect{t}_n \mid \vect{0},I_{d_t}) \nonumber\\
		&p(\vect{y}_n^{(1)} \mid \vect{t}_n,\vect{x}_n;\mat{W}_{xk}^{(1)}, \mat{W}_{tk}^{(1)}, \vect{\mu}_k^{(1)}, \mat{\Psi}_k^{(1)},z_{nk})  \\
		& \quad  = \mathcal{N} (\vect{y}_n^{(1)} \mid \mat{W}_{xk}^{(1)} \vect{x}_n + \mat{W}_{tk}^{(1)} \vect{t}_n + \vect{\mu}_k^{(1)}, \mat{\Psi} _k^{(1)})^{z_{nk}} \nonumber\\
		&p(\vect{y}_n^{(2)} \mid \vect{t}_n,\vect{x}_n;\mat{W}_{xk}^{(2)}, \mat{W}_{tk}^{(2)}, \vect{\mu}_k^{(2)}, \mat{\Psi} _k^{(2)},z_{nk})  \\
		& \quad = \mathcal{N} (\vect{y}_n^{(2)} \mid \mat{W}_{xk}^{(2)} \vect{x}_n + \mat{W}_{tk}^{(2)} \vect{t}_n + \vect{\mu}_k^{(2)}, \mat{\Psi} _k^{(2)})^{z_{nk}} \nonumber\\
		&p( z_{nk};\pi_k) = \mathrm{Multinomial}(z_{nk} \mid \pi_k) \nonumber \\
		&\mathcal{N}(\vect{x} \mid \vect{\mu} , \mat{\Psi}) =
				\frac{1}{\sqrt{ \mid 2\pi \mat{\Psi} \mid }}
				\exp\left[-\frac{1}{2}\{(\vect{x}-\vect{\mu})^T\mat{\Psi}^{-1}(\vect{x}-\vect{\mu})\}\right] \nonumber \\
	\end{aligned}
	\label{ep:MPPCCA generative model}
\end{eqnarray}
The model $k$ in \eqref{ep:MPPCCA generative model} describes one linear causal relationship
between $\vect{y}^{(1)}$, $\vect{y}^{(2)}$ and $x$. This description 
is equivalent to the PPCCA proposed by
Mukuta and Harada \cite{mukuta2014probabilistic}

The latent variable vector $\vect{t}_n$ represents the common factor
among the observed variables $\vect{y}_{n}^{(1)}$ and $\vect{y}_{n}^{(2)}$.
The transformation matrices ($\mat{W}_{tk}^{(1)}$, $\mat{W}_{tk}^{(2)}$) and
Their respective variances ($\mat{\Psi}_{tk}^{(1)}$, $\mat{\Psi}_{tk}^{(2)}$) determine
how well the common factor $\vect{t}_{n}$ relates to $\vect{y}_{n}^{(1)}$ and $\vect{y}_{n}^{(2)}$.
Meanwhile, the transformation matrix $\mat{W}_{xk}$ and its variance $\mat{\Psi}_{xk}$ determine
the relationships from $\vect{x}_n$ to $\vect{y}_n^{(1)}$ and
from $\vect{x}_n$ to $\vect{y}_n^{(2)}$.
The above variables can be represented as follows:
\begin{eqnarray}
	\vect{y}_n &=& \begin{pmatrix}
		\vect{y}_n^{(1)}\\
		\vect{y}_n^{(2)}
	\end{pmatrix}\in \mathbb R^{(d_1+d_2)},\nonumber\\
	\vect{\mu}_k &=& \begin{pmatrix}
		 \vect{\mu}_k^{(1)}\\
		 \vect{\mu}_k^{(2)}
	\end{pmatrix}\in \mathbb R^{(d_1+d_2)},\nonumber\\
	\mat{W}_{xk} &=&\begin{pmatrix}
		\mat{W}_{xk}^{(1)}\\
		\mat{W}_{xk}^{(2)}
	\end{pmatrix}\in \mathbb{R}^{(d_1+d_2) \times d_x },\nonumber\\
	\mat{W}_{tk} &=& \begin{pmatrix}
		\mat{W}_{tk}^{(1)}\\
		\mat{W}_{tk}^{(2)}
	\end{pmatrix}\in \mathbb{R}^{(d_1+d_2) \times d_t},\nonumber\\
	\mat{\Psi}_{k} &=& \begin{pmatrix}
		\mat{\Psi}^{(1)}_k & \mat{0}\\
		\mat{0}         & \mat{\Psi}^{(2)}_k
	\end{pmatrix}\in \mathbb R^{(d_1+d_2) \times (d_1+d_2)}.
\end{eqnarray}
Based on the above descriptions,
\eqref{ep:MPPCCA generative model} can be summarized as follows:
\begin{eqnarray}
	\begin{aligned}
		&p(\vect{y}_n \mid \vect{t}_n,\vect{x}_n;\mat{W}_{xk}, \mat{W}_{tk}, \vect{\mu}_k, \mat{\Psi} _k,z_{nk})\\
		&\quad = \mathcal{N} (\vect{y}_n \mid \mat{W}_{xk} \vect{x}_n + \mat{W}_{tk} \vect{t}_n + \vect{\mu}_k, \mat{\Psi} _k)^{z_{nk}}
	\end{aligned}
\end{eqnarray}

\vspace{0mm}
\subsection{Marginalization of the latent variables}
\vspace{0mm}
We now marginalize the generative model 
proposed by Mukuta and Harada 2014\cite{mukuta2014probabilistic} in terms of $\vect{t}$.
To this end, we integrate the probability density function over all possible states.
In the first step of marginalization, we sum the probabilities over $z$:
\begin{eqnarray}
	\begin{aligned}
		&p(\vect{y}_n \mid  \vect{t}_n, \vect{x}_n; \mat{W}_{xk}, \mat{W}_{tk}, \vect{\mu}, \mat{\Psi} ,\pi) \\
		&\quad = \sum_{\vect{z}} \sum_{k=1}^{K}p(z_{nk} \mid \pi_k)
				p(\vect{y}_n \mid \vect{t}_n, \vect{x}_n;\mat{W}_{xk}, \mat{W}_{tk}, \vect{\mu}_k, \mat{\Psi} _k,z_{nk}) \nonumber\\
		&\quad= \sum_{k=1}^{K}\pi_{k} \mathcal{N} (\vect{y}_n \mid \mat{W}_{xk} \vect{x}_n + \mat{W}_{tk} \vect{t}_n + \vect{\mu}_k, \mat{\Psi} _k)
	\end{aligned}
\end{eqnarray}

The next step marginalizes the model in terms of $\vect{t}_n$.
\begin{eqnarray}
	\begin{aligned}
		&p(\vect{y}_n \mid \vect{x}_n;\mat{W}_{x}, \mat{W}_{t}, \vect{\mu}, \mat{\Psi} ,\pi) \\
		&= \int_{-\infty}^{\infty} p(\vect{y}_n \mid \vect{t}_n,\vect{x}_n;\mat{W}_{x}, \mat{W}_{t}, \vect{\mu}, \mat{\Psi} ,\pi)p(\vect{t}_n)dt_n \nonumber\\
		&= \sum _{k=1}^{K} \vect{\pi}_k \frac{1}{\sqrt{ \mid 2 \pi C_k \mid }} \\
		&\quad \exp \left[ -\frac{1}{2} \{ (\vect{y}_n -\mat{W}_{xk} \vect{x}_n - \vect{\mu}_k)^T C_k^{-1} (\vect{y}_n -\mat{W}_{xk} \vect{x}_n - \vect{\mu}_k) \} \right] \nonumber\\
		&= \sum_{k=1}^{K}\pi_{k} \mathcal{N} (\vect{y}_n \mid \mat{W}_{xk} \vect{x}_n + \vect{\mu}_k, C_k)\\
		&\mat{C}_{k} = \mat{\Psi}_{k} + \mat{W}_{tk}\mat{W}_{tk}^{T}
	\end{aligned}
\end{eqnarray}

The log likelihood of $\vect{y}$ is
\begin{eqnarray}
	\begin{aligned}
		&\ln p(\vect{y} \mid \vect{x};\mat{W}_{x}, \mat{W}_{t}, \vect{\mu}, \mat{\Psi})\\
		&\quad = \sum_{n=1}^{N} \ln \sum_{k=1}^{K}
				\pi_k \mathcal{N}(\vect{y}_n \mid \mat{W}_{xk} \vect{x}_n + \vect{\mu}_k, C_k).
	\end{aligned}
	\label{eq:likelihood}
\end{eqnarray}

\vspace{0mm}
\subsection{EM algorithm}
\vspace{0mm}
By applying the EM algorithm to MPPCCA, we can determine
the latent variables of the model, including $\vect{z}$.
The likelihood function in \eqref{eq:likelihood} increases by iterating the E- and M-steps.
The log likelihood of $\{\vect{y}, \vect{z}\}$ is given by
\begin{eqnarray}
	\begin{aligned}
		&\ln p(\vect{y, z} \mid \Theta)\\
		& \quad = \sum_{n=1}^{N} \sum_{k=1}^{K}z_{nk}
			\ln \pi_k \mathcal{N}(\vect{y}_n \mid \mat{W}_{xk} \vect{x}_n + \vect{\mu}_k, C_k),
		\end{aligned}
\end{eqnarray}
where $\Theta = [\pi_k, \vect{\mu}_k, \mat{\Psi}_k \mat{W}_{xk}, \mat{W}_{tk}]$
For current parameters $\Theta^{old}$, the contribution ratio $r_{nk}$ is defined as follows:
\begin{eqnarray}
	\begin{aligned}
		r_{nk}&=\mathbb{E}\left[z_{nk} \middle | \Theta^{old}\right] \\
			&\equiv p(z_{nk} \mid \vect{y}_n, \Theta^{old}) \\
			&= \frac{p(z_{nk} \mid \Theta^{old}) p(\vect{y}_n \mid z_{nk}, \Theta^{old})}{p(\vect{y}_n \mid \Theta^{old})} \\
			&= \frac{\pi_k \mathcal{N} (\vect{y}_n \mid \mat{W}_{xk} \vect{x}_n +
				\vect{\mu}_k, C_k)}{\sum_{j=1}^{K}\pi_j \mathcal{N} (\vect{y}_n \mid \mat{W}_{xj} \vect{x}_n + \vect{\mu}_j, C_j)}
 \end{aligned}
\end{eqnarray}

The E-step calculates the contribution ratio $r_{nk}$
using the parameters $\Theta^{old}$.
The M-step then seeks the parameters $\Theta$
that ultimately maximize the log likelihood of $Q(\Theta,\Theta^{old})$.
\begin{eqnarray}
	\begin{aligned}
		&Q(\Theta^{new},\Theta^{old})\\
		&\quad = \mathbb{E}_{\vect{z}}\left[\sum_{n=1}^{N} \sum_{k=1}^{K}z_{nk} \ln \pi_k \mathcal{N}(\vect{y}_n \mid \mat{W}_{xk} \vect{x}_n + \vect{\mu}_k, C_k)\right]  \\
		&\quad= \sum_{n=1}^{N} \sum_{k=1}^{K}  r_{nk} \ln \pi_k \mathcal{N}(\vect{y}_n \mid \mat{W}_{xk} \vect{x}_n + \vect{\mu}_k, C_k )  \nonumber\\
		& \quad= \sum_{n=1}^{N} \sum_{k=1}^{K}  r_{nk} \{\ln \pi_k + \mathcal{N}(\vect{y}_n \mid \mat{W}_{xk} \vect{x}_n + \vect{\mu}_k, C_k )\}
	\end{aligned}
\end{eqnarray}

The update equations are as follows.
\begin{eqnarray}
	\begin{aligned}
		r_{nk}&= \frac{\pi_k \mathcal{N} (\vect{y}_n \mid \mat{W}_{xk}
			\vect{x}_n + \vect{\mu}_k, C_k)}{\sum_{j=1}^{K}\pi_j
			\mathcal{N} (\vect{y}_n \mid \mat{W}_{xj} \vect{x}_n + \vect{\mu}_j, C_j)} \\
		\vect{\mu}_k &= \frac{\sum_{n=1}^{N} r_{nk} (\vect{y}_n - \mat{W}_{xk} \vect{x}_n)}{\sum_{n=1}^{N} r_{nk}} \\
		\pi_k &= \frac{1}{N}\sum_{n=1}^{N}r_{nk} \\
		\mat{W}_{xk} &= \left(\sum_{n=1}^{N}r_{nk}\tilde{\vect{y}}_{nk}\tilde{\vect{x}}_{nk}^T\right)
			\left(\sum_{n=1}^{N}r_{nk}\tilde{\vect{x}}_{nk}\tilde{\vect{x}}_{nk}^T\right)^{-1} \\
		\mat{W}_{tk} &= U_k(\Lambda_k - \mat{\Psi}_k)^{\frac{1}{2}}R \\
		\mat{\Psi}_{k} &= S_k - \mat{W}_{tk} {\mat{W}_{tk}}^T
	\end{aligned}
\end{eqnarray}

\begin{eqnarray}
	\tilde{\vect{y}}_{nk} = \vect{y}_n - \frac{\sum_{n=1}^N
		r_{nk}\vect{y}_n}{\sum_{n=1}^N r_{nk}},\tilde{\vect{x}}_{nk} =
		\vect{x}_n - \frac{\sum_{n=1}^{N}r_{nk}\vect{x}_n}{\sum_{n=1}^{N}r_{nk}}\\
	\mat{S}_k = \frac{\sum_{n=1}^{N}r_{nk}(\tilde{\vect{y}}_{nk}-\mat{W}_{xk}
		\tilde{\vect{x}}_{nk})(\tilde{\vect{y}}_{nk}-\mat{W}_{xk}\tilde{\vect{x}}_{nk})^T}{\sum_{n=1}^{N}r_{nk}}.
\end{eqnarray}
The matrices $\mat{U}_k$ and $\Lambda_k$ contain the eigenvectors and eigenvalues of $\mat{S}_k$, respectively.
$\Lambda_k$ is a diagonal matrix, and
$R$ is an arbitrary orthogonal matrix.

\vspace{0mm}
\subsection{Regularization}
\vspace{0mm}
If the sample size is small or if multicollinearity occurs (i.e., if two or more data are strongly correlated), the covariance matrix becomes ill-conditioned. Therefore, inverting this matrix in the EM algorithm risks destabilizing the algorithm.
To preserve the stability of the computation,
we incorporate the ridge regression method into our model.

\begin{eqnarray}
	C_k&=& \mat{\Psi}_k + \mat{W}_{tk}\mat{W}_{tk}^T + \eta_{C} I_{(d_1+d_2)}
	\label{eq:normalize_c}
\end{eqnarray}
\begin{eqnarray}
	\begin{aligned}
		&\mat{W}_{xk} = \\
		&\left(\sum_{n=1}^{N}r_{nk}\tilde{\vect{y}}_{nk}\tilde{\vect{x}}_{nk}^T\right) \left(\sum_{n=1}^{N}r_{nk}\tilde{\vect{x}}_{nk}\tilde{\vect{x}}_{nk}^T + \eta_{\mat{W}_x} I_{(d_x+d_x)} \right)^{-1} \label{eq:normalize_Wx},
	\end{aligned}
\end{eqnarray}
where I is the identity matrix.
As $\eta_c$ and $\eta_{\mat{W}_x}$ are both non-zero, small positive values,
the estimation is stable.

\vspace{0mm}
\subsection{Clustering based on MPPCCA and k-means}
\vspace{0mm}
In MPPCCA, the calculated $z_{nk}$ is only the statistical expectation.
Therefore, to achieve deterministic clustering by MPPCCA,
we determine the cluster $k$ of data $n$ using $\argmax_k \left(z_{nk} \right)$.

To evaluate the capacity of MPPCCA,
we apply basic k-means clustering to
$\{ \vect{y}^{(1)}, \vect{y}^{(2)}, \vect{x} \}$ or
$\{x_{t-1},y_{t-1},y_t\}$.

\vspace{0mm}
\section{Experiments}
\label{sec:toy_data}
\vspace{0mm}
The MPPCCA was evaluated on
synthetic data generated by probabilistic models. For this purpose, we designed two experiments. 
\begin{itemize}
	\item Experiment 1: Synthetic time series containing multiple causal relationships. The cluster size is the size of the causal relationships.
	\item Experiment 2: Synthetic time series with and without causal relationships. The cluster size is irrelevant.
\end{itemize}

\vspace{0mm}
\subsection{Exp. 1: Synthetic time series with multiple causal relationships having the same cluster size as the size of causal relationships.}
\label{sec:toy_time}
\vspace{0mm}
MPPCCA is expected to separate the time-series data
into multiple causal relationships with no prior knowledge,
and to quantify the causalities within the separated patterns.

\begin{figure}[tbp]
	\centering
	\subfigure[Synthetic time series with multiple causal relationships.]{
		\includegraphics[width=0.75\columnwidth]{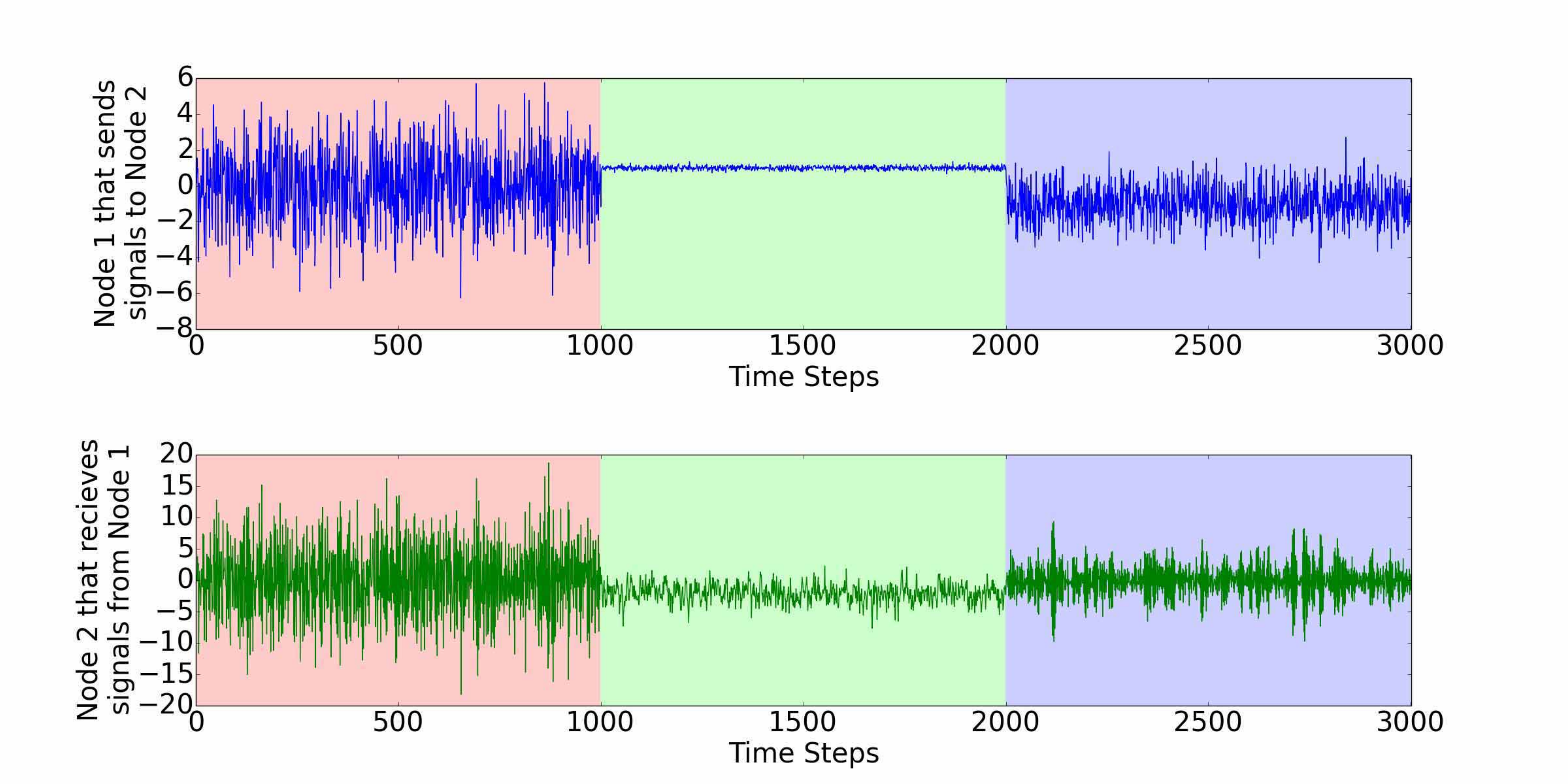}
		\label{fig:toy_gc_time_data_redundant}
	}
	\subfigure[Clustering estimation by MPPCCA.]{\includegraphics[clip, width=0.75\columnwidth]{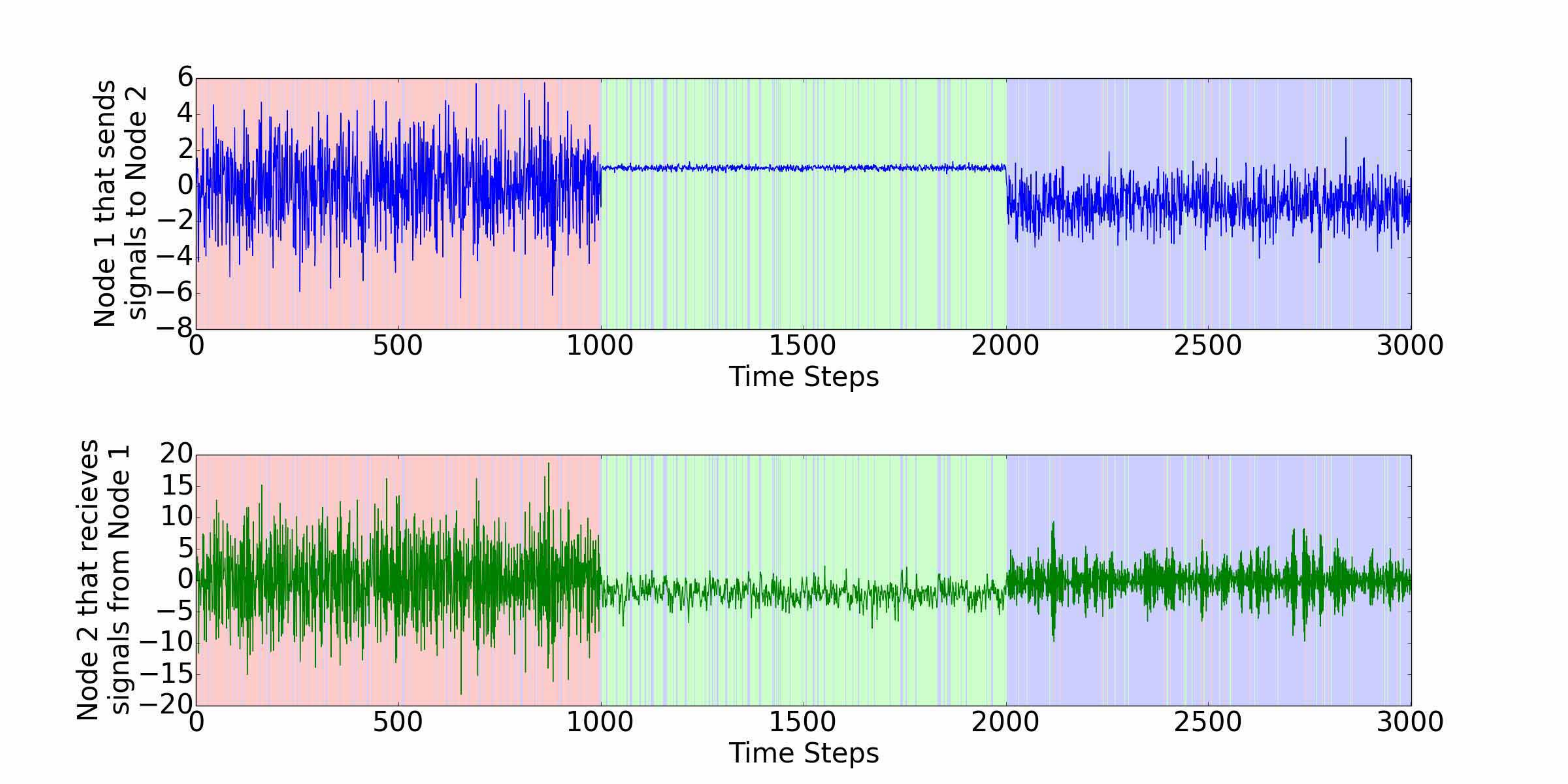}}
	\subfigure[Clustering estimation by k-means.]{
		\includegraphics[clip, width=0.75\columnwidth]{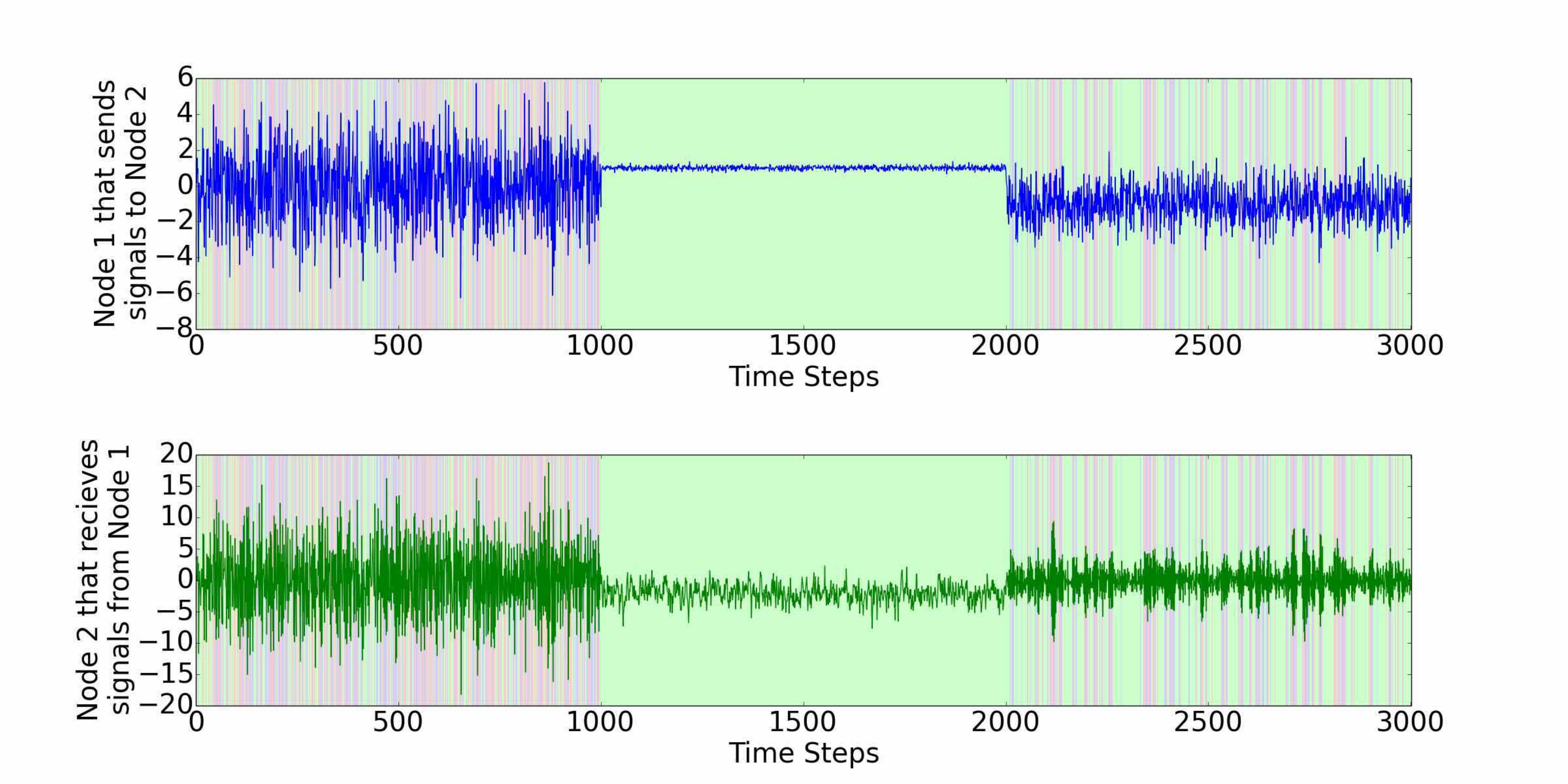}
	}
	\caption{Synthetic data (a) and the clustering 
		results of the synthetic time series estimated by MPPCCA (b) and k-means (c)
		with multiple causal relationships in the time domain (Exp. 1).
		The red, green and blue regions delineate the models that generated the samples in panel (a), and 
		the estimated causal clusters in panel (b).
		The cause and effect sides are denoted Node1 and Node 2, respectively.}
	\label{fig:straight_results_redundant}
\end{figure}

In this experiment, the time series included multiple causal relationships.
The different causal patterns should be separated out as clusters.
The synthetic time series was generated by the following equations (the parameters are listed in \tabref{table:gc_params}).
\begin{eqnarray}
	y_t &=& \mathcal{N}(a_k y_{t-1} + b_k x_{t-1}, \Psi_{yk}) \label{eq:y_t} \\
	x_t &=& \mathcal{N}(\mu_{xk}, \Psi_{xk})  \label{eq:x_t}\\
	&&x_t,y_t,a_k,b_k,\Psi_{yk},\mu_{xk},\Psi_{xk} \in \mathbb{R}^{1}
\end{eqnarray}
The size of the PPCCA model was $K=3$, and each cluster generated 1000 successive samples.
\begin{table}[tbp]
 \centering
 \caption{Parameters of the synthetic time series with multiple causal relationships (Exp. 1).}
 \begin{tabular}{|l||l|l|l|l|l|} \hline
  k&$a_k$ &$b_k$ &$\mu_{xk}$ &$\Psi_{xk}$ &$\Psi_{yk}$ \\ \hline
  1& -0.5 & 2.5 &  0.0& 2.0 & 0.2\\ \hline
  2& 0.5  & -1.0&  1.0& 0.1 & 1.3\\ \hline
  3& -0.9 & 0.2 & -1.0& 1.0 & 1.3\\ \hline
 \end{tabular}
 \label{table:gc_params}
\end{table}
In this case,
all samples $y_t$ from all generative models were determined from $x_{t-1}$ and $y_{t-1}$ rather than from $y_{t-1}$ alone.
The data generated in different models with different values of the causality-strength parameter $b$ are presented in \figref{fig:toy_gc_time_data_redundant}.

\begin{figure*}[tbp]
	\centering
	\subfigure[Ground truth]{
		\includegraphics[clip, width=5.0cm]{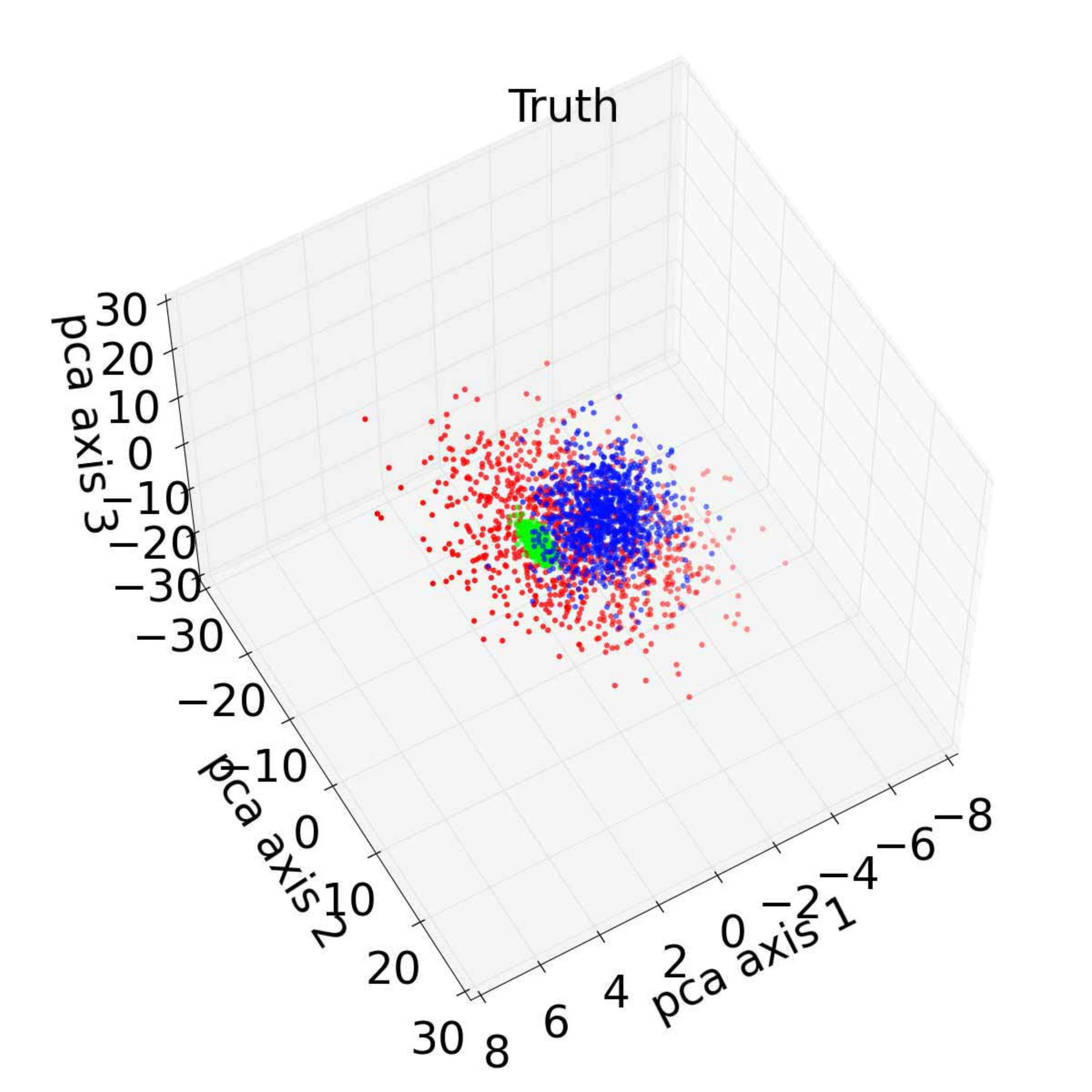}
	}
	\subfigure[MPPCCA]{
		\includegraphics[clip, width=5.0cm]{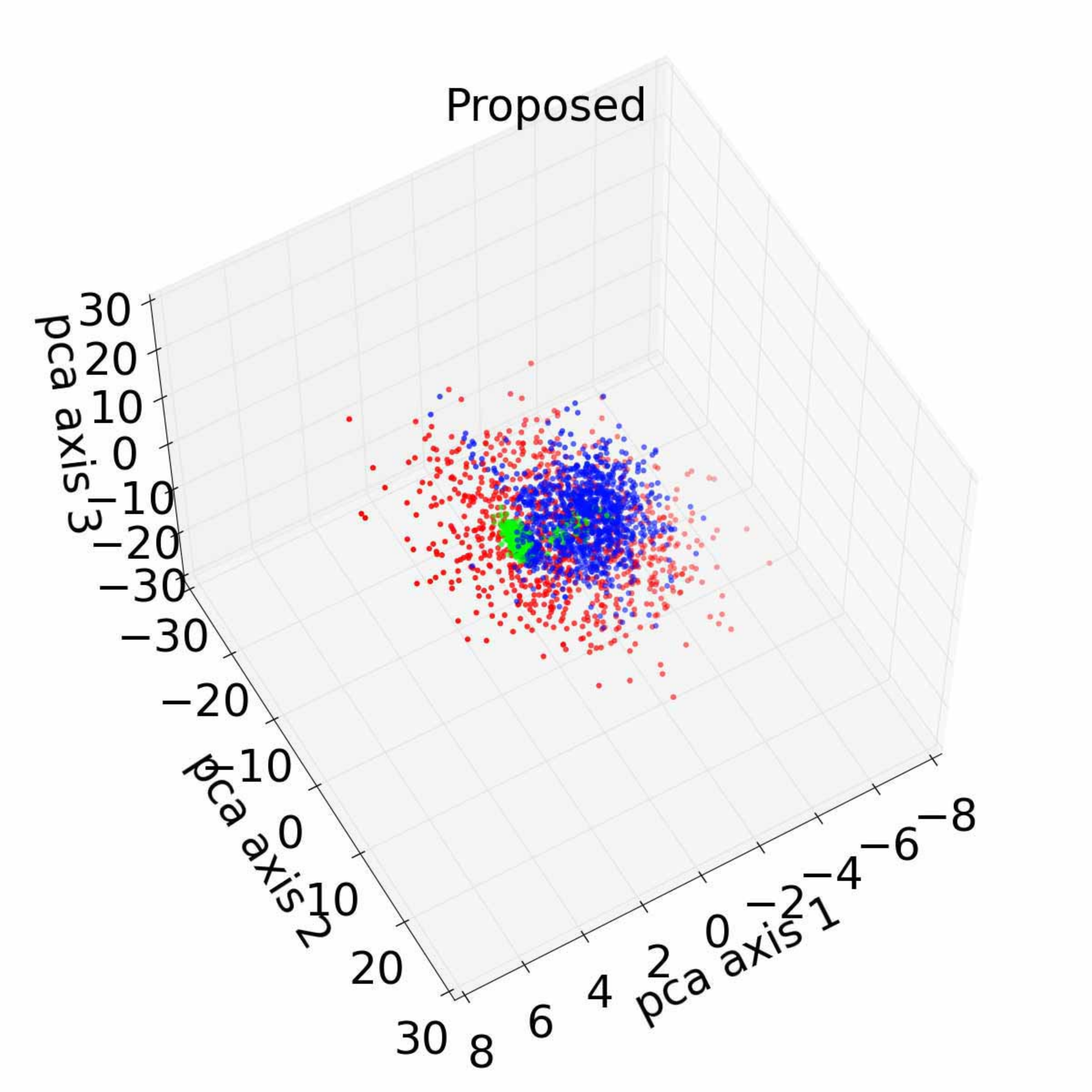}
	}
	\subfigure[k-means]{
		\includegraphics[clip, width=5.0cm]{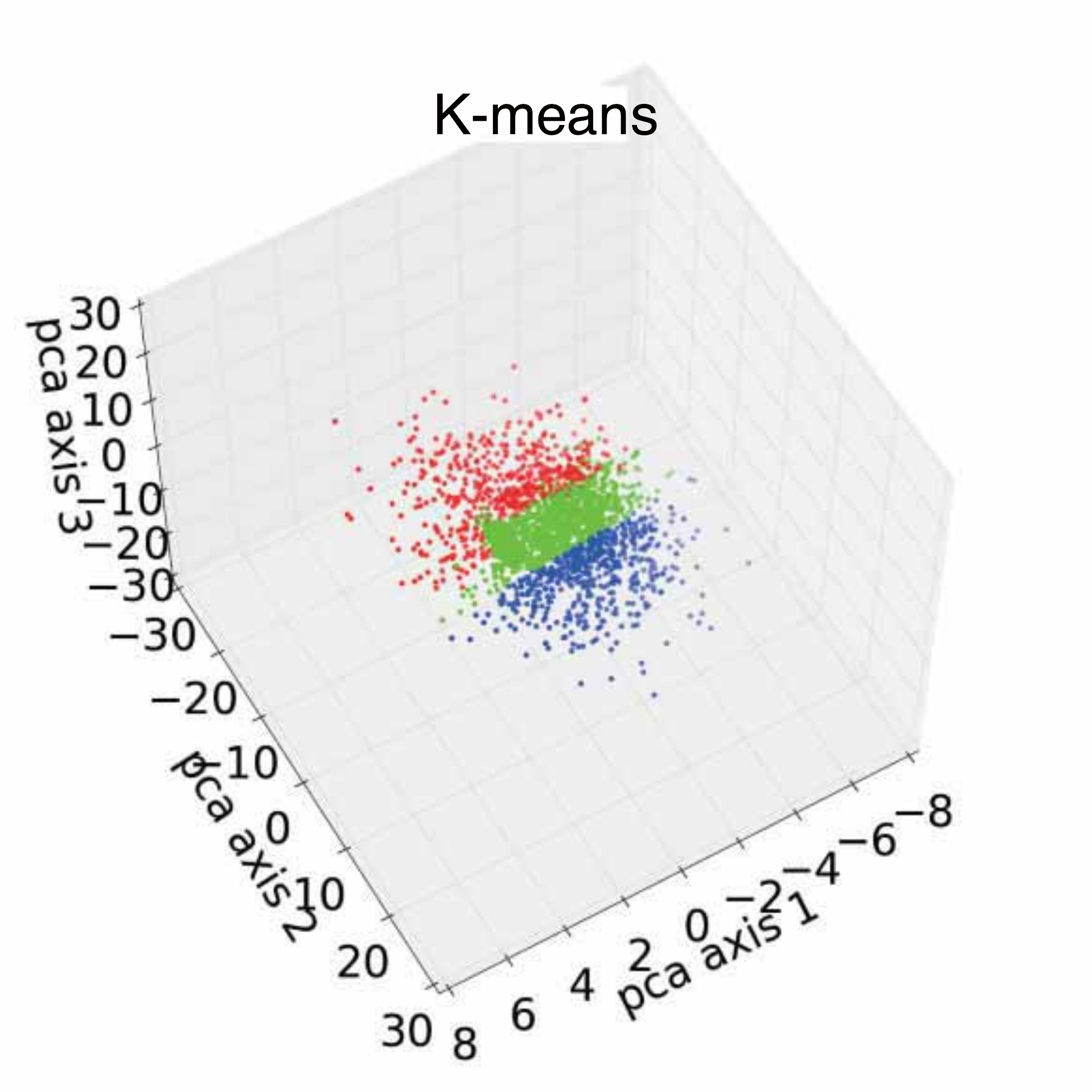}
	}
	\caption{Clustering results of the synthetic time series with multiple causal relationships
	in the PCA space $\{x_{t-1},y_{t-1},y_t\}$ (Exp. 1).}
\label{fig:toy_gc_results}
\end{figure*}

In the mixed-causality by MPPCCA, we selected
$y_t$ and $x_{t-1}$ as the target multivariate variables
and eliminated the effects of $y_{t-1}$ from the targets.
\figref{fig:toy_gc_results} shows the state space
of $\{x_{t-1},y_{t-1},y_t\}$ after reduction to three dimensions by PCA.

\begin{figure}[tbp]
	\begin{center}
	\includegraphics[width=0.8\columnwidth]{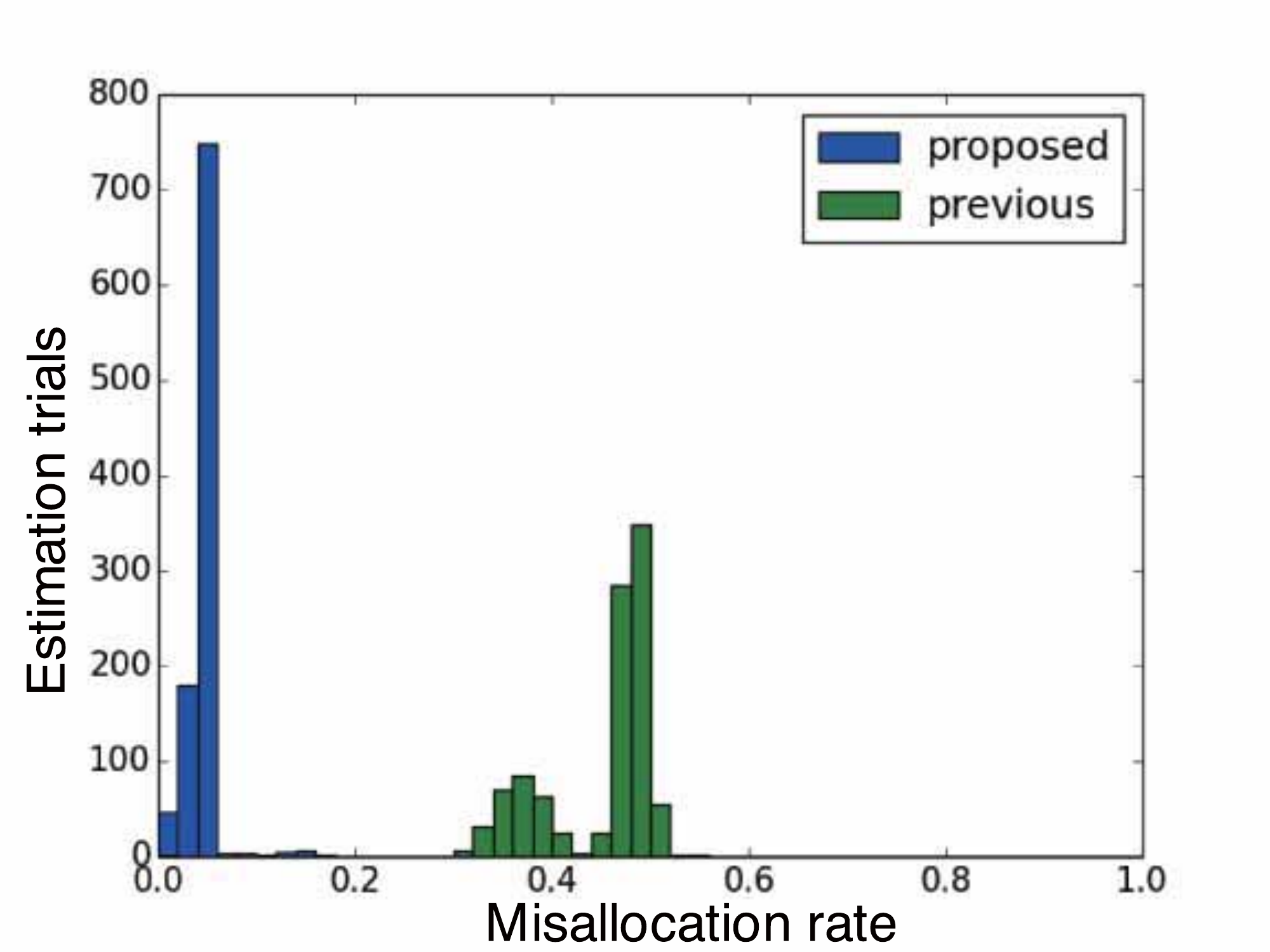}
	\caption{Histogram of misallocation rates when all clusters have a causal relationship (Exp. 1).
		MPPCCA correctly estimates more than 90 \% of the samples, but
		k-means misallocates 30-50 \% of the samples.}
  \label{fig:toy_gc_hist}
 \end{center}
 \begin{center}
  \includegraphics[width=0.8\columnwidth]{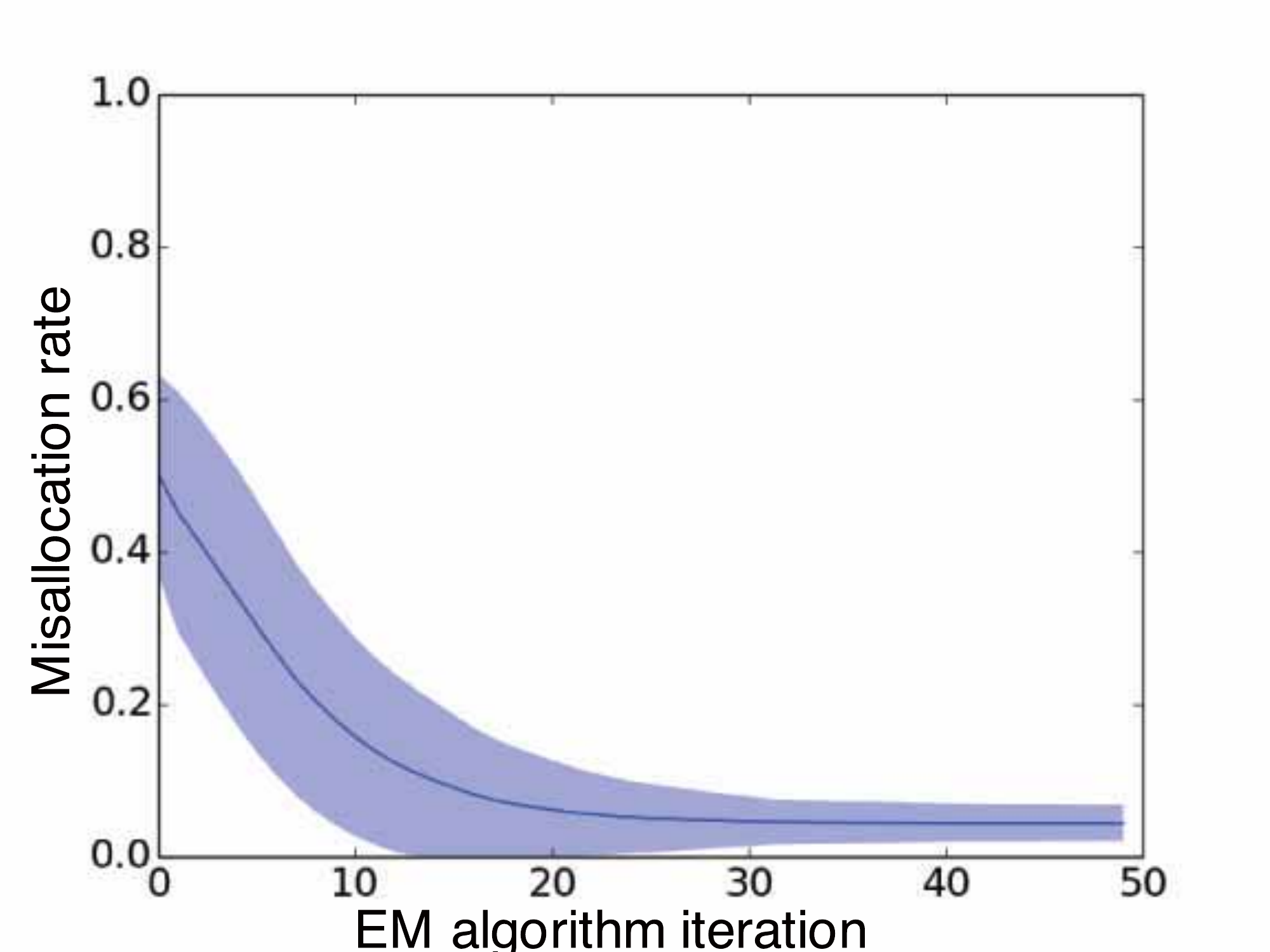}
  	\caption{Misallocation rate
	in the synthetic time series with one-directional causality
	versus number of EM steps (Exp. 1).
	Solid line and band represent the average and standard deviation of 1000 trial estimations, respectively.}
  \label{fig:toy_gc_error}
 \end{center}
\end{figure}

To evaluate the clustering,
we define the misallocation rate as
the ratio of the minority models ($k$) that generate the data in a cluster.
\ref{fig:toy_gc_hist} is a histogram of the misallocation rates in 1000 clustering trials by MPPCCA and k-means.

Whereas MPPCCA estimated the correct clusters in more than 90 \%
of the trials, the k-means method placed different causal patterns into the same cluster.
\figref{fig:toy_gc_error} shows how the average and standard deviation of the misallocation rate
reduce with increasing number of EM learning steps.
The error converges after 30 EM steps.
Panels (b) and (c) of \figref{fig:toy_gc_results}
show examples of cluster estimation by MPPCCA and k-means, respectively.
The k-means estimated the clusters incorrectly,
because it does not explicitly handle the causal relationships.

Panels (a) and (b) of \figref{fig:straight_results_redundant} show the clusters in the time-series
estimated by the proposed MPPCCA and k-means, respectively.
Whereas the proposed methods correctly clustered the time-series data,
the k-means method clustered them incorrectly.
\tabref{table:gc_causal} shows the GC indices
of the clusters estimated by MPPCCA and k-means, and the GC index of the entire time series.
The ground truth indices were more closely matched by MPPCCA than by k-means.
\begin{table}[tbp]
 \centering
 \caption{Estimated granger causality indexes.}
 \begin{tabular}{|l|c|c|c|}\hline
  Method                & Cluster 1              & Cluster 2              & Cluster 3              \\ \hline
  Ground Truth    & 4.59   & $1.50 \times 10^{-2}$ & $6.95 \times 10^{-3}$ \\ \hline
  MPPCCA  & 4.62   & $1.62 \times 10^{-2}$ & $8.05 \times 10^{-4}$ \\ \hline
  K-means  & 1.05  & $8.52 \times 10^{-1}$ & $2.08 \times 10^{-3}$ \\ \hline
  Whole      & \multicolumn{3}{c|}{$1.80 \times 10^{-1}$ }\\ \hline
 \end{tabular}
 \label{table:gc_causal}
\end{table}

\vspace{0mm}
\subsection{Exp. 2: Synthetic time series with and without causal relationships and redundant cluster size.}
\vspace{0mm}
In the second experiment, we evaluated
whether MPPCCA can separate patterns with causal relationships from those without causal relationships.
The samples were generated by the following equation:
\begin{eqnarray}
  y_t &=&
   \begin{cases}
    \mathcal{N}(a y_{t-1} + b x_{t-1}, \Psi_{y}) \; \; \; &(1300 < t < 1700) \label{eq:y_t1} \\
    \mathcal{N}(\mu_{y_r}, \Psi_{y_r}) \; \; \; &\mathrm{otherwise} \label{eq:y_t2} \\
   \end{cases}
\end{eqnarray}
\begin{eqnarray}
 x_t &=& \mathcal{N}(\mu_{x}, \Psi_{x})  \label{eq:x_t1}
\end{eqnarray}
The first and last 1300 samples $y_t$ were generated with no causal relation to $x_t$. The middle 400 samples were causally related to $x_t$. 
The parameters $x_t,y_t,a_k,b_k,\Psi_{yk},\mu_{xk},\Psi_{xk} \in \mathbb{R}^{1}$ were
those of the previous section.
The values of the remaining parameters are given in \tabref{table:manyrandom_params}
\begin{table}[tbp]
 \centering
 \caption{Parameters of the synthetic time series data.}
 \begin{tabular}{|l|l|l|l|l|l|l|l|l|} \hline
  $a$ &$b$   &$\mu_{x}$ &$\mu_{y}$&$\Psi_{x}$ &$\Psi_{y}$ &$\Psi_{y_r}$ \\ \hline
   0.5 & -1.0&  1.0     & 0.0     & 0.1       & 1.3       &1.3          \\ \hline
 \end{tabular}
 \label{table:manyrandom_params}
\end{table}

\begin{figure}[tbp]
	\centering
	\subfigure[Synthetic time series with causal and non-causal relationships.]{
		\includegraphics[width=0.75\columnwidth]{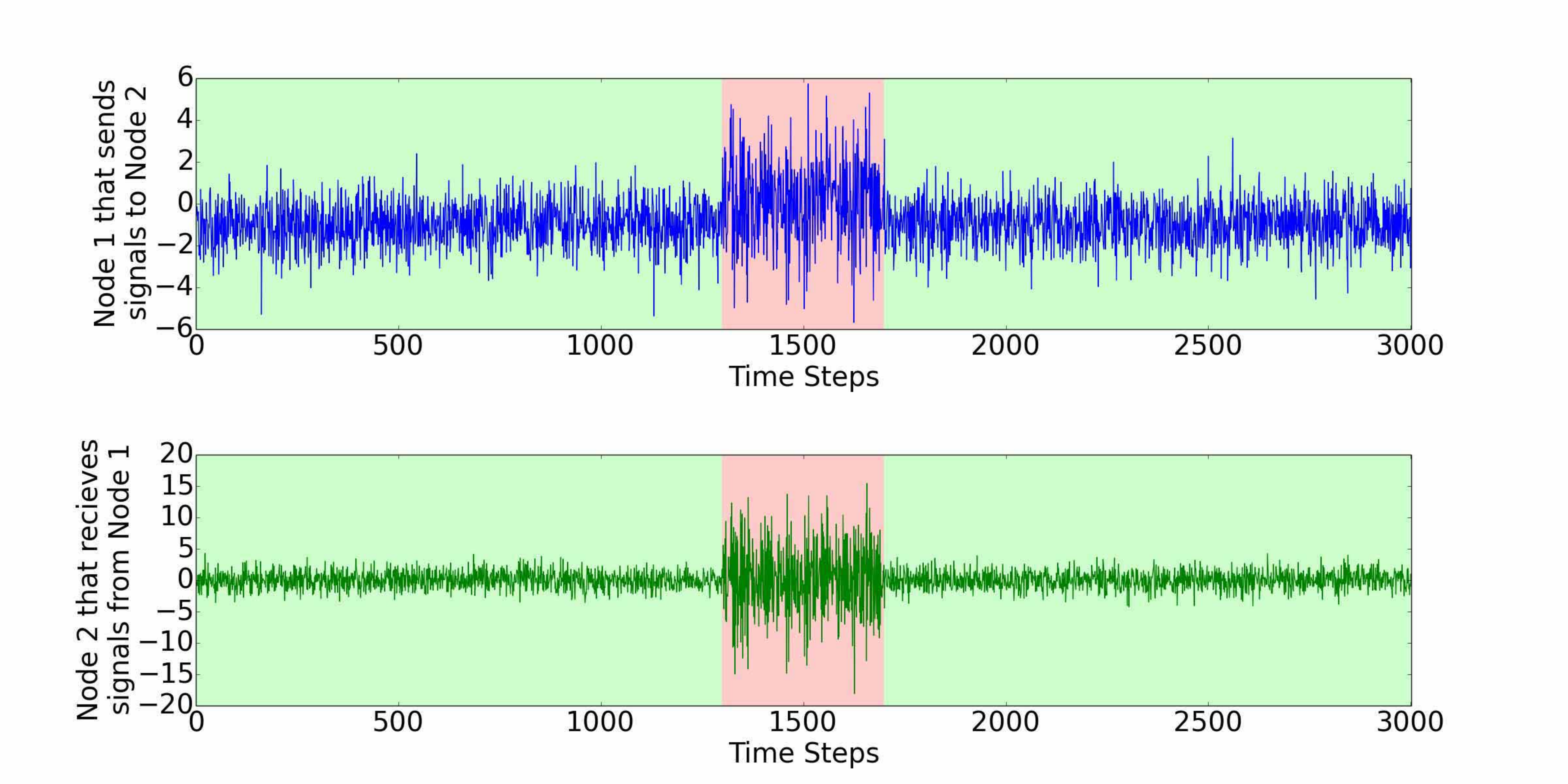}
		\label{fig:toy_gc_time_data_causal_and_non-causal}
	}
	\subfigure[Clustering estimation by MPPCCA.]{\includegraphics[clip, width=0.75\columnwidth]{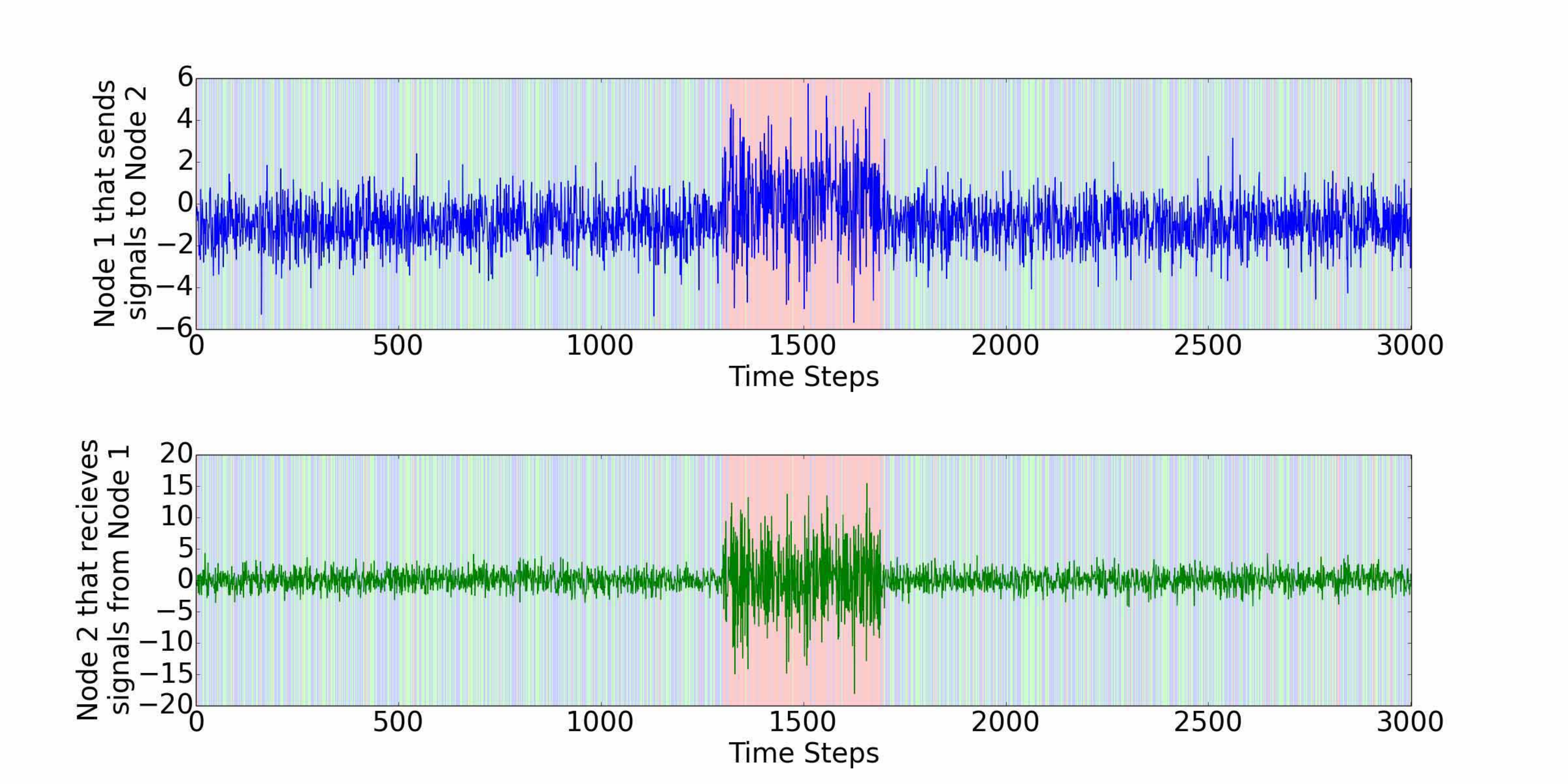}}
	\subfigure[Clustering estimation by k-means.]{
		\includegraphics[clip, width=0.75\columnwidth]{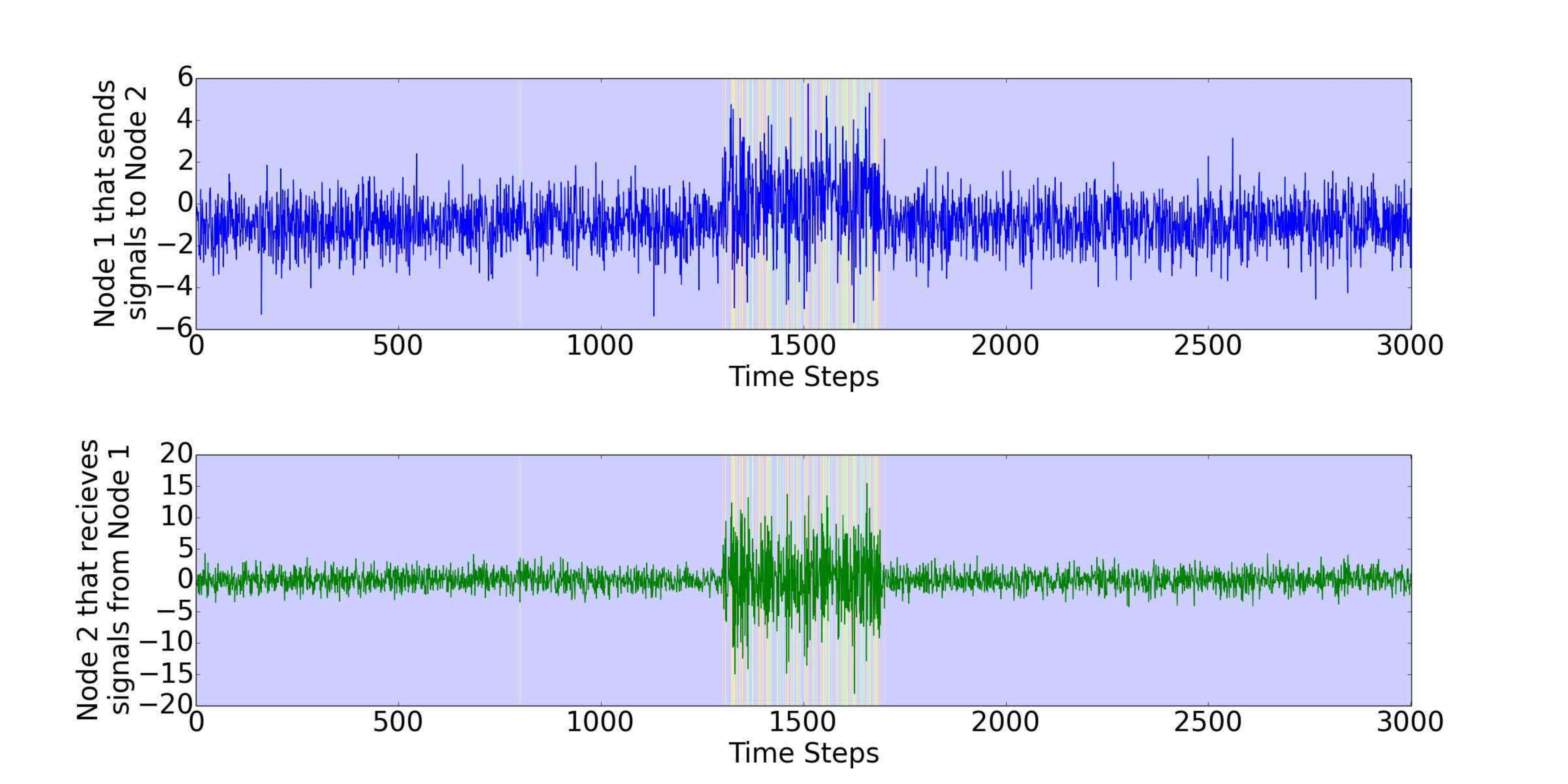}
	}
	\caption{Synthetic data of combined causal and non-causal time-series (a)
		and the cluster results
		of the synthetic time series estimated by MPPCCA (b) and k-means (c) with redundant cluster size $K=3$
		and multiple causal relationships in the time domain (Exp. 1).
		The red, green and blue regions delineate the models that generated the samples in panel (a), and
		the estimated causal clusters in panels (b) and (c).
		The cause and effect sides are denoted by Node1 and Node 2, respectively.}
	\label{fig:results_causal_and_non-causal}
\end{figure}

\begin{figure*}[tbp]
	\centering
	\subfigure[Ground truth]{
		\includegraphics[clip, width=5.0cm]{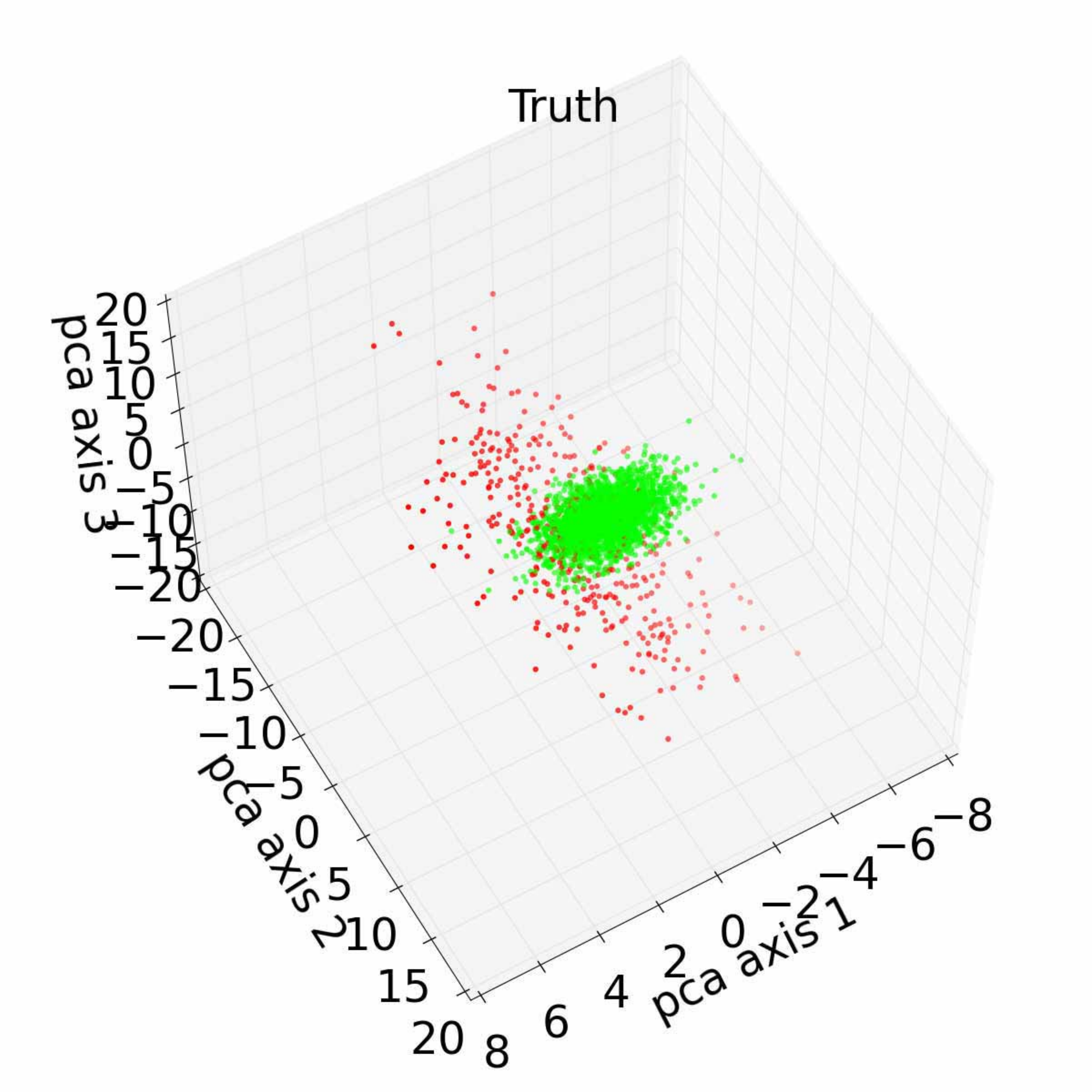}
	}
	\subfigure[MPPCCA]{
		\includegraphics[clip, width=5.0cm]{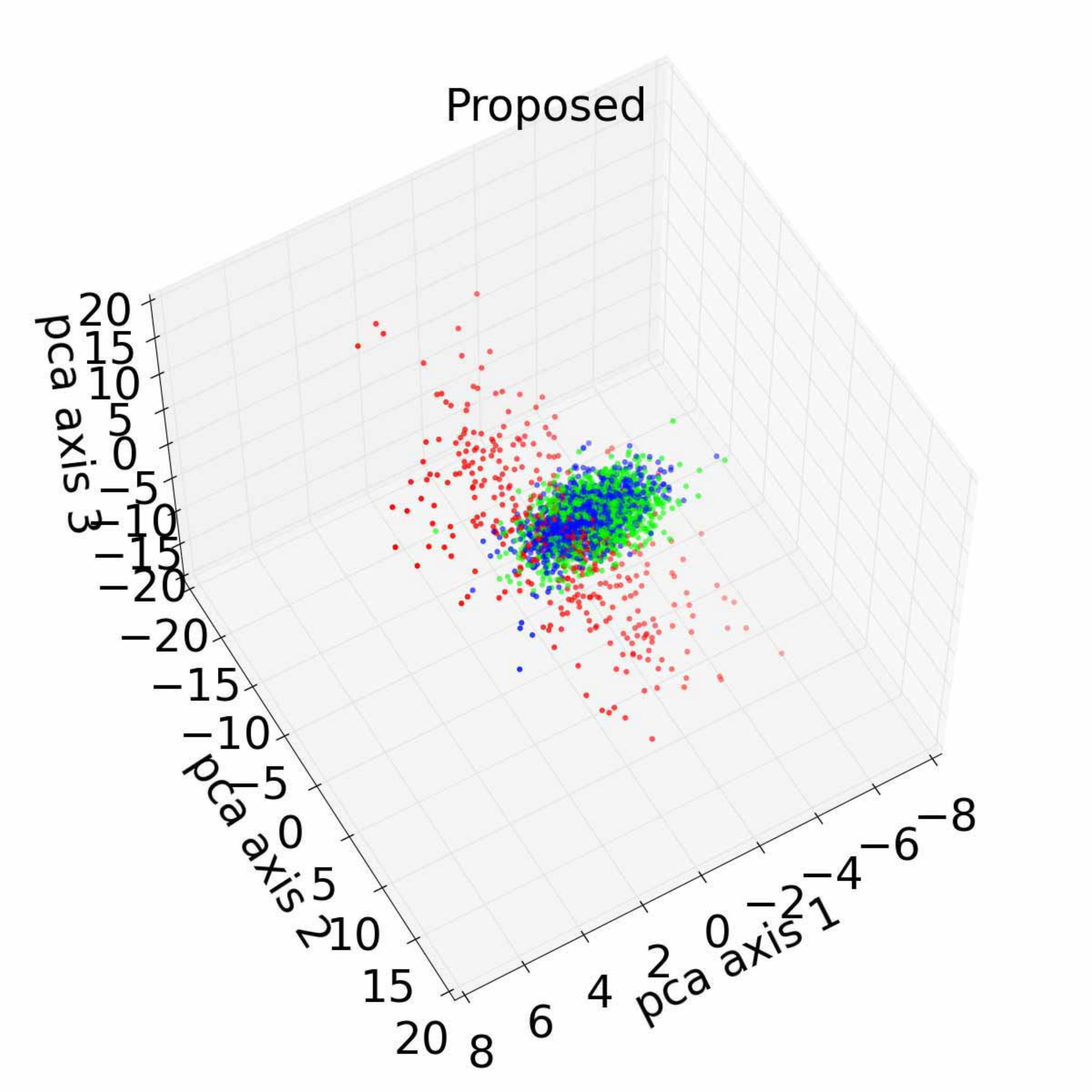}
	}
	\subfigure[k-means]{
		\includegraphics[clip, width=5.0cm]{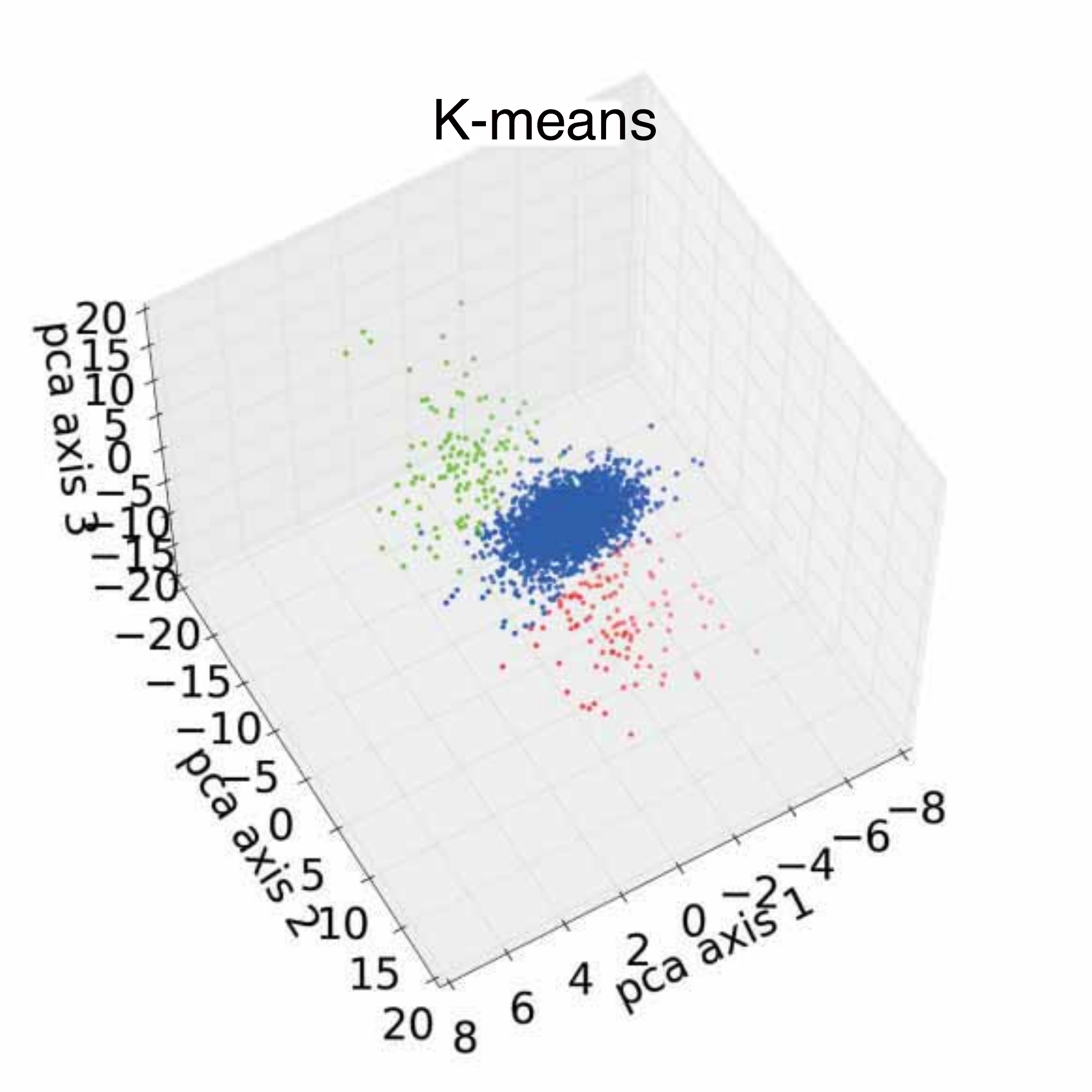}
	}
	\caption{Cluster estimation of
		synthetic time-series with and without causal relationships (Exp. 2).}
\label{fig:manyrandom_results}
\end{figure*}

Panel (a) of \figref{fig:results_causal_and_non-causal}
shows the synthetic time series and 
the colors of the figure represent the true clusters.
\figref{fig:manyrandom_results} shows
the state space of $\{x_{t-1},y_{t-1},y_t\}$, compressed into three dimensions by PCA.

Next, we applied MPPCCA and k-means to mixed causal and non-causal data with $K=3$.
Panels (b) and (c) of \figref{fig:manyrandom_results} present the clustering results of MPPCCA and k-means, respectively.
Whereas MPPCCA correctly estimated the clusters with causal relationships, 
the k-means method separated one causal relationship into two clusters. 
The variance of the synthetic data was higher in clusters with causal relationships than in clusters without causal relationships.

Panels (a) and (b) of \figref{fig:manyrandom_results} visualize the clusters in the time series.
MPPCCA estimates a cluster with causal relationship in the middle of the graph.
The result indicates that MPPCCA extracted the partial dataset containing causal relationships from the complete dataset.
In contrast,
k-means failed to distinguish between causally and non-causally related data.

\tabref{table:manyrandom_causal} gives the GC indexes of clusters
in the ground truth, and those estimated by MPPCCA and k-means, over the entire time series.
Consistent with the ground truth, the GC of MPPCCA was high in Cluster 1 and low in Clusters 2 and 3.
However, the k-means method estimated relatively high GC indices for two clusters (Clusters 1 and 2),
although only Cluster 1 has a causal relationship.
\begin{table}[tbp]
	\centering
	\caption{ Granger causality values of the estimated clusters (Exp. 2).}
	\begin{tabular}{|l|c|c|c|}\hline
	Method                 & Cluster 1              &  Cluster 2              & Cluster 3              \\ \hline
	Ground truth   & 4.58   & \multicolumn{2}{c|}{$7.10 \times 10^{-26}$ }  \\ \hline
	MPPCCA & 3.93   & $2.31 \times 10^{-2}$ & $3.60 \times 10^{-4}$ \\ \hline
	K-means            & 3.95   & $2.34   $        & $1.47 \times 10^{-2}$ \\ \hline
	Entire analysis   & \multicolumn{3}{c|}{$1.80 \times 10^{-1}$ }\\ \hline
 \end{tabular}
 \label{table:manyrandom_causal}
\end{table}

\vspace{0mm}
\section{Real data analysis}\label{chap:real_data}
\vspace{0mm}
Finally, we determine whether our proposed methods can extract communication patterns from real data.
In the target task, two players alternately throw and catch a ball, and sometimes fake a random throw.
As expected, this simple ball game generated patterns with high GC indices,
because the actions of the two subjects were causally and physically related through the ball.
Meanwhile, the fake throws were excluded from the causal patterns
because the actions of the random thrower did not affect the other subject.

\vspace{0mm}
\subsection{Acquisition of motion capture data}
\vspace{0mm}
To measure the action scenes of the two players throwing and catching a ball, we recorded the action by a motion-capture system.
The upper-body parts involved in the throwing and catching motions were marked by
seven points the Cartesian coordinates $(x, y, z)$. The selected parts were the
abdomen, both shoulders, both elbows, and both wrists,
giving a 21-dimensional dataset for each person.
The motions were measured for 1000 s at 60 fps, providing 36001 frames of data.
The measured data were transformed into coordinate systems
with origin at the central abdominal region of each person.

\begin{figure}[tbp]
 \begin{center}
  \includegraphics[width=\columnwidth]{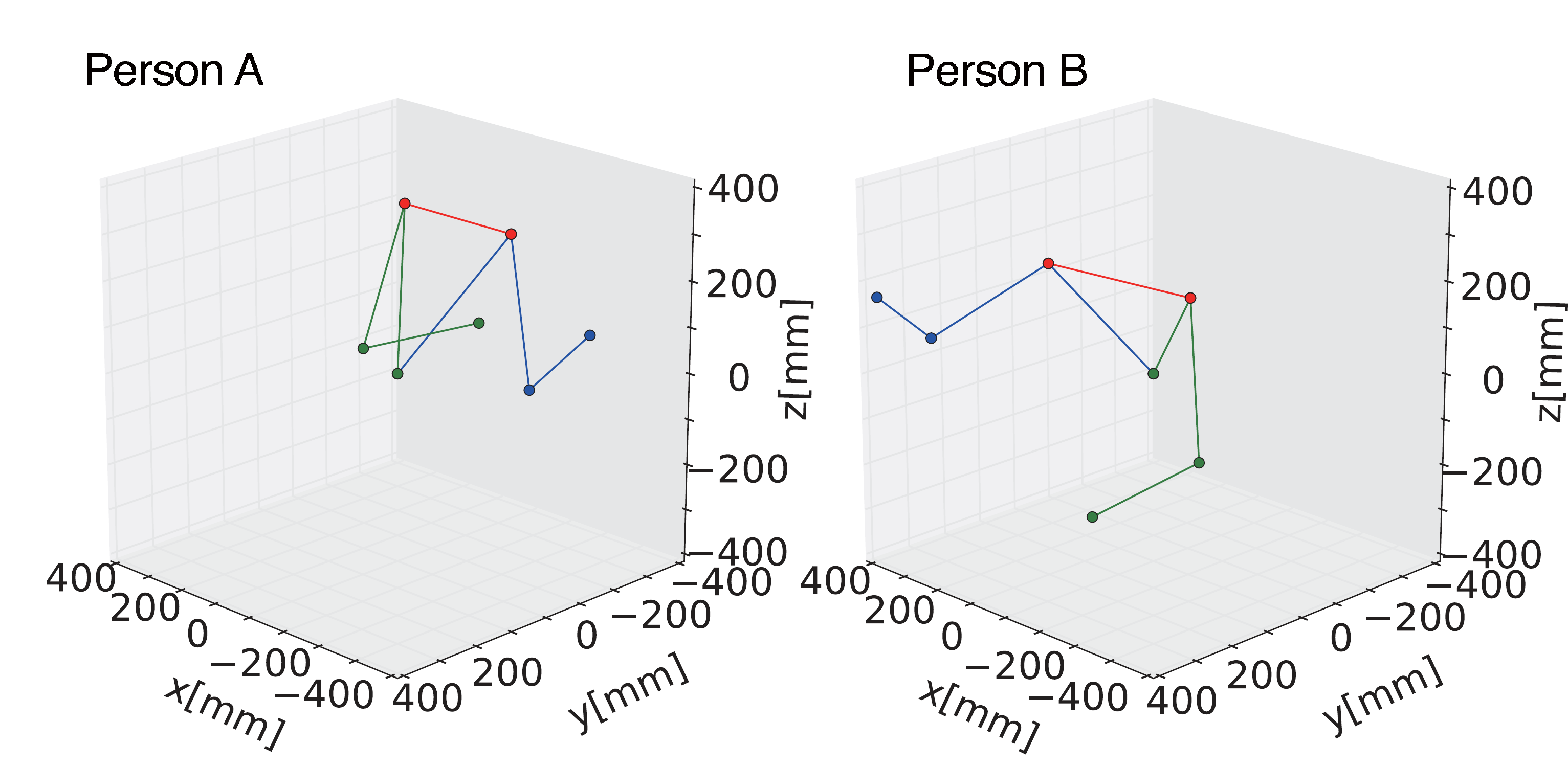}
  \caption{Data captured by the motion-capture system.
  The blue and green lines represent the right and left sides, respectively, of each subject.}
  \label{fig:motion_data}
 \end{center}
\end{figure}

The motion-capture data are displayed in \figref{fig:motion_data}.
In \figref{fig:motion_data}, the persons on the left and right sides are designated as A and B, respectively.
The proposed methods correctly clustered the data based on the causality from B to A.

\begin{figure}[tbp]
 \centering
 \subfigure{\includegraphics[clip, width=0.31\columnwidth]{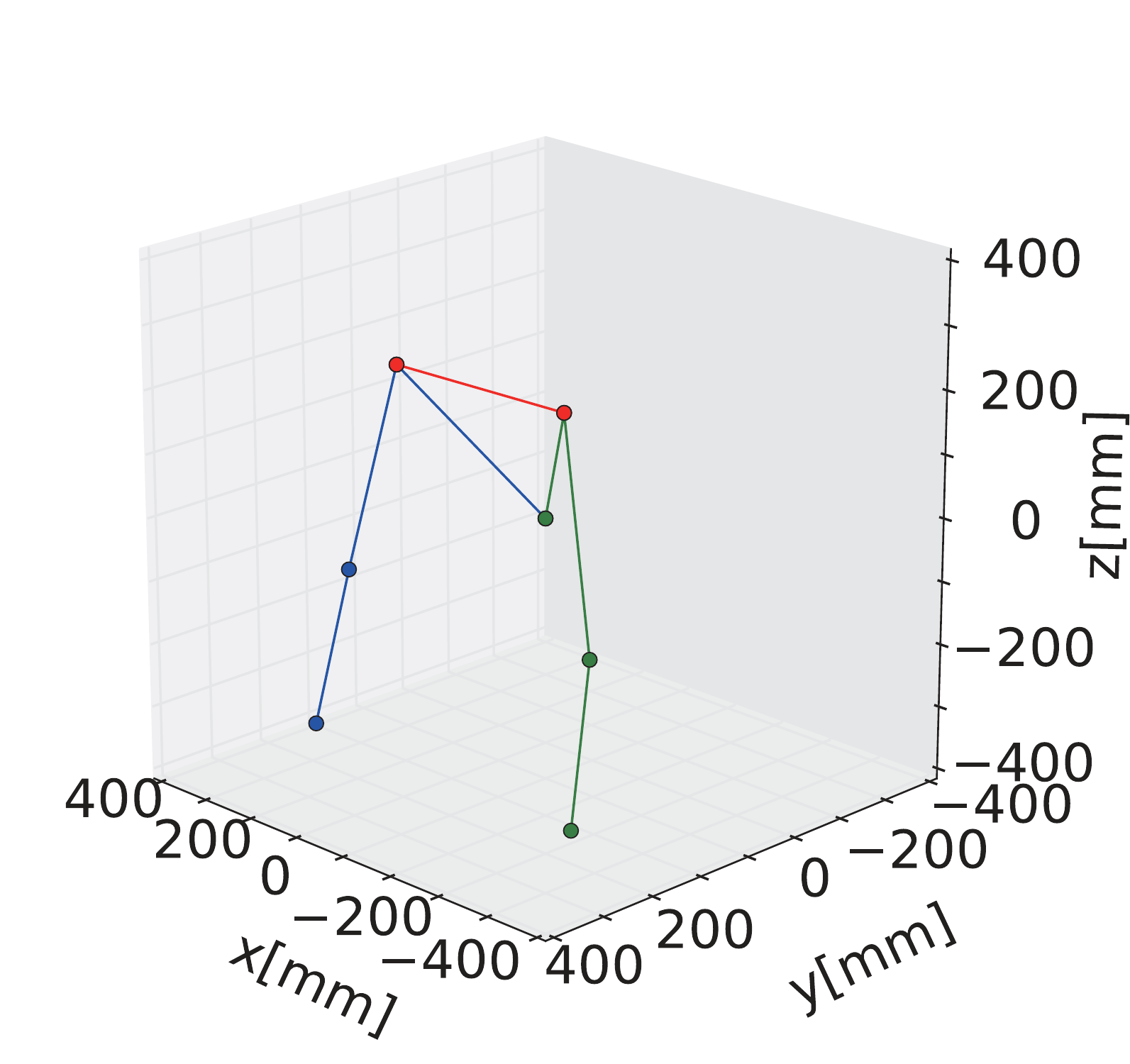} }
 \subfigure{\includegraphics[clip, width=0.31\columnwidth]{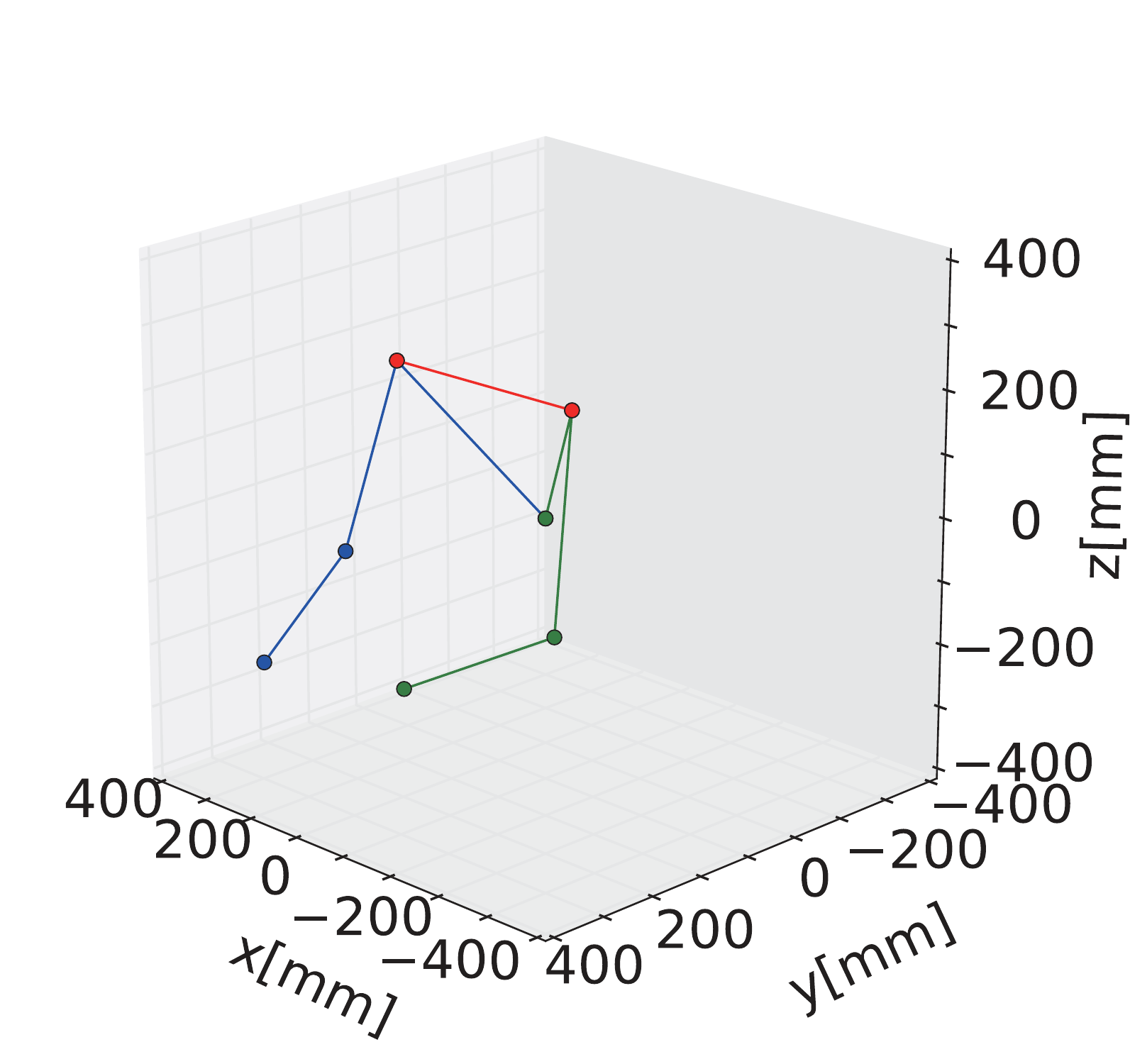} }
 \subfigure{\includegraphics[clip, width=0.31\columnwidth]{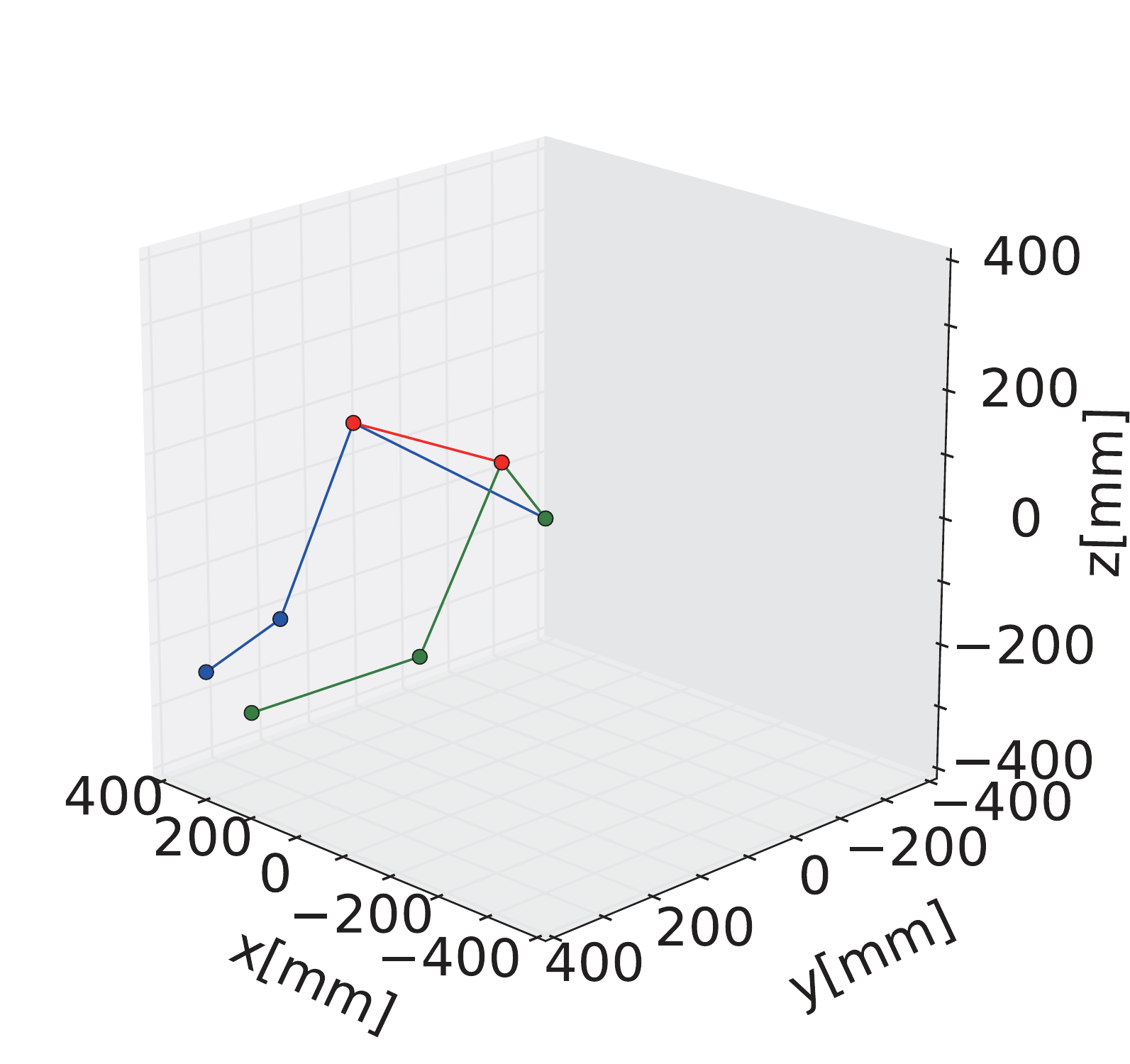} }
  \caption{Receiving the ball}
 \label{fig:catch}
\end{figure}

\begin{figure}[tbp]
 \centering
 \subfigure{\includegraphics[clip, width=0.31\columnwidth]{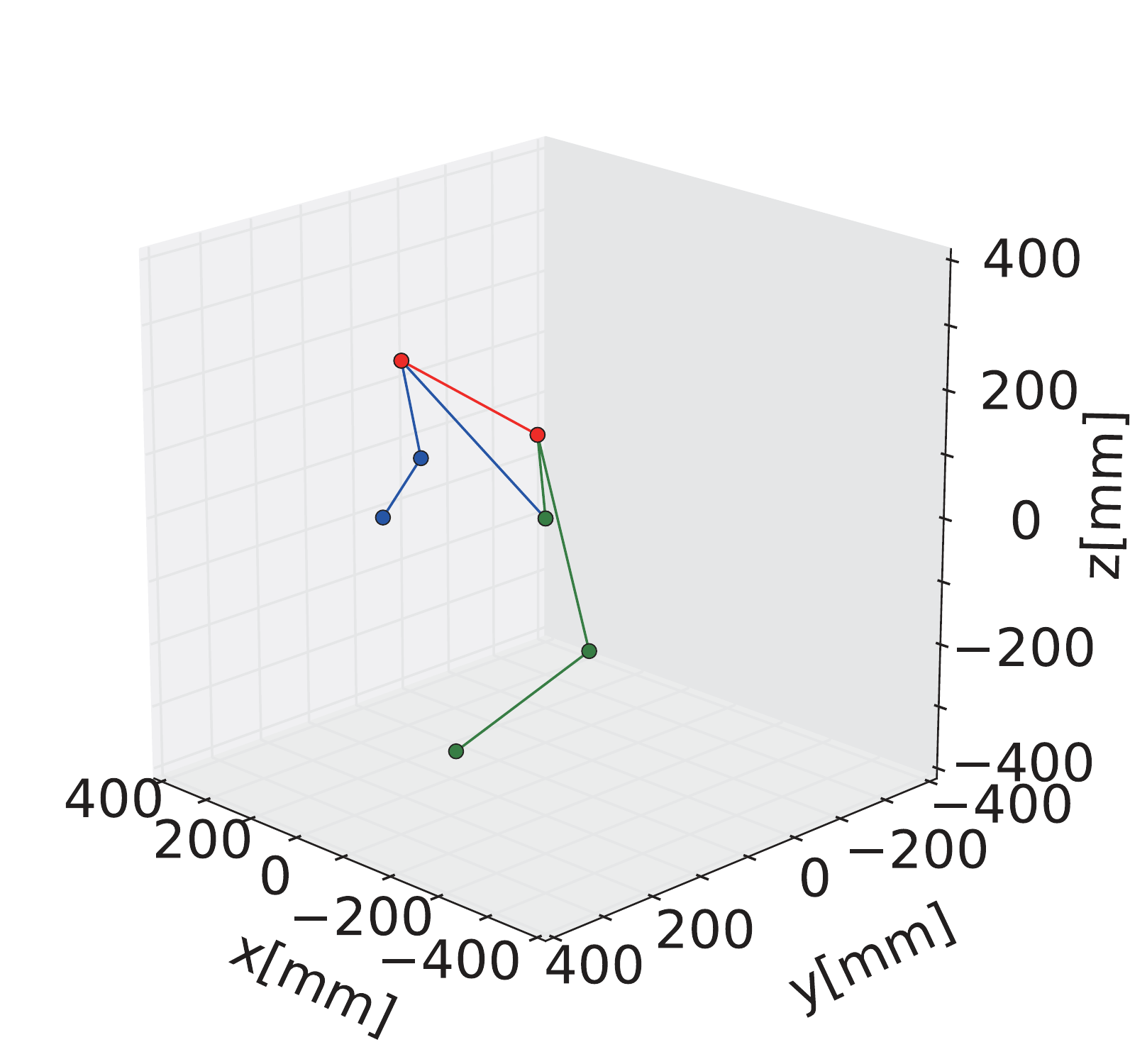} }
 \subfigure{\includegraphics[clip, width=0.31\columnwidth]{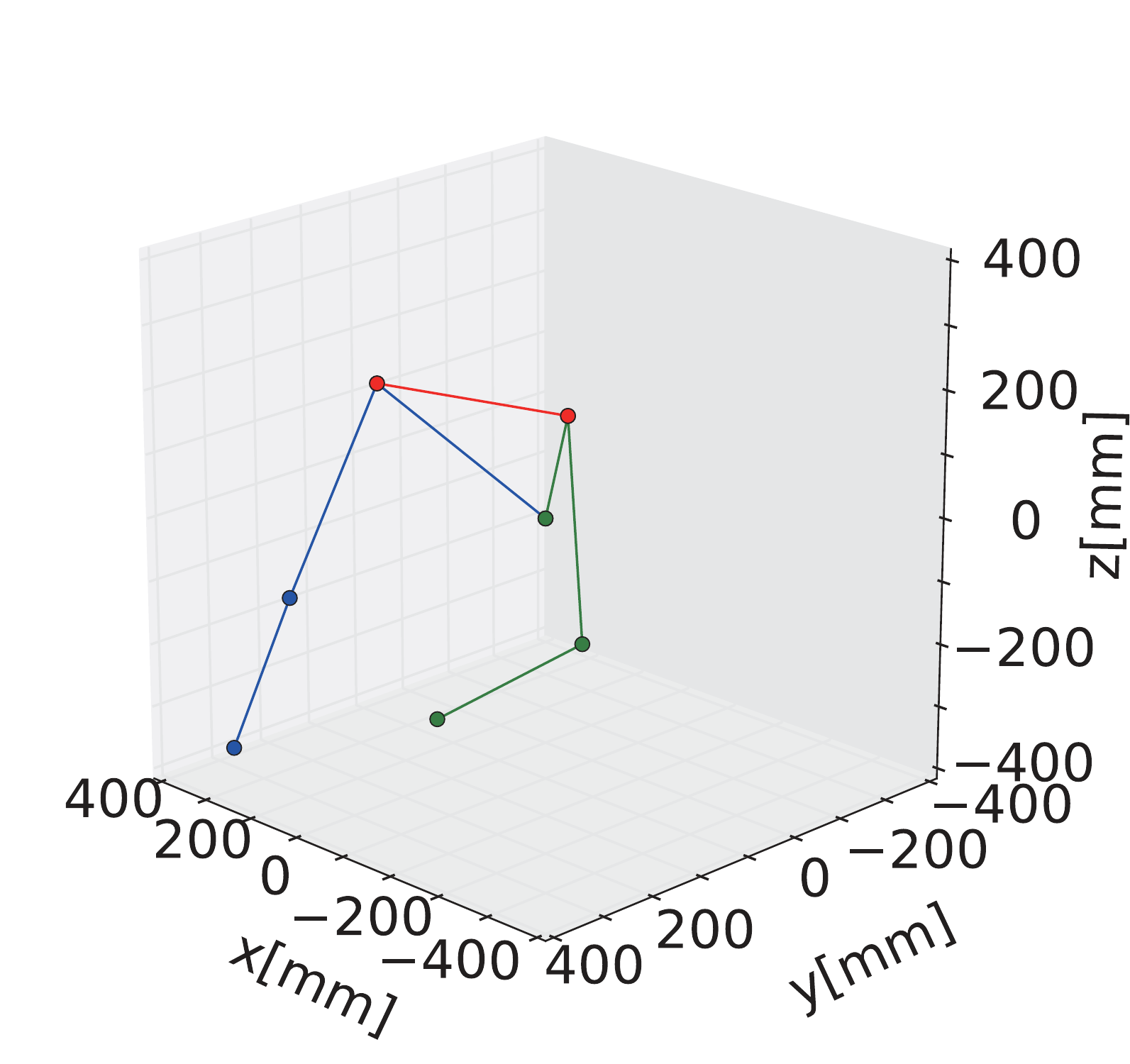} }
 \subfigure{\includegraphics[clip, width=0.31\columnwidth]{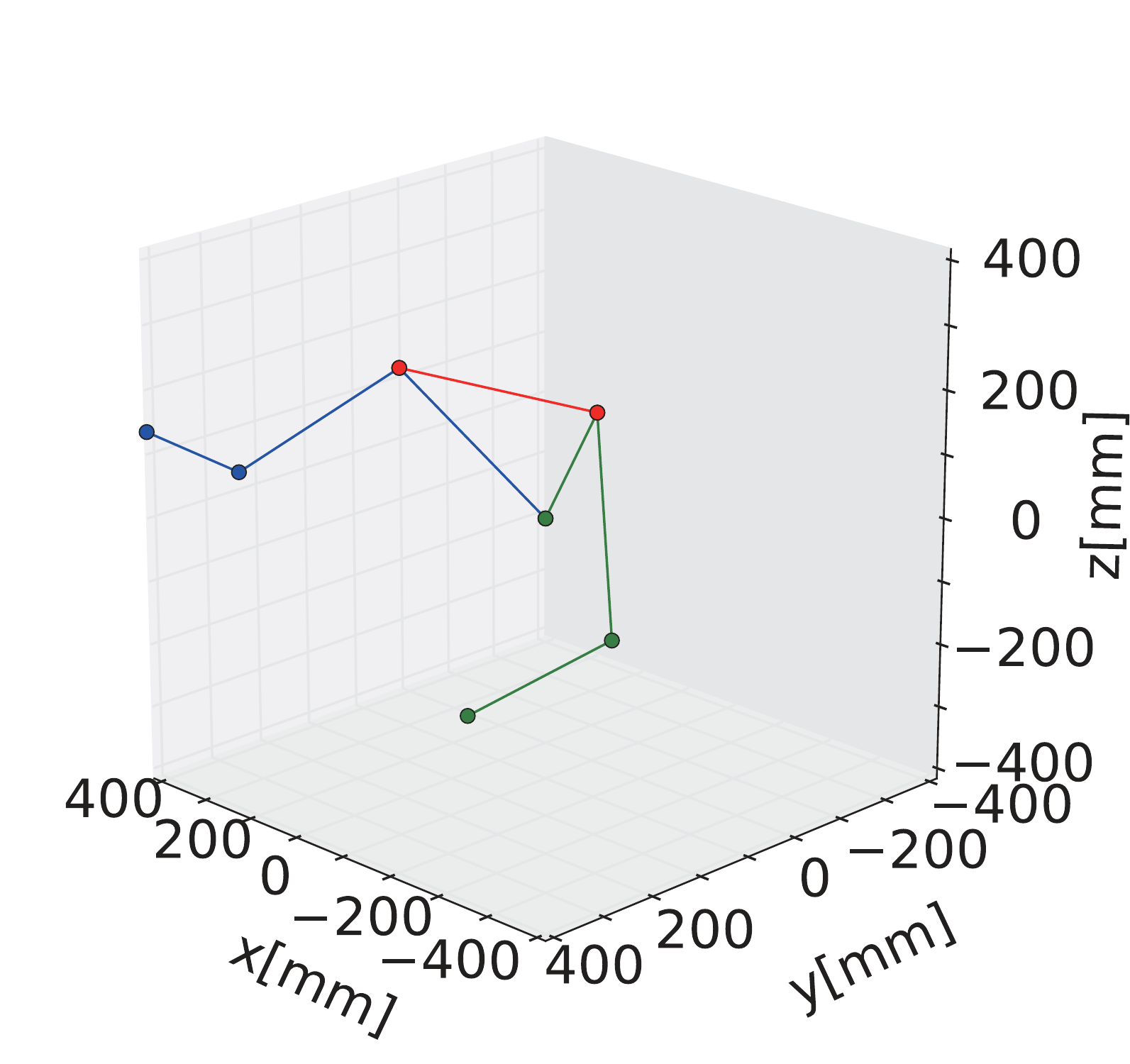} }
  \caption{Throwing the ball underhand}
 \label{fig:sitanage}
\end{figure}

\begin{figure}[tbp]
 \centering
 \subfigure{\includegraphics[clip, width=0.31\columnwidth]{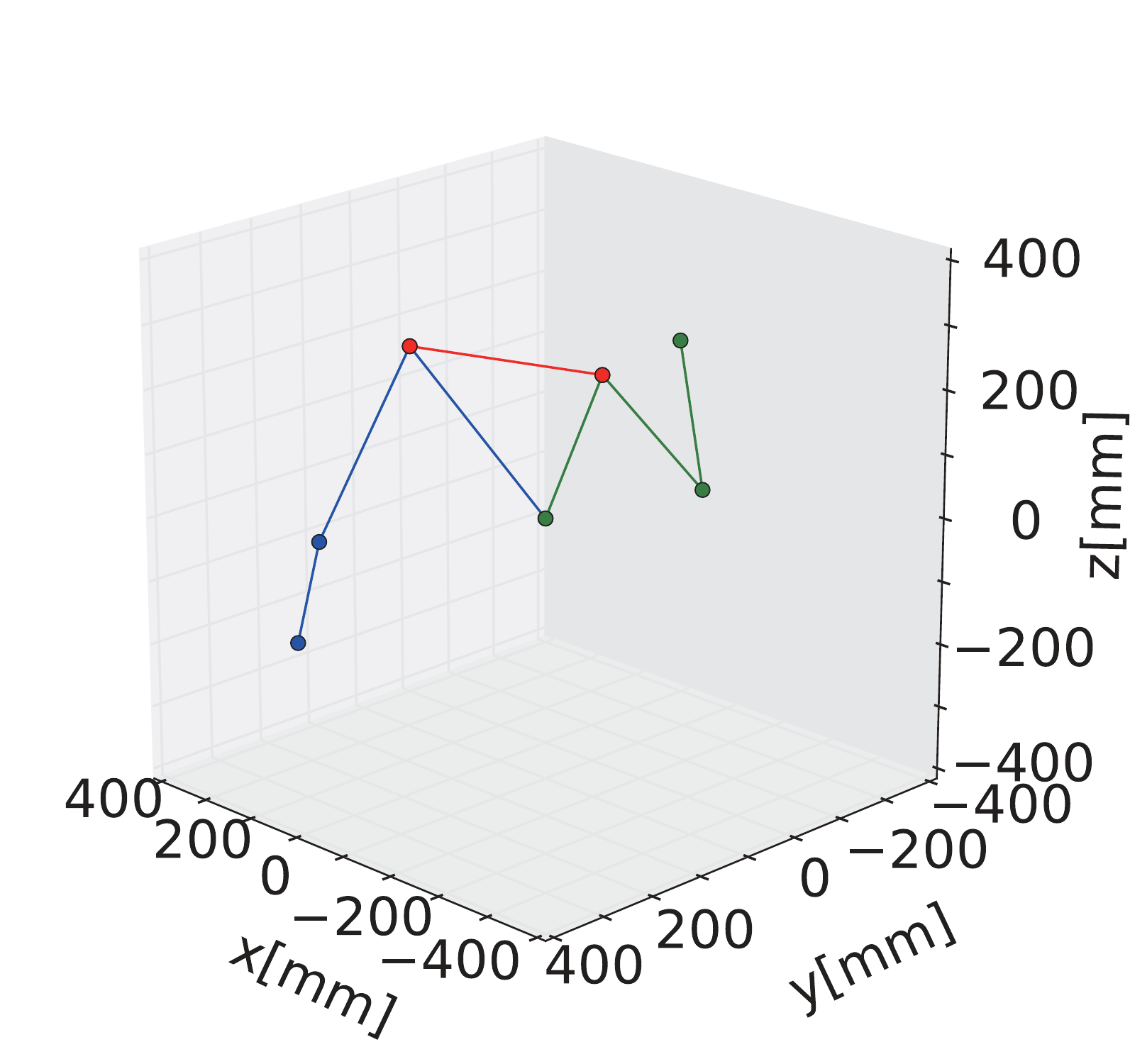} }
 \subfigure{\includegraphics[clip, width=0.31\columnwidth]{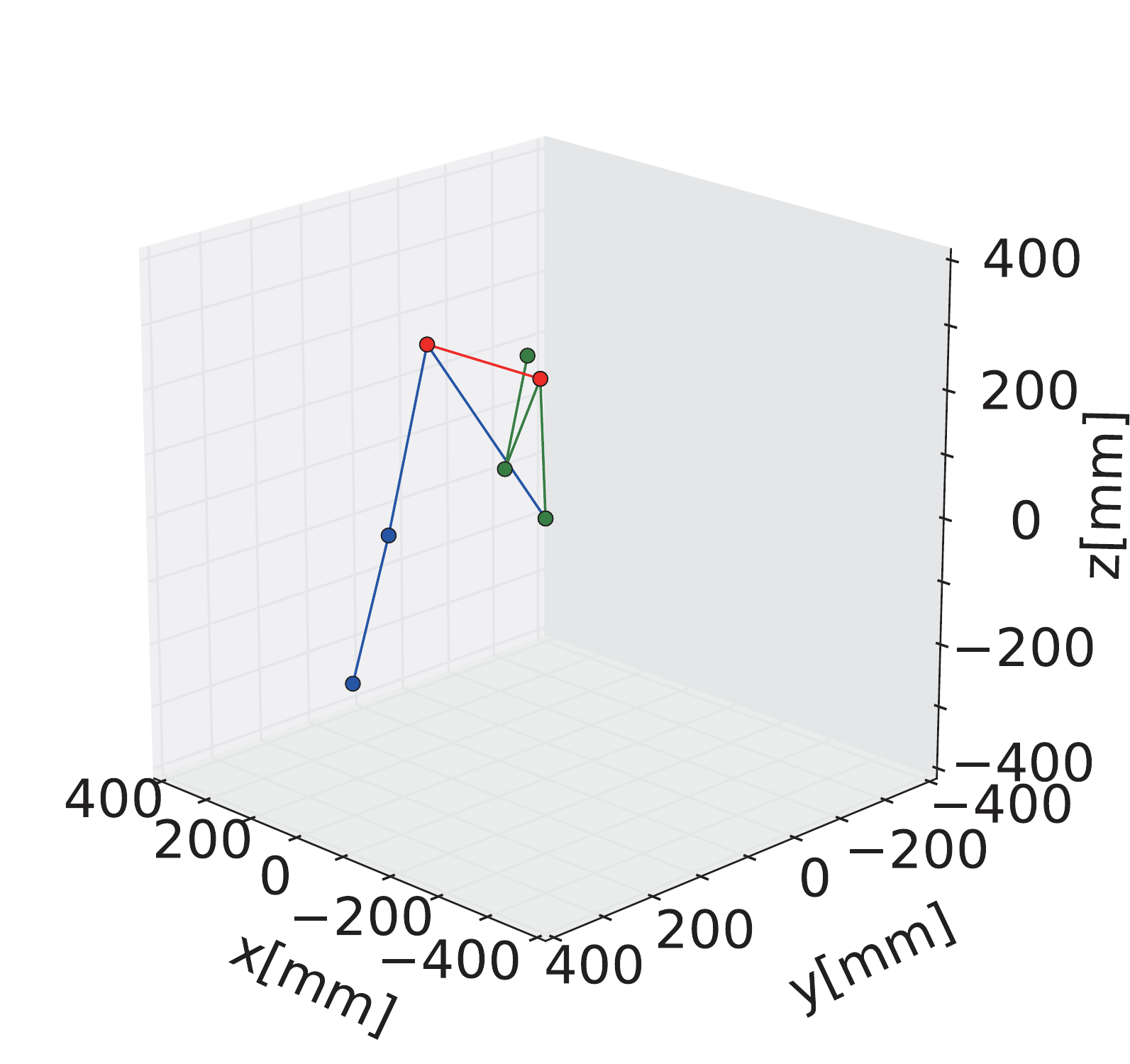} }
 \subfigure{\includegraphics[clip, width=0.31\columnwidth]{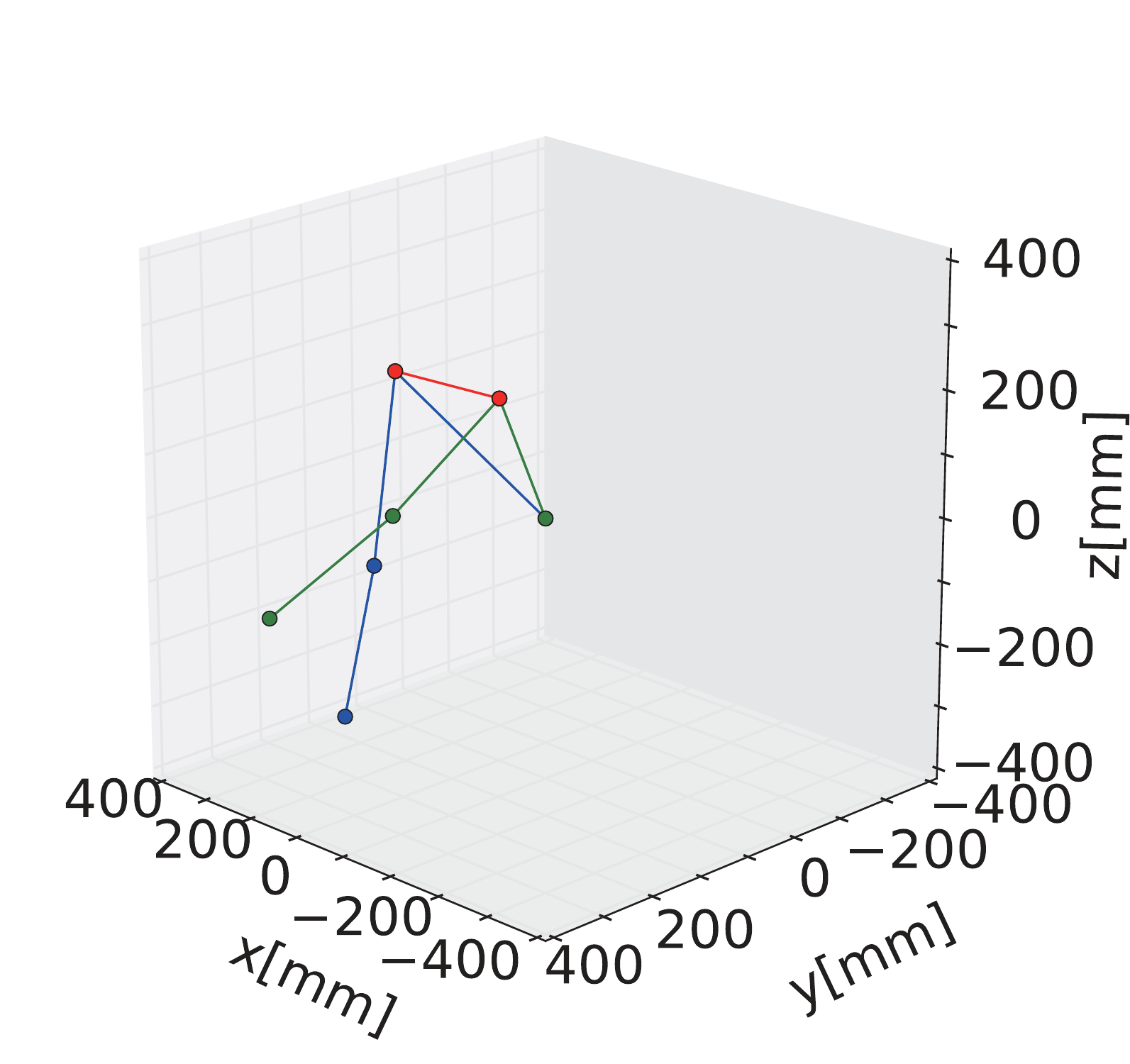} }
  \caption{Throwing the ball overhand}
 \label{fig:uenage}
\end{figure}

\begin{figure}[tbp]
 \centering
 \subfigure{\includegraphics[clip, width=0.31\columnwidth]{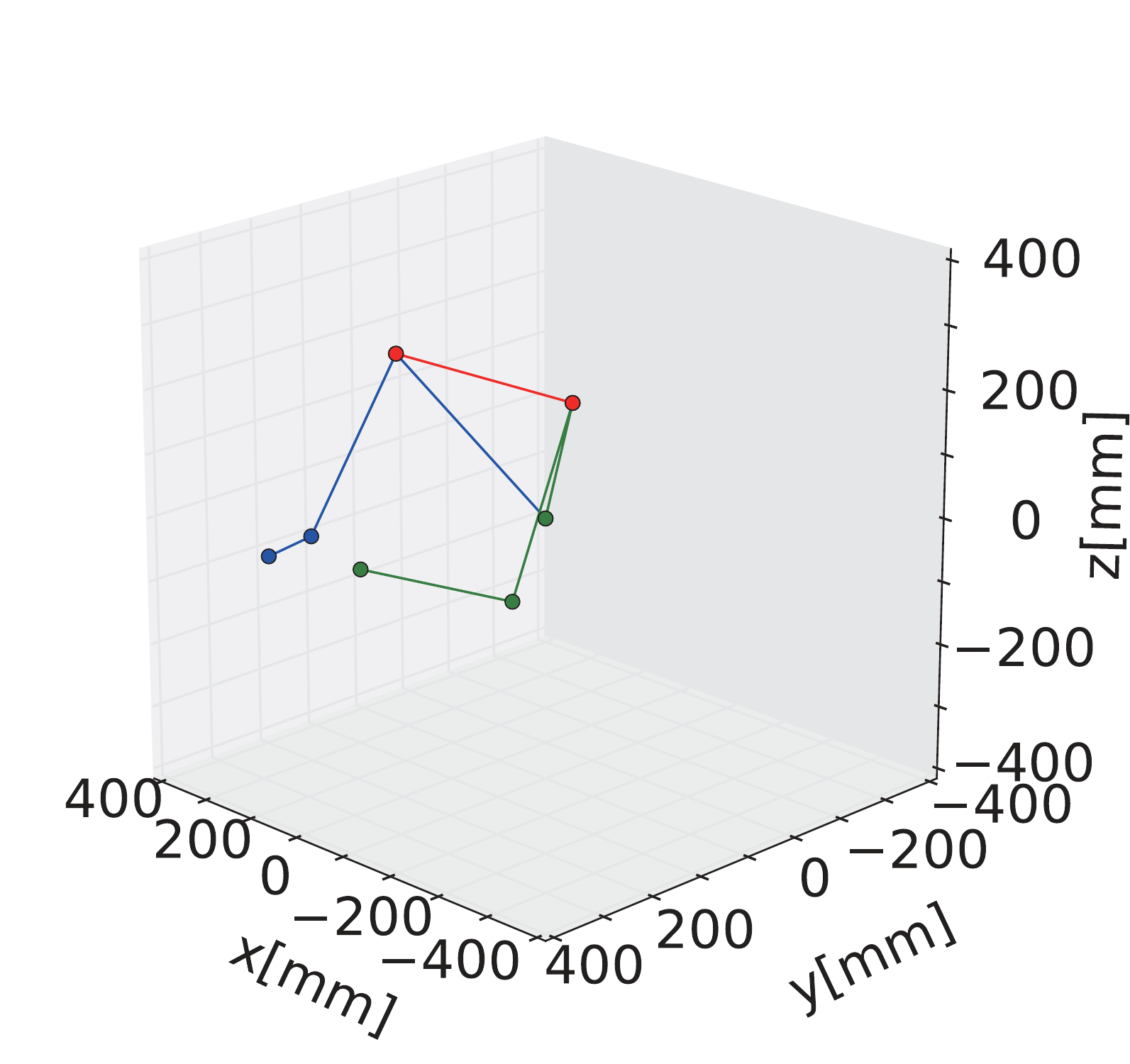} }
 \subfigure{\includegraphics[clip, width=0.31\columnwidth]{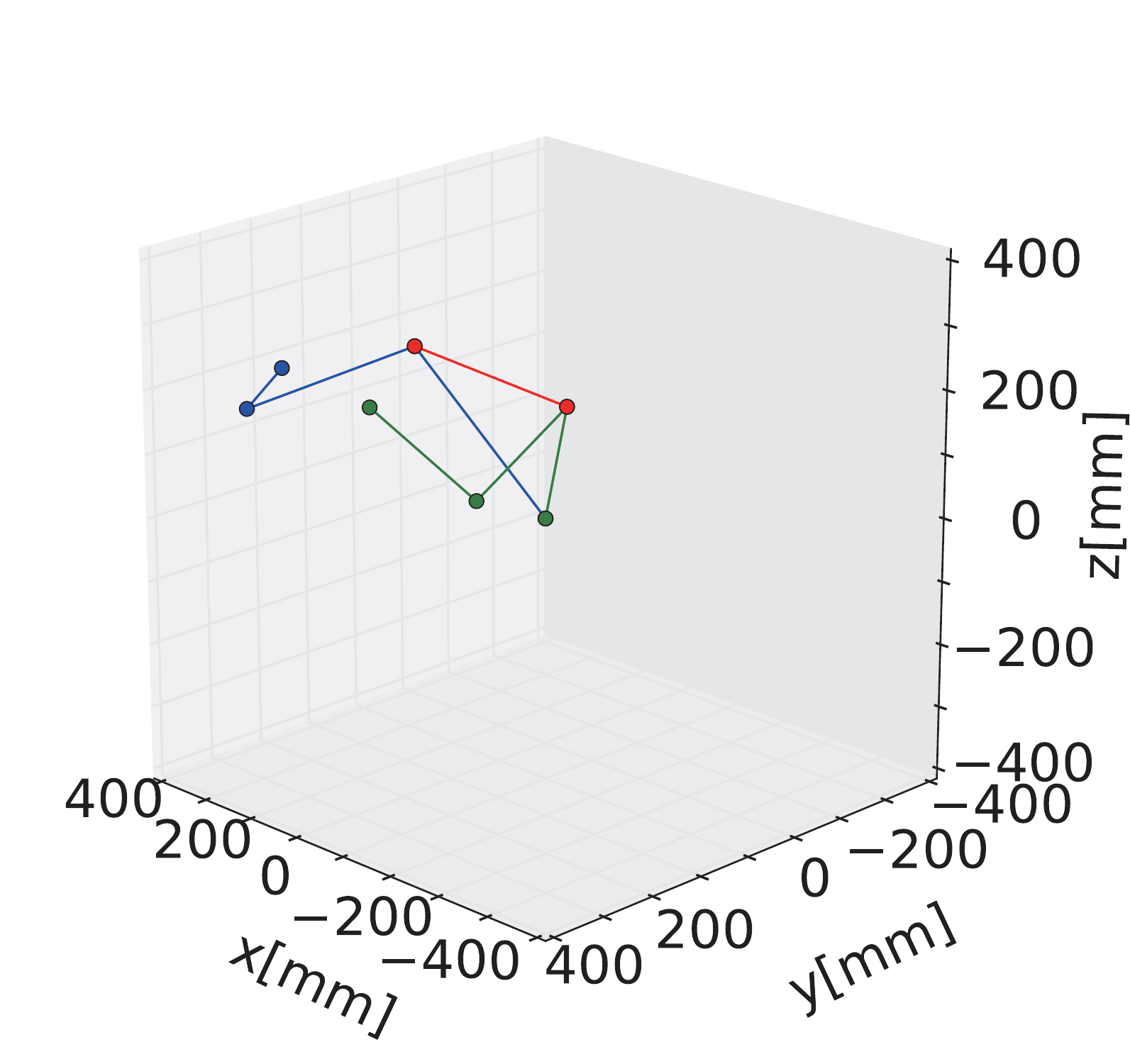} }
 \subfigure{\includegraphics[clip, width=0.31\columnwidth]{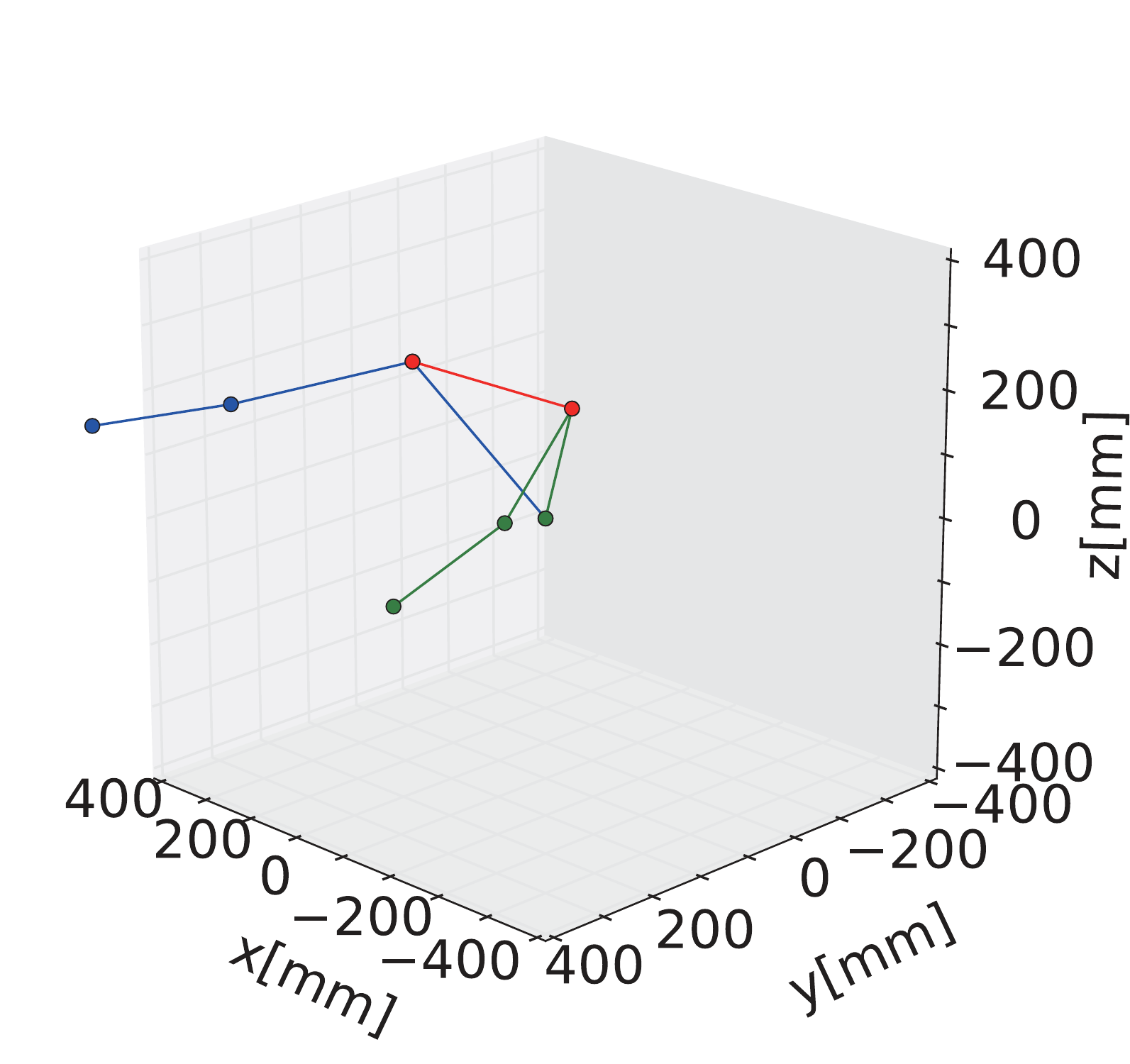} }
  \caption{Throwing the ball with both hands}
 \label{fig:ryoutenage}
\end{figure}

\begin{figure}[tbp]
 \centering
 \subfigure{\includegraphics[clip, width=0.31\columnwidth]{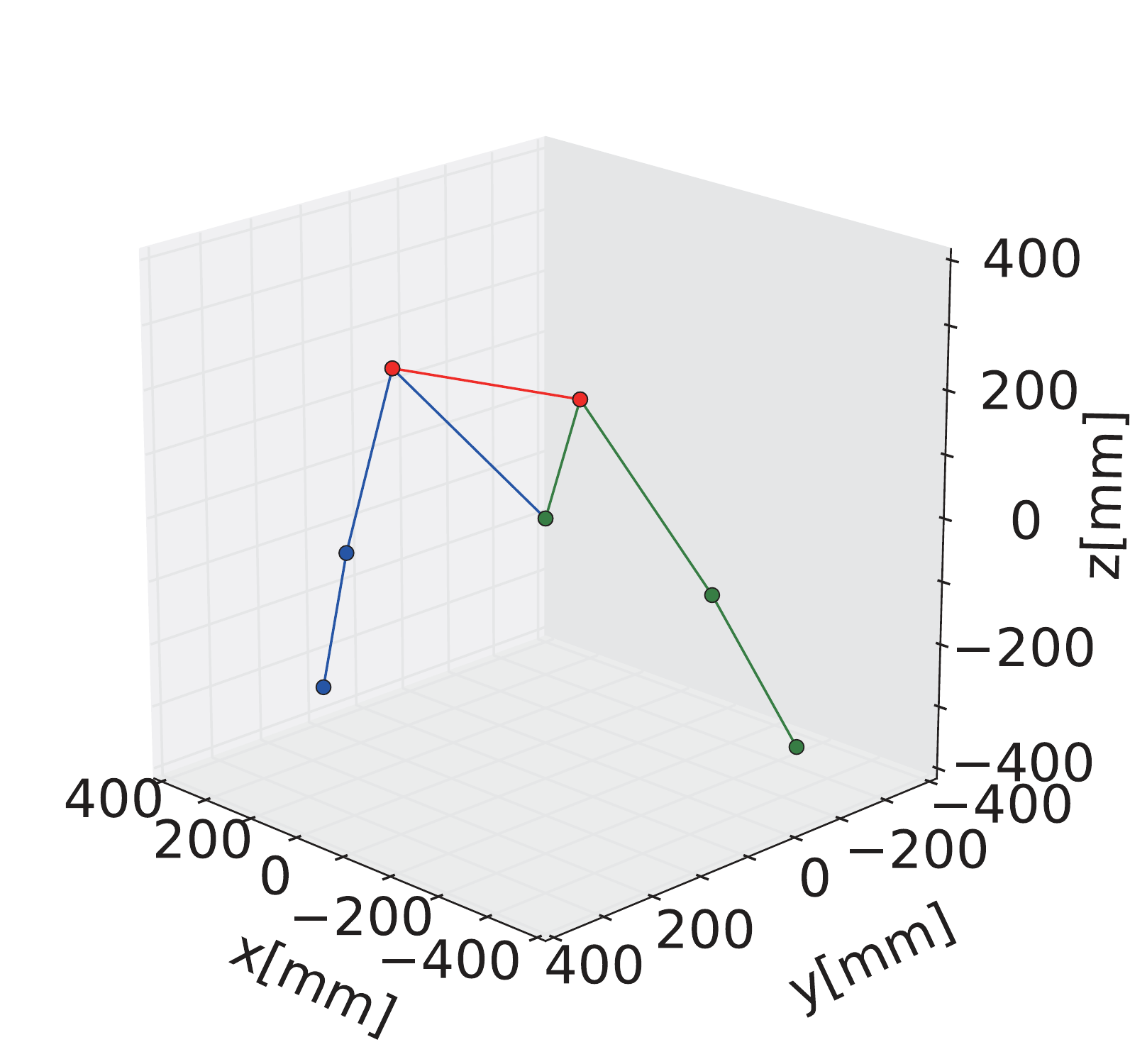} }
 \subfigure{\includegraphics[clip, width=0.31\columnwidth]{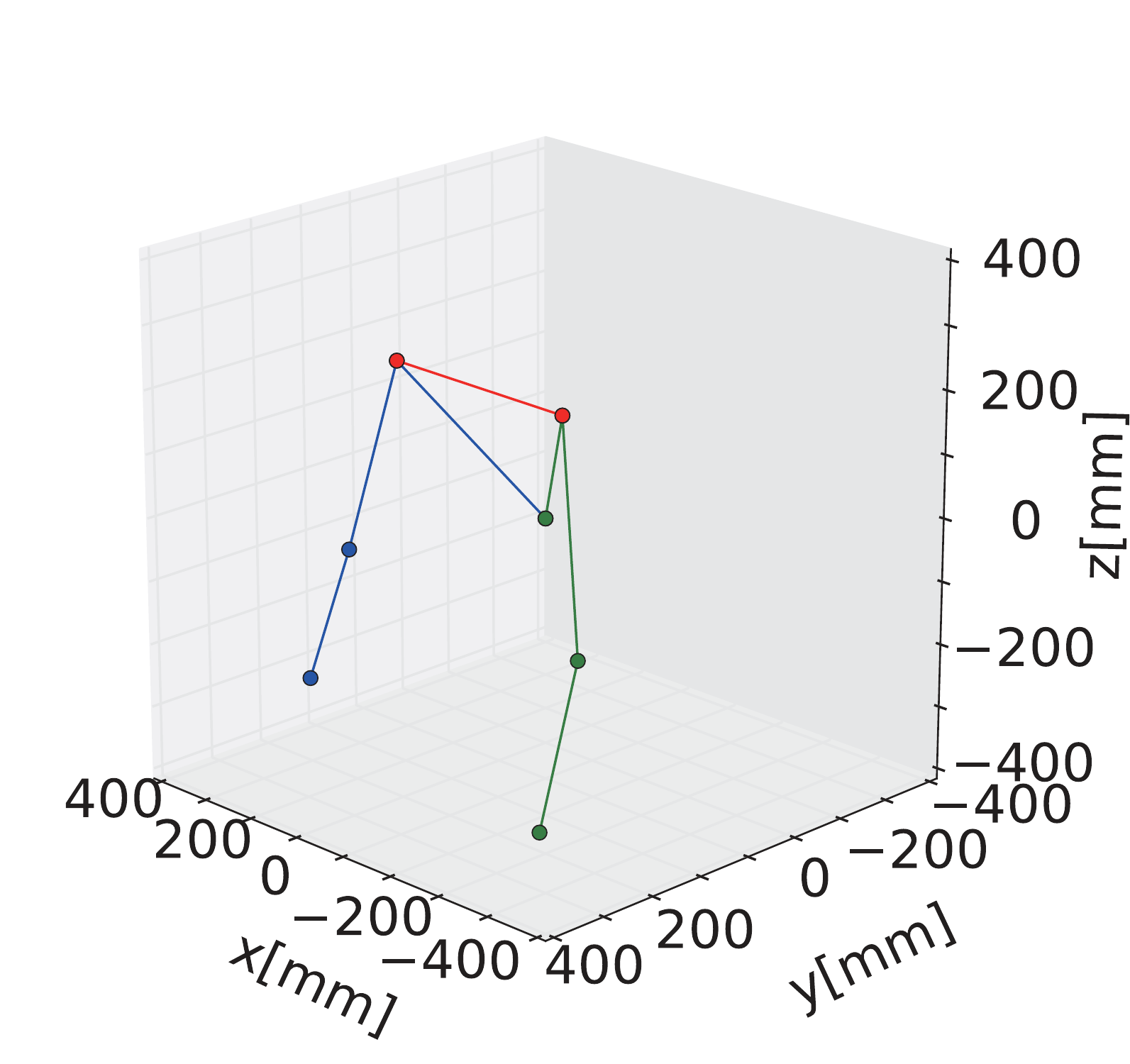} }
 \subfigure{\includegraphics[clip, width=0.31\columnwidth]{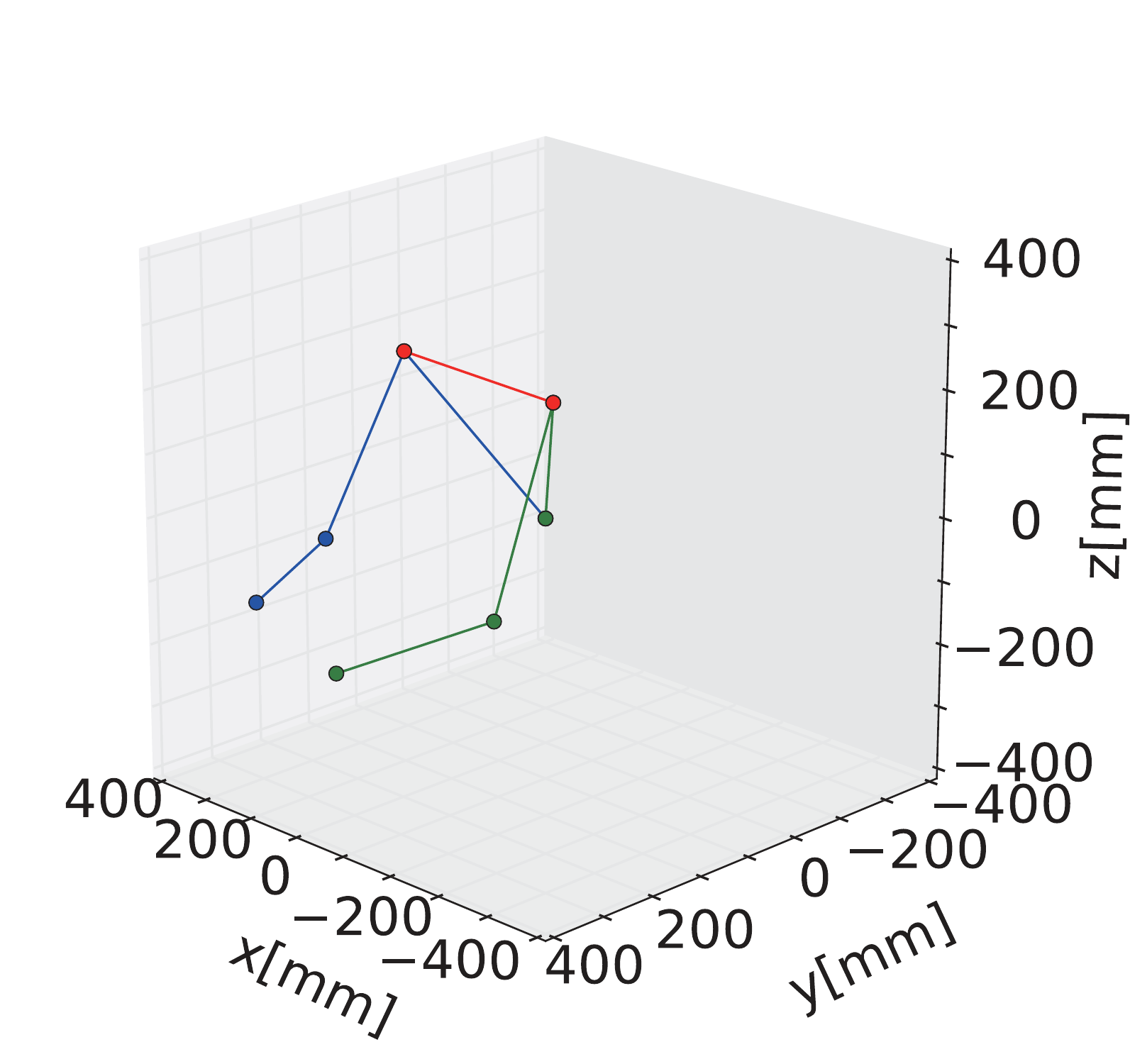} }
  \caption{Fake throw of the ball}
 \label{fig:faint}
\end{figure}

\begin{figure}[tbp]
 \centering
 \subfigure{\includegraphics[clip, width=0.31\columnwidth]{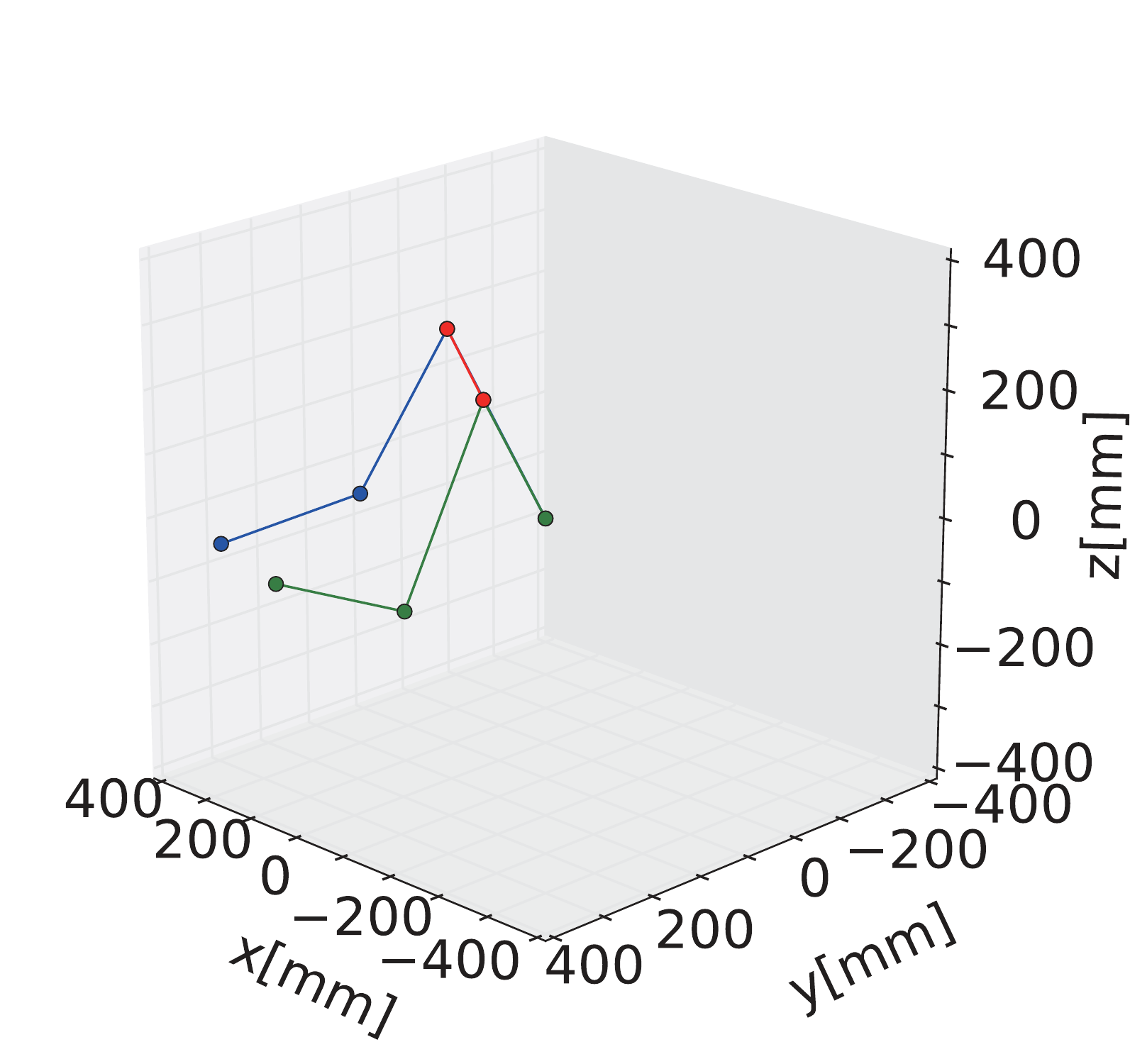} }
 \subfigure{\includegraphics[clip, width=0.31\columnwidth]{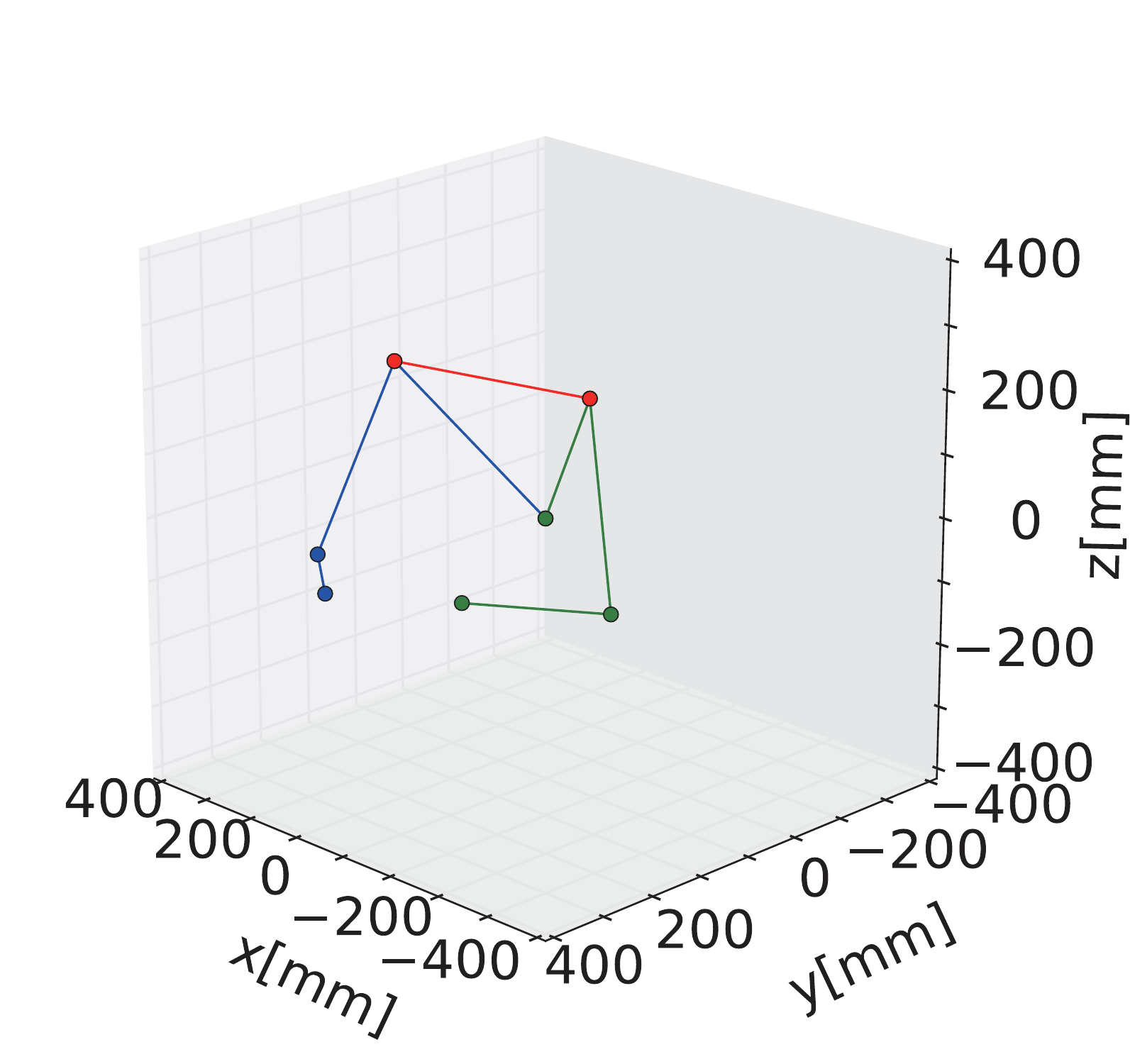} }
 \subfigure{\includegraphics[clip, width=0.31\columnwidth]{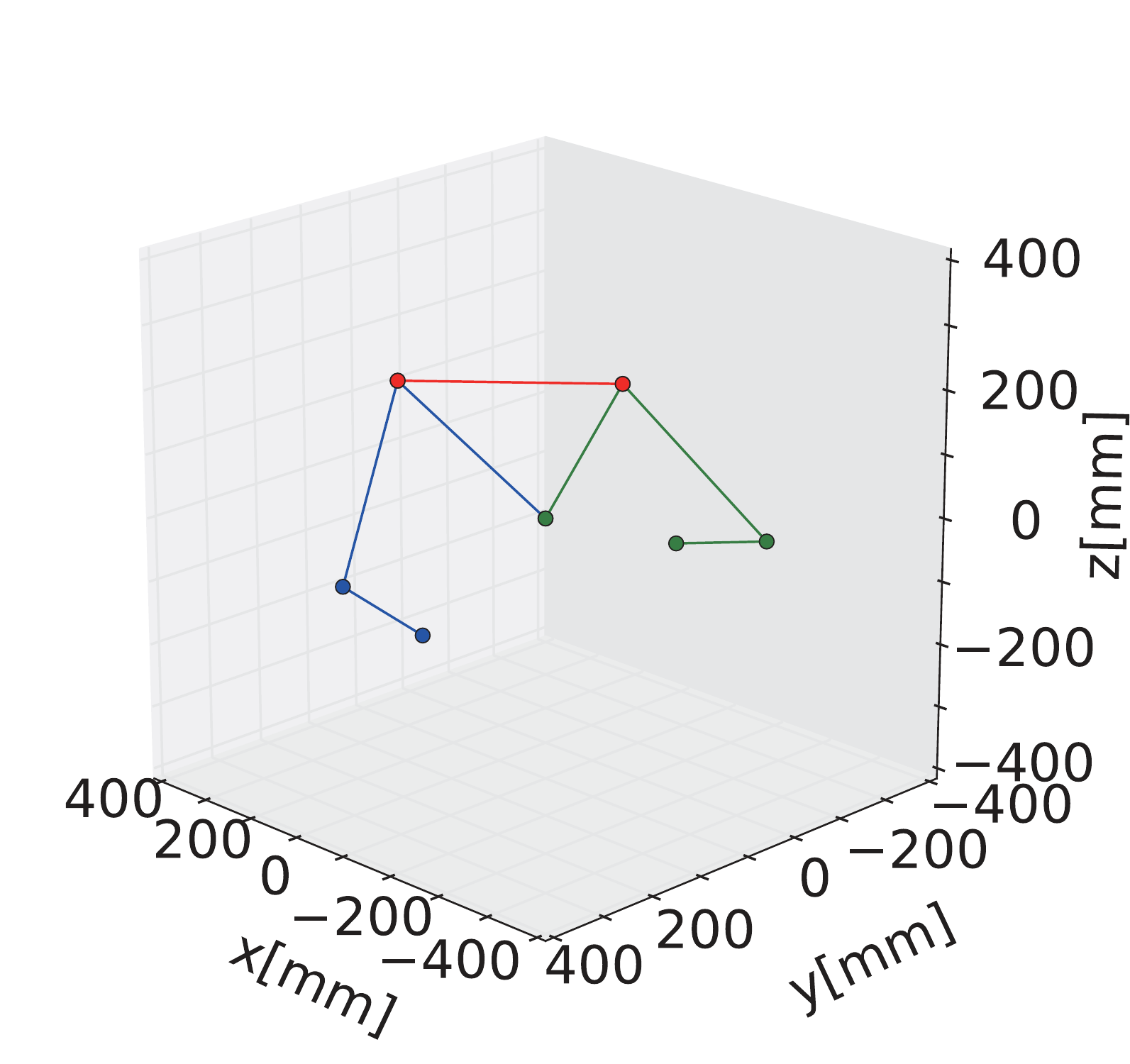} }
  \caption{Passing the ball from either hand to the opposite hand}
 \label{fig:yoko}
\end{figure}

We the instructed the subject holding the ball to randomly perform 
the following maneuvers (which are typical ball-handling behaviors).
\begin{itemize}
 \item Throw the ball underhand.
 \item Throw the ball overhand.
 \item Throw the ball with both hands.
 \item Fake a throw
 \item Pass the ball from either hand to the opposite hand
 \item Receive the ball with both hands.
\end{itemize}
The captured behaviors are displayed in \figref{fig:catch} - \figref{fig:yoko}
The opponent subject, who did not hold the ball,
was instructed to watch the ball and to catch it when thrown.


The feature vector was constructed by collecting all position vectors
$\mathrm{position}^m(t) \in \mathbb{R}^{21}$ as follows:
\begin{eqnarray}
 \mathrm{position}^m(t) &=&
  \begin{pmatrix}
   \mathrm{position}_{\mathrm{belly}_x}^m (t)\\
   \mathrm{position}_{\mathrm{belly}_y}^m (t)\\
   \mathrm{position}_{\mathrm{belly}_z}^m (t)\\
   \mathrm{position}_{\mathrm{left-shoulder}_x}^m (t)\\
   \vdots \\
   \mathrm{position}_{\mathrm{right-wrist}_z}^m (t)
  \end{pmatrix} \label{eq:position}
\end{eqnarray}
where $m \in \{ A,B \}$ represents subject A or B,
and t denotes time.

\vspace{0mm}
\subsubsection{Preprocessing}
\vspace{0mm}
For causal relationship detection, MPPCCA requires the past and present data on the result side, and the past data on the causal side.
In the proposed method, clustering is based on the biased correlation between the present information on the result side and the past information on the causal side, excluding the past information on the result side.
This approach is expected to reveal whether the clustering can be explained by one causal relation.

We denote the present and past information vectors on the result side by $A(t)$,
and $A_\mathrm{past}(t)$ respectively, and the past information vector on the causal side by $B_\mathrm{past}(T)$.
In the next subsection, we derive these three vectors from the position vector $\mathrm{position}^m (t)$ obtained from the motion capture data.

\vspace{0mm}
\subsubsection{Structure of Feature Vector}
\vspace{0mm}
To create the feature vector, we combine the position vector acquired by the motion capture with the velocity vector obtained from the position vector.
The velocity vector $\mathrm {velocity} ^ m (t) $ is obtained from the position vector $\mathrm{position}^m(t)$ as follows.
\begin{eqnarray}
	\mathrm{velocity}^m(t) & = & \mathrm{position}^m(t) - \mathrm{position}^m(t - 1)
\end{eqnarray}
As mentioned above, causality analysis uses the past information of the causal and result sides and the present information of the result side.
Clustering in the proposed method is based on PCCA.
The clusters are estimated by a GC-based criterion that determines whether the data can be predicted by a linear causal relationship.

\vspace{0mm}
\subsubsection{Feature vector}
\vspace{0mm}
The feature vector $\mathrm{feature}^m(t) \in \mathbb{R}^{42}$ combines
the $\mathrm{velocity}^m(t)$ and $\mathrm{position}^m(t)$ vectors:
\begin{eqnarray}
 \mathrm{feature}^m(t) &=&
  \begin{pmatrix}
   \mathrm{position}^m(t) \\
   \mathrm{velocity}^m(t)
  \end{pmatrix},
\end{eqnarray}
where
\begin{eqnarray}
 \mathrm{velocity}^m(t) &=& \mathrm{position}^m(t) - \mathrm{position}^m(t-1).
\end{eqnarray}

\vspace{0mm}
\subsubsection{Embedded time-series vector}
\vspace{0mm}
We now define the embedding vectors $\mathrm{embedding}^m(t) \in \mathbb{R}^{42\tau/s}$
of the feature vectors proposed in the previous section.
by \eqref{eq:embedding}
\begin{eqnarray}
 \mathrm{embedding}^m(t) &=&
  \begin{pmatrix}
   \mathrm{feature}^m(t-d)   \\
   \mathrm{feature}^m(t-d-s) \\
   \mathrm{feature}^m(t-d-2s) \\
   \vdots \\
   \mathrm{feature}^m(t-d-\tau+1) \\
  \end{pmatrix}\label{eq:embedding},
\end{eqnarray}
where $d,\tau, and s$ denote the delay-frame size,
the embedding-frame size and the sampling timeframe, respectively.
The delay frame size is

In this analysis, we set $d=10$, $s=5$ and $\tau=100$.
The feature vector $\mathrm{feature}^A(t)$, the
embedding vector $\mathrm {embedding}^A(t)$, and the embedding vector 
$\mathrm{embedding}^B(t)$
correspond to the present information on the result side, the past information on the result side, and the past information on the causal side, respectively.

\begin{sidewaystable}
 \centering
 \caption{Granger causality index of each cluster in the ball play experiment}
  \begin{tabular}{|c|c|c|c|} \hline
   k & GC index & Action by Subject A (Result side) & Action by Subject B (Causal side) \\
   \hline
   \hline
   0 & 1.55 & Receiving the ball & Throwing the ball by left overhand\\
     &      &                    & and underhand and by both hands. \\
   \hline
   1 & 1.59 & Receiving the ball then moving the ball & Throwing the ball \\
     &      & between right and left.                 & then moving arm down. \\
   \hline
   2 & 0.785& Meaningless movement & Receiving the ball then moving the ball \\
     &      &                      & between right and left. \\
   \hline
   3 & 0.331& Meaningless movement & Moving the ball randomly. \\\hline
   4 & 1.90 & Receiving the ball & Throwing the ball by right overhand and underhand. \\\hline
   5 & 0.730&  Passing the ball from either hand to other hand & No movement (finally receiving the ball) \\\hline
   6 & 0.989&  Passing the ball from either hand to other hand & No movement \\\hline
   7 & 1.11 & Any type of throwing & Receiving the ball \\\hline
   8 & 0.626& Meaningless movement & Moving the ball randomly \\\hline
   9 & 0.971& Any type of throwing & Receive the ball and move it to the right \\\hline
  \end{tabular}
 \label{table:real_granger}
\end{sidewaystable}

\vspace{0mm}
\subsubsection{Principle component analysis}
\vspace{0mm}
The embedding vectors proposed in the previous section are high-dimensional and their elements are highly correlated. Under these circumstances, the computation will become unstable.
Therefore, prior to causal pattern analysis by MPPCCA, we process the embedding vectors by PCA.
PCA reduces the dimensionality of the vectors and converts them into totally uncorrelated vectors.
In this analysis, we adopted the minimum basis in which the cumulative contribution ratio reaches 90\%.
The feature vectors and the embedding vectors were transformed with respect to this basis, providing an input vector to the MPPCCA.

\begin{eqnarray}
 A(t)               &=& \mathrm{PCA} \left( \mathrm{feature}^A(t) \right) \nonumber \\
 A_\mathrm{past}(t) &=& \mathrm{PCA} \left( \mathrm{embedding}^A(t)\right) \nonumber \\
 B_\mathrm{past}(t) &=& \mathrm{PCA} \left( \mathrm{embedding}^B(t)\right) \nonumber
\end{eqnarray}
To cluster the MPPCCA by causal patterns, we input MPPCCA with the present and past state vectors ($A(t)$ and $A_\mathrm{past}(t)$ respectively) of the result side, and with the 
past state vector $B_\mathrm{past}(t)$ of the causal side.

\vspace{0mm}
\subsection{Result}
\vspace{0mm}
The proposed method and k-means clustering were applied to the preprocessed data vectors in the previous section
with cluster size K = 10.
We then calculated the GC index of the generated clusters.
The causal patterns divided into the various clusters, and the GC indices of the clusters,
are described in \tabref{table:real_granger}.

We found three types of causal patterns with the three highest GC indeces,
as described below.
\begin{description}
 \item[Cluster 4 (GC=1.90)]~\\
At the result side (A), MPPCCA extracted the ball-catching patterns. .
At the causal side (B), it extracted the actions of throwing the ball with the right hand (both overhand and underhand). 
Different movement patterns (different feature amounts)
were classified in this cluster. Although the hand trajectories differed between overhand and underhand throwing,
the causal relationships among the forward speeds of the right hand at the causal side (B) were consistent with those of moving both
hands at the result side (A)..
 \item[Cluster 0 (GC=1.55)]~\\
At the result side (A), MPPCCA extracted the patterns of receiving the ball.
At the causal side (B), it extracted the actions of throwing the ball with the left hand (both overhand and underhand), and of throwing the ball with both hands. Right hand movements were not assigned to this cluster.
 \item[Cluster 1 (GC=1.59)]~\\
At the causal side (B), MPPCCA extracted the arms-down movement after the subject had thrown the ball. At
the result side (A), it detected the movement of the ball between left and right after the subject had received the ball.
We consider that this cluster differs from the above clusters 
because the behavioral pattern differs after receiving the ball,
although the causal relationships appear very similar to those of Clusters 4 and 0.
\end{description}

\begin{figure}[tbp]
 \centering
 \subfigure{\includegraphics[clip, width=11.0cm]{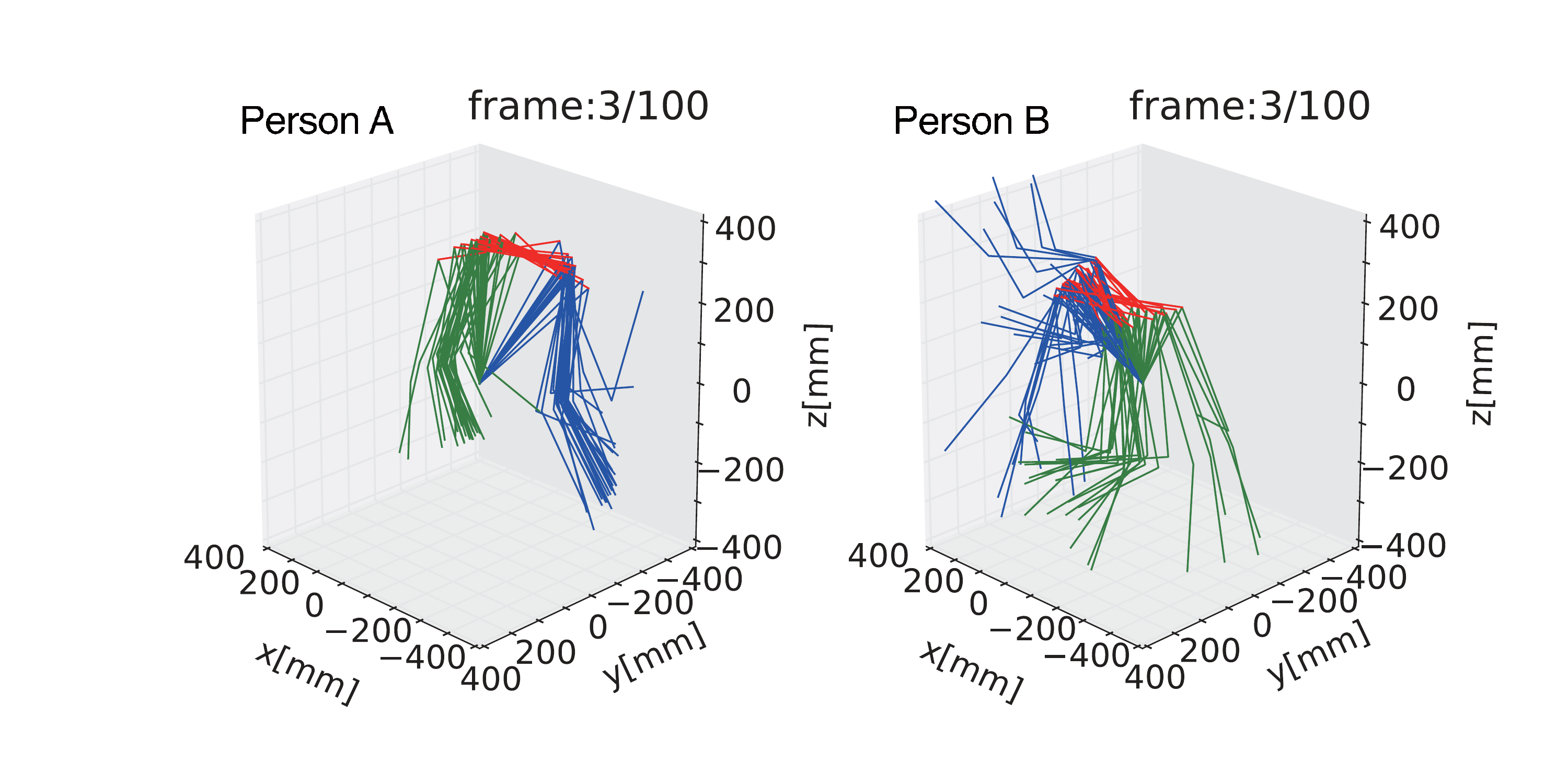} }
 \subfigure{\includegraphics[clip, width=11.0cm]{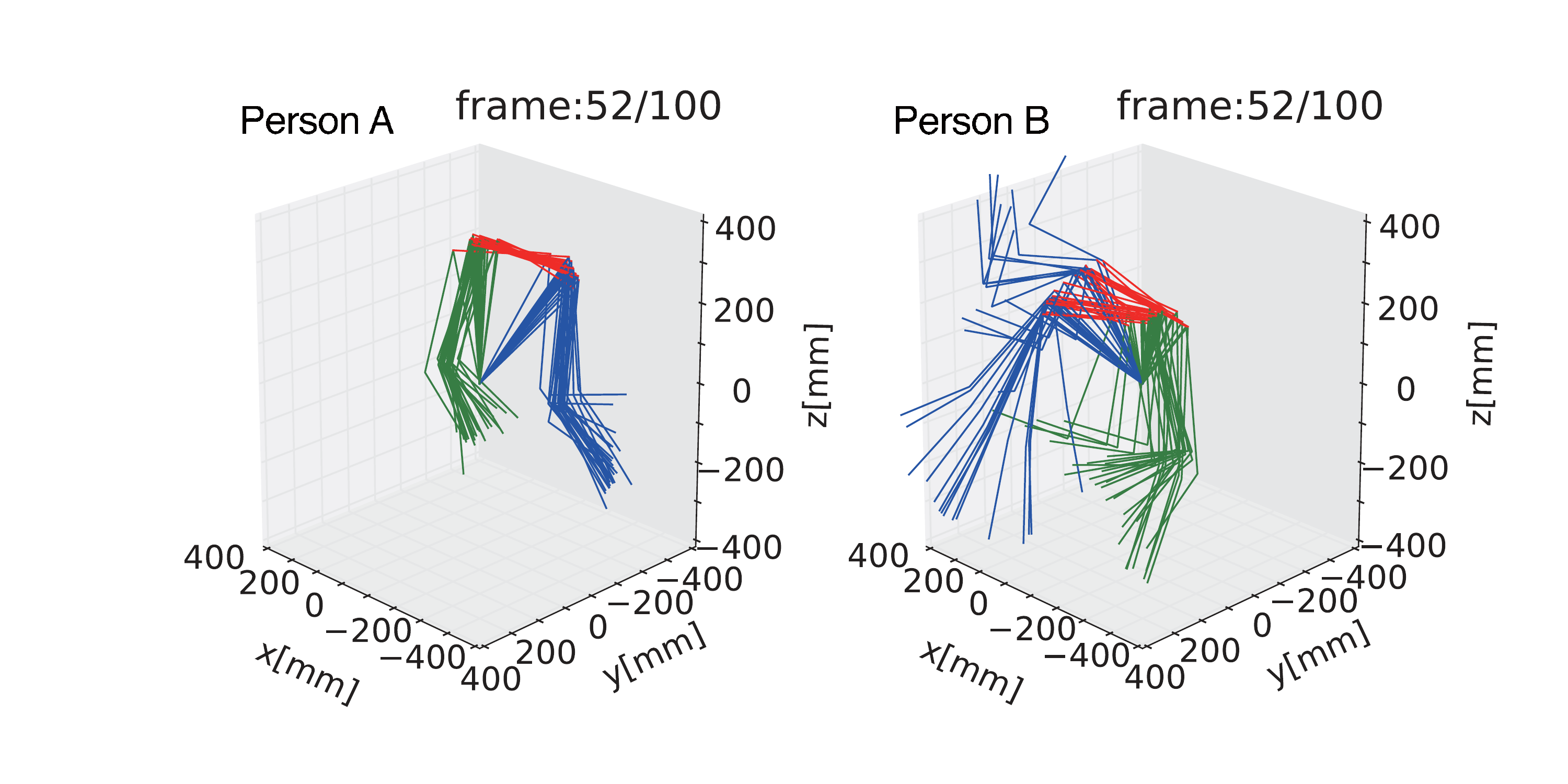} }
 \subfigure{\includegraphics[clip, width=11.0cm]{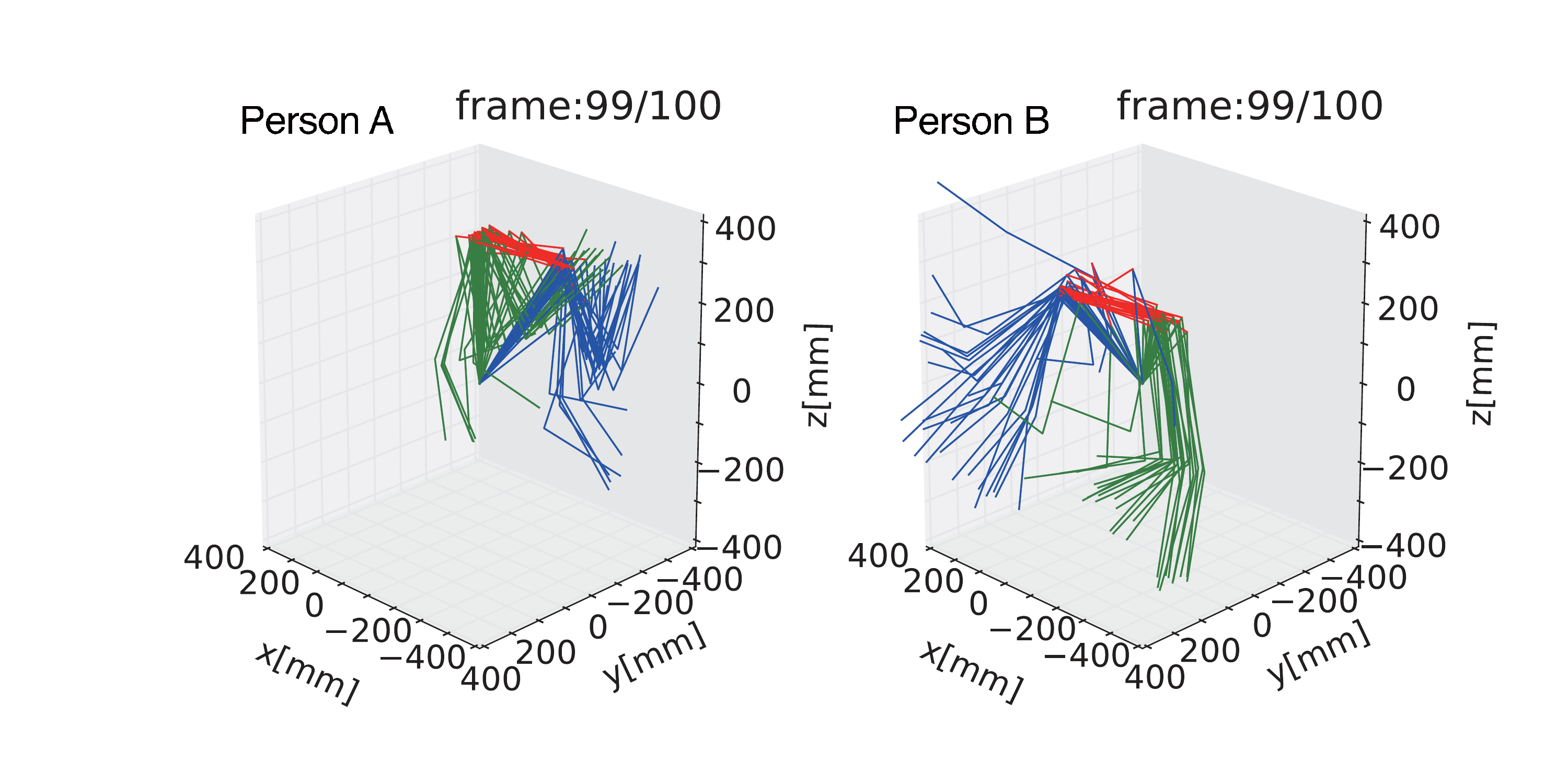} }
  \caption{$k=4$ A: Receiving the ball B: Throwing the ball by right overhand and right underhand.}
 \label{fig:cluster_4}
\end{figure}

\begin{figure}[tbp]
 \centering
 \subfigure{\includegraphics[clip, width=11.0cm]{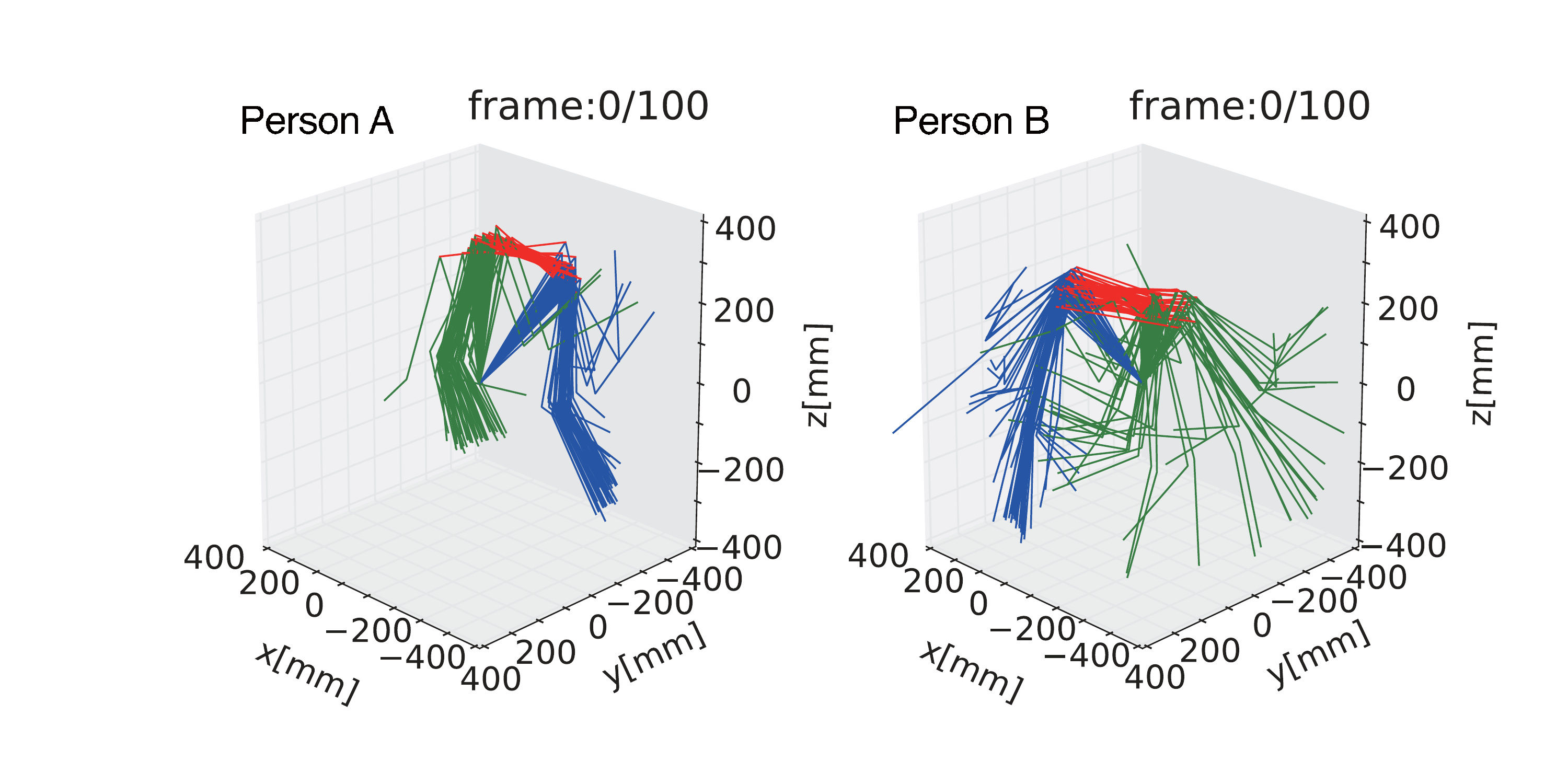} }
 \subfigure{\includegraphics[clip, width=11.0cm]{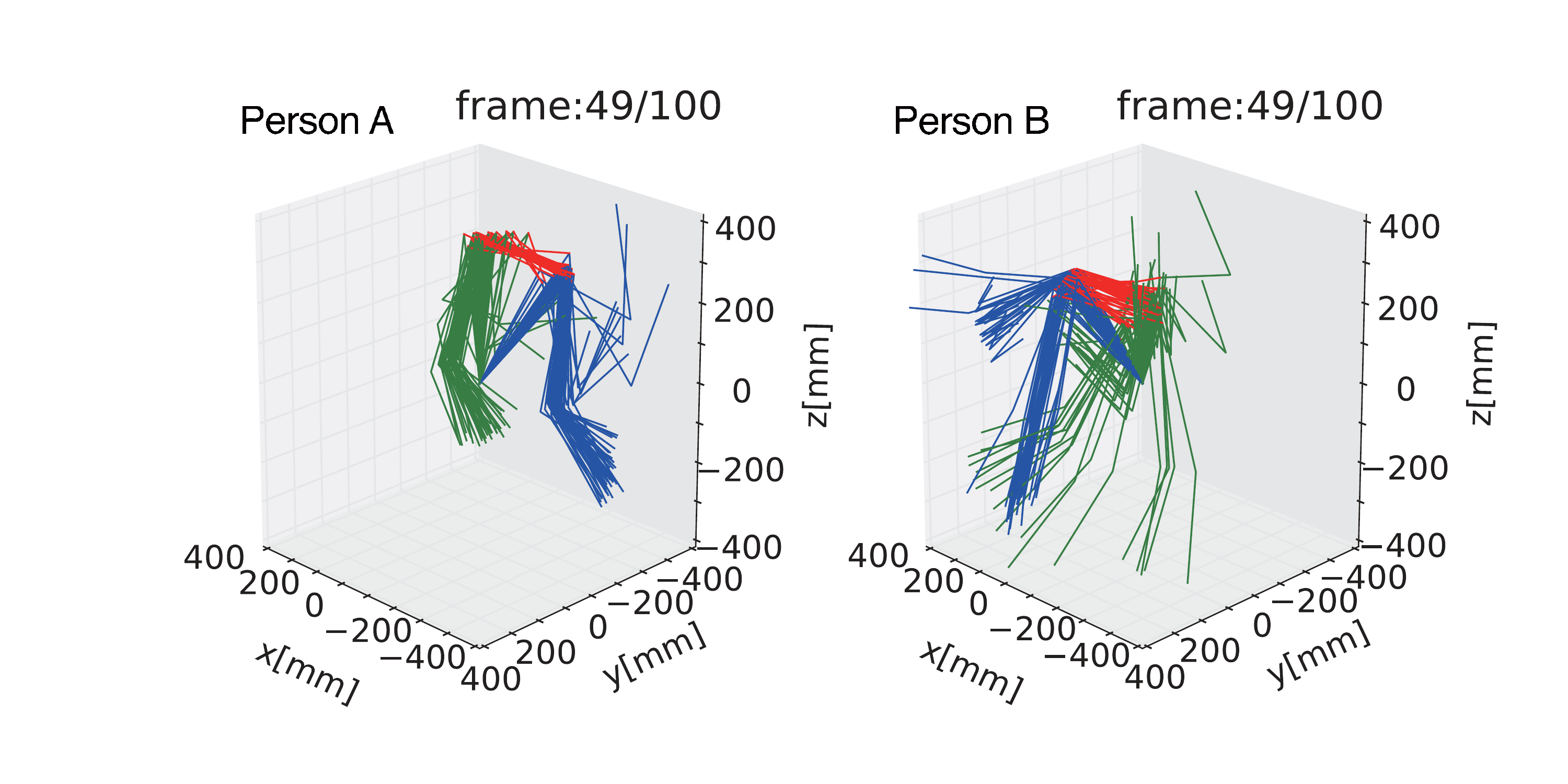} }
 \subfigure{\includegraphics[clip, width=11.0cm]{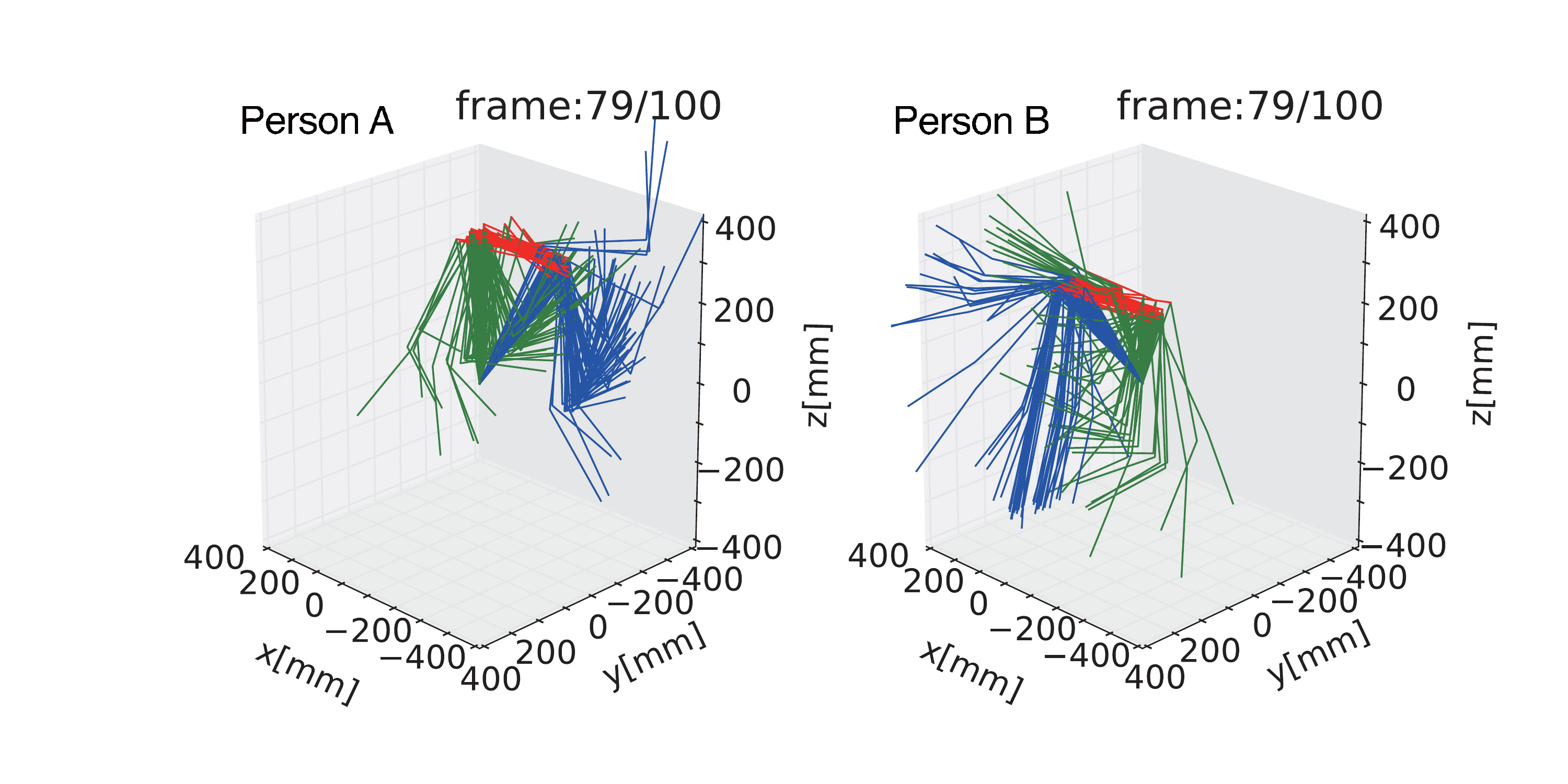} }
 \caption{$k=0$. A: Receiving the ball. B: Throwing the ball
 by left overhand, left underhand and both hands.}
 \label{fig:cluster_0}
\end{figure}

\begin{figure}[tbp]
 \centering
 \subfigure{\includegraphics[clip, width=11.0cm]{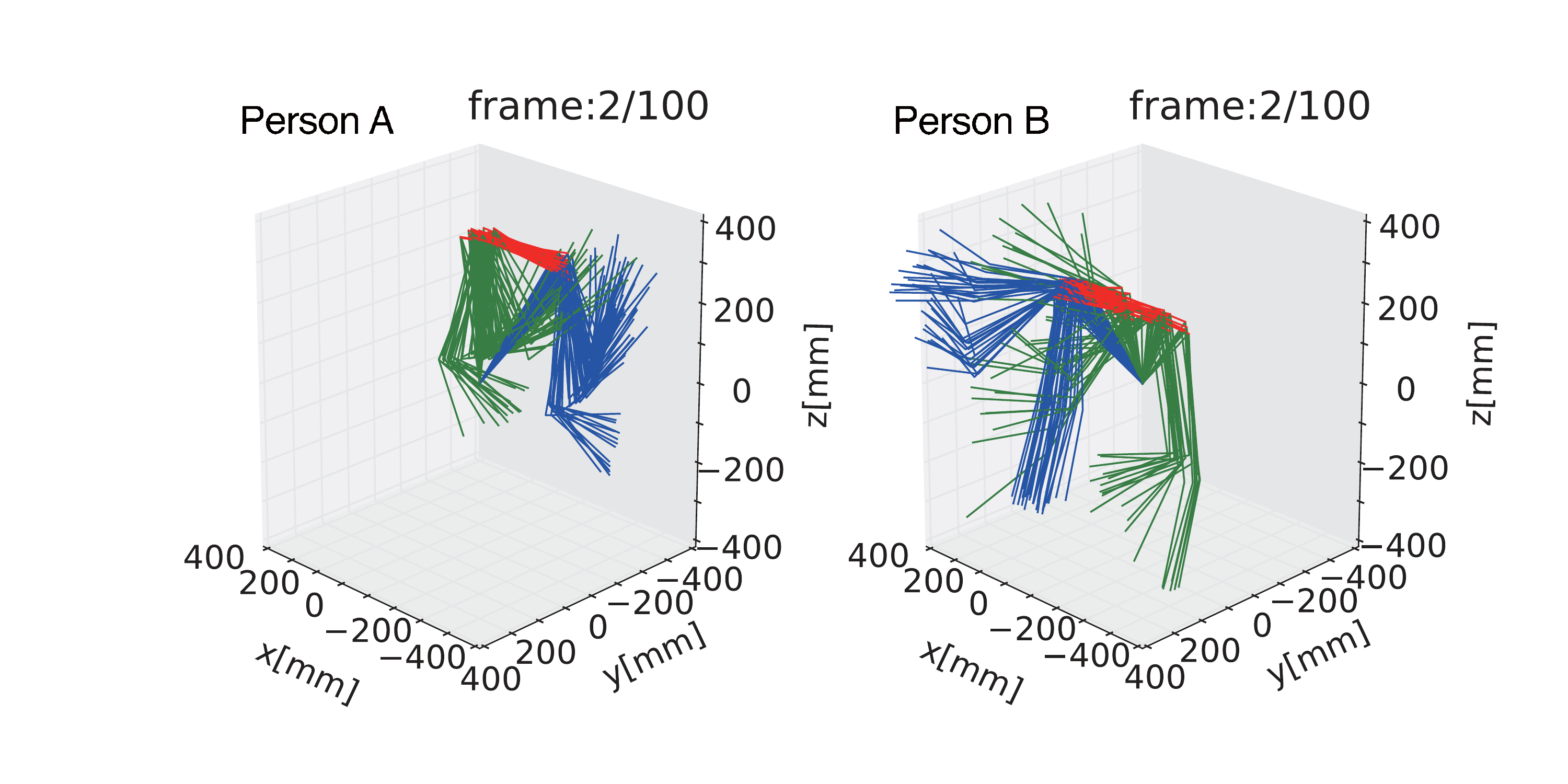} }
 \subfigure{\includegraphics[clip, width=11.0cm]{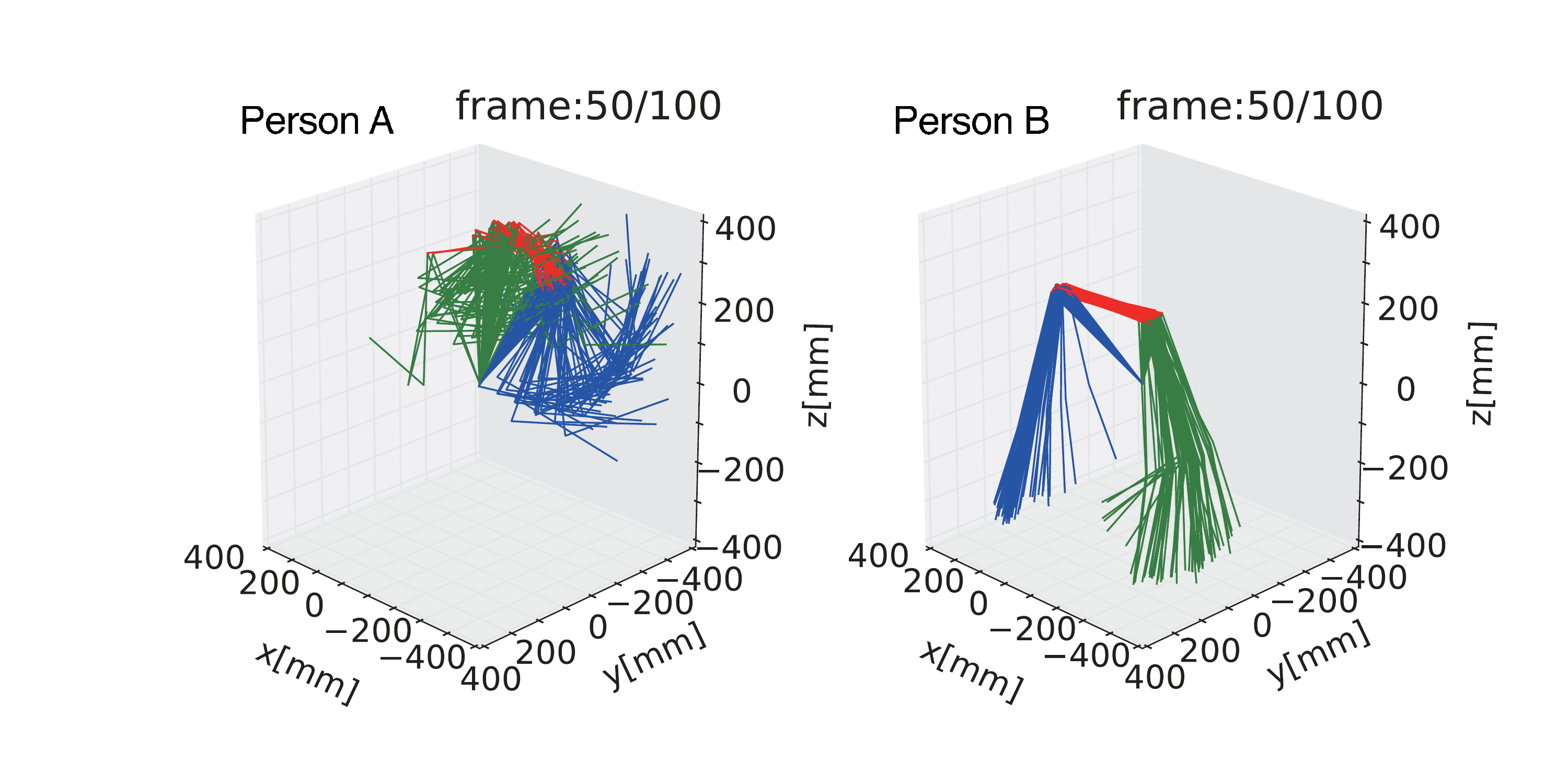} }
 \subfigure{\includegraphics[clip, width=11.0cm]{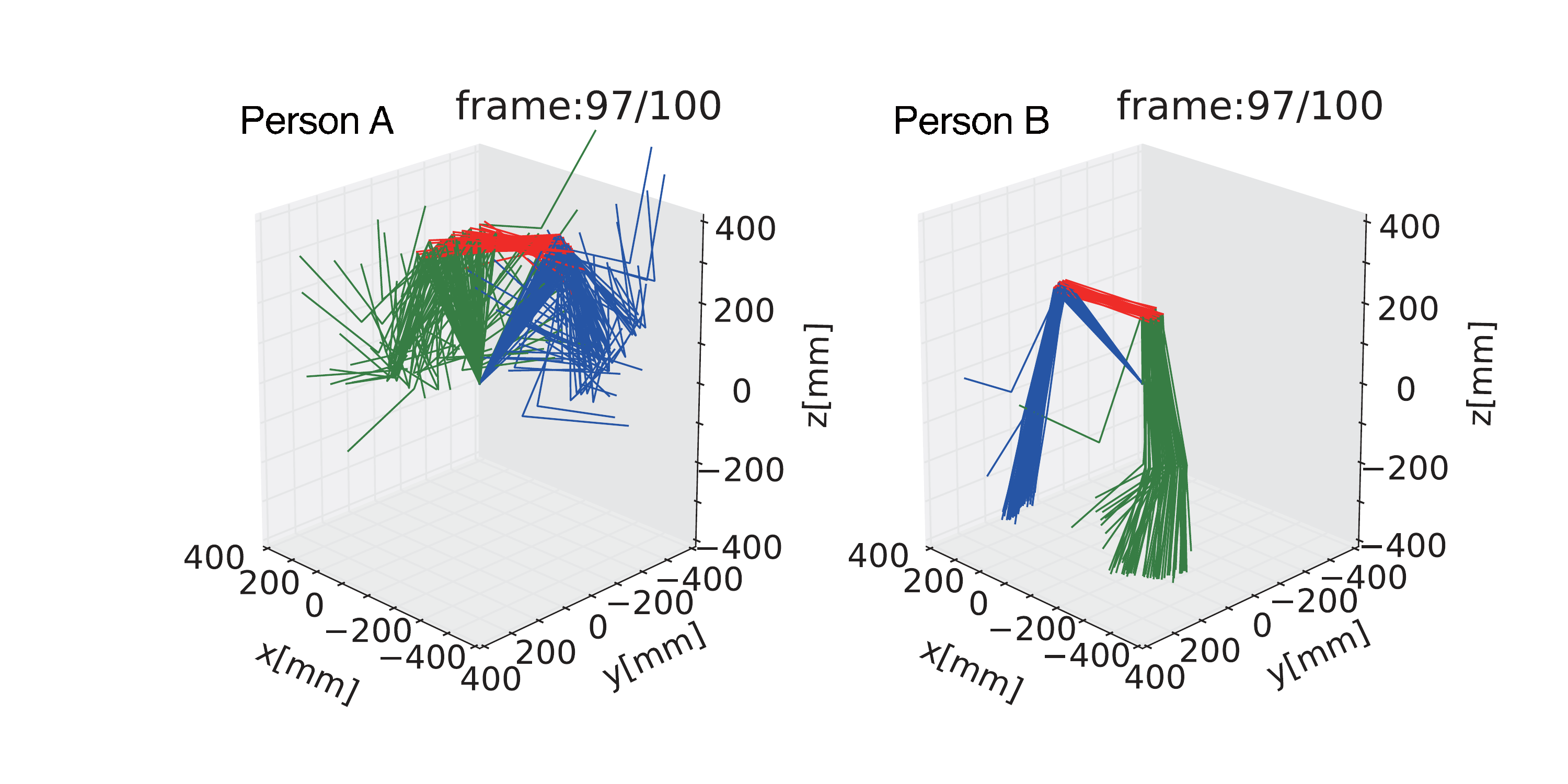} }
 \caption{$k=1$. A: Receiving the ball, then moving it to right or left
 	B: Throwing the ball, then moving both hands down.}
 \label{fig:cluster_1}
\end{figure}

The averages of the feature vectors in each cluster are not interpretable. Therefore, to visualize each cluster obtained in the proposed method,
we defined representative data in each cluster
as a feature vector on the mid-point of the successive time-series related to that cluster.
No intermediate behaviors such as horizontal throwing
were observed between overhand throwing and underhand tossing in the data set.
Therefore, in this analysis,
we selected representative data to visualize the action pattern extracted in each cluster.
Here, the representative data were the central data
in the time-series of data in each cluster.
\figref{fig:cluster_4}, \ref{fig:cluster_1} and \ref{fig:cluster_0} show
the extracted causal patterns between the two subjects
as representative data of the three clusters.

As we expected,
the patterns with higher GC indices were
combined behaviors with throwing actions in the causal side
and receiving actions in the result side.
However, the patterns with the fourth GC index value ($k=7$) are
the opposite behaviors regarding throwing and receiving.
There are two possibilities that can explain the result of the analysis.
The first is that the actions of the subject in causal side evoke
the throwing actions by the subject in result side.
The second is that the method is unstable to estimate parameters
in the case that .

\vspace{0mm}
\subsection{Discussion}
\vspace{0mm}
Proposed method can extracted the scenes where the ball is thrown as clusters with high GC indices.
Among the extracted clusters,
the right hand throw and the left hand throw were distinguished
while the overhand throwing and the underhand tossing were not distinguished.
In order to disscuss the reason why such a clustering result was obtained by our proposed method,
we should consider the mapping spaces for which each correlation coefficients are found.
In the partial canonical correlation analysis,
we derive a mapping space that maximizes the correlation coefficient in the mapping space.
The mapping space is obtained for each cluster and
data points are classified into respective clusters.

In the case of ball throwing and receiving,
to maximize the correlation,
the method should determine the axis picking up the horizontal velocity of a arm throwing in causal side
because the vertical position and velocity do not affect the receiving action in result side pretty much.
Also, the more velocity of horizontal axis of arm throwing the ball,
the more rapid response of receiving action by the opponent.

On the other hand,
right-handed throwing and left-handed throwing were distinguished in terms of the clusters where they belong.
This is unlike the case of the overhand throwing and the underhand tossing.
As described above, since the position and velocity in the anteroposterior direction in the any one-handed throwing action are common,
the linear mapping try to compress the vertical variabilities.
However, among right-handed and left-handed throwing actions,
there is no common elements in the feature
unless we introduce nonlinear feature such as multivariate polynomial combining the right and the left variables.

In the future study,
we will make the estimation of the parameters stable
and will construct a model to determine the cluster size automatically.
The former can be realized by MAP estimation with appropriate prior distribution of the parameters.
The latter can be realized by non-parametric Bayesian modeling.

\vspace{0mm}
\section{Conclusion}
\vspace{0mm}
We proposed our MPPCCA model for extracting causal patterns from time-series data, and evaluated it in experiments on synthetic and real datasets.
MPPCCA correctly clustered the data in terms of causal relationships
rather than simultaneous correlations,
and extracted the causal patterns in real data without supervising signals.

\section*{Resource}
This article was accepted by and presented
in IEEE Data Science and Advanced Analytics 2017 (DSAA2017) in Tokyo.
In the case that anyone refer this article,
please put the following reference information.
\begin{quote}
Hiroki Mori, Keisuke Kawano and Hiroki Yokoyama,
``Causal Patterns: Extraction of multiple causal relationships
by Mixture of Probabilistic Partial Canonical Correlation Analysis,''
IEEE Data Science and Advanced Analytics 2017, pp.744--754, 2017
 (DOI 10.1109/DSAA.2017.60)
\end{quote}

We open python source code to examine MPPCCA at\\
\url{https://github.com/kskkwn/mppcca} .\\
We hope that the reader use the code to analyze your problem.

\bibliographystyle{plain}
\bibliography{ms}

\begin{thebibliography}{10}

\bibitem{barnett2009granger}
Lionel Barnett, Adam Barrett, and Anil Seth.
\newblock Granger causality and transfer entropy are equivalent for gaussian
  variables.
\newblock {\em Physical review letters}, 103(23):238701, 2009.

\bibitem{bishop:2006:PRML}
Christopher~M. Bishop.
\newblock {\em Pattern Recognition and Machine Learning}.
\newblock Springer, 2006.

\bibitem{fujita2010identification}
Andr{\'e} Fujita, Joao~Ricardo Sato, Kaname Kojima, Luciana~Rodrigues Gomes,
  Masao Nagasaki, Mari~Cleide Sogayar, and Satoru Miyano.
\newblock Identification of granger causality between gene sets.
\newblock {\em Journal of Bioinformatics and Computational Biology},
  8(04):679--701, 2010.

\bibitem{ladroue2009beyond}
Christophe Ladroue, Shuixia Guo, Keith Kendrick, and Jianfeng Feng.
\newblock Beyond element-wise interactions: identifying complex interactions in
  biological processes.
\newblock {\em PLoS One}, 4(9):e6899, 2009.

\bibitem{mukuta2014probabilistic}
Yusuke Mukuta and Tatsuya Harada.
\newblock Probabilistic {P}artial {C}anonical {C}orrelation {A}nalysis.
\newblock In {\em Proceedings of the 31st International Conference on Machine
  Learning}, pages 1449--1457, 2014.

\bibitem{rao1969partial}
B.~Raja Rao.
\newblock Partial canonical correlations.
\newblock {\em Trabajos de estadistica y de investigaci{\'o}n operativa},
  20(2):211--219, 1969.

\bibitem{roebroeck2005mapping}
Alard Roebroeck, Elia Formisano, and Rainer Goebel.
\newblock Mapping directed influence over the brain using granger causality and
  {fMRI}.
\newblock {\em Neuroimage}, 25(1):230--242, 2005.

\bibitem{sato2010analyzing}
Jo{\~a}o~R. Sato, Andr{\'e} Fujita, Elisson~F. Cardoso, Carlos~E. Thomaz,
  Michael~J. Brammer, and Edson Amaro~Jr.
\newblock Analyzing the connectivity between regions of interest: an approach
  based on cluster granger causality for {fMRI} data analysis.
\newblock {\em Neuroimage}, 52(4):1444--1455, 2010.

\bibitem{stock1999business}
James~H. Stock and Mark~W. Watson.
\newblock Business cycle fluctuations in us macroeconomic time series.
\newblock {\em Handbook of macroeconomics}, 1:3--64, 1999.

\bibitem{yamashita2011causal}
Yuya Yamashita, Tatsuya Harada, and Yasuo Kuniyoshi.
\newblock Causal flow.
\newblock In {\em IEEE International Conference on Multimedia and Expo}, pages
  1--6. IEEE, 2011.

\end{thebibliography}

\end{document}